\documentclass[12pt, headings=standardclasses]{scrartcl} 
\usepackage{graphicx}
\usepackage{amsmath,amssymb,amsfonts}%
\usepackage{booktabs}
\usepackage{appendix}
\usepackage{url}
\usepackage{pdflscape}
\usepackage{setspace}
\usepackage{enumerate}
\usepackage{rotating}
\usepackage[hidelinks]{hyperref}
\usepackage{enumitem}
\usepackage{longtable}
\usepackage{arydshln}
\usepackage{placeins}
\usepackage{adjustbox}
\usepackage[round]{natbib}
\title{The kinematics of global warming: Semiparametric analysis of warming level, rate, and acceleration}

\author{Mikkel Bennedsen\thanks{
Aarhus Center for Econometrics (ACE) and Center for Research in Energy: Economics and Markets (CoRE),
Aarhus University,
Universitetsbyen 51,
8000 Aarhus C, Denmark
E-mail:\
\href{mailto:mbennedsen@econ.au.dk}{\nolinkurl{mbennedsen@econ.au.dk}}.
},
J. Eduardo Vera-Valdés\thanks{
Department of Mathematical Sciences,
Aalborg University,
Skjernvej 4A,
9210 Aalborg Øst, Denmark, and CoRE,
E-mail:\
\href{mailto:eduardo@math.aau.dk}{\nolinkurl{eduardo@math.aau.dk}}.
}}

\begin{document}

\maketitle

\vspace{-1.0cm}
\begin{abstract}
We develop a semiparametric kinematic state-space framework for estimating the latent dynamics of global warming from noisy monthly temperature observations. The framework estimates the warming level, warming rate, and warming acceleration jointly, while incorporating covariates, measurement error, serial dependence, and time-varying error variance directly in the model.  Applying the framework to five major global temperature records, and comparing it with nonparametric kernel and parametric benchmark estimators, we show that the inferred recent warming dynamics depend crucially on the parametric restrictions imposed on the warming path. The current underlying warming level is robustly estimated across datasets and methods, with estimates close to \(1.42\,^{\circ}\mathrm{C}\) above pre-industrial levels. Estimates of the current warming rate differ materially across estimator classes: flexible local estimators imply substantially higher current rates than the parametric benchmarks. Across the five records, the mean warming rates are \(0.45\) and \(0.50\,^{\circ}\mathrm{C} \, \textnormal{decade}^{-1}\) for the state-space and local-polynomial estimators, respectively, compared with \(0.28\) and \(0.35\) for the quadratic and changepoint benchmarks. Endpoint acceleration estimates from the flexible estimators are positive across all five records, although their statistical significance varies across records and methods. In the preferred state-space model, nominal \(90\%\) intervals for average acceleration over the past decade lie above zero in all five records, with a cross-dataset mean estimate of approximately \(0.17\,^{\circ}\mathrm{C} \, \textnormal{decade}^{-2}\). Our results suggest that global warming is proceeding faster, and has accelerated more strongly in the recent period, than conventional estimates imply.
\end{abstract}

\newpage

\sloppy
\section{Introduction}
\label{sec:introduction}
Global mean surface temperature is the most widely used summary of the state of
the climate system. The observed temperature record, however, is a noisy and
imperfect guide to the underlying warming process. Short-run internal
variability, measurement differences across datasets, and changes in
observational coverage can all obscure the slow-moving warming signal that is
most relevant for scientific assessment and policy interpretation. Three
 features of this latent warming process are of particular interest:
the current warming level, which determines proximity to policy thresholds such
as \(1.5^\circ\mathrm{C}\) and \(2.0^\circ\mathrm{C}\)
\citep{ClarkeRichardson2021,ThorneEtAl2026,ForsterEtAl2026,KirchengastPichler2025};
the warming rate, increasingly reported as a key climate indicator in its own
right \citep{SamsetEtAl2023,ForsterEtAl2026}; and the warming acceleration, that is,
whether the warming rate itself is changing
\citep{FR2011,Richardson2022,JenkinsEtAl2022,BeaulieuEtAl2024,FR2026}.

A convenient starting point for analyzing observed temperatures \(y(t)\) is the model
\citep[e.g.,][]{FR2026}
\begin{align}
y(t) = x(t) + \beta' z(t) + \varepsilon(t),
\label{eq:y}
\end{align}
where \(x(t)\) denotes the latent underlying warming level, \(z(t)\) is a vector
of observed covariates used as controls, \(\beta\) is a vector of coefficients,
and \(\varepsilon(t)\) captures remaining short-run variability, persistence,
and measurement error. In this representation, \(x(t)\) is the latent warming
level, its first derivative \(\dot{x}(t)=v(t)\) is the warming rate, and its
second derivative \(\ddot{x}(t)=a(t)\) is the warming acceleration.
The empirical problem is therefore to use observed temperatures $y(t)$ to recover the latent warming path \(x(t)\) and to conduct
inference on the current values and recent evolution of \(v(t)\) and \(a(t)\) in
the presence of covariates, serial dependence, and time-varying uncertainty.

The climate-monitoring literature has increasingly argued that smooth local or
adaptive methods are better suited than traditional long-window summaries for
estimating current climate conditions in a non-stationary world
\citep{ClarkeRichardson2021,WuEtAl2011,RigalEtAl2019,ScherrerEtAl2024,NicklasEtAl2025}.
Much of this work, however, has focused primarily on the current climate mean or
warming level, rather than on the joint assessment of the current warming level,
warming rate, and acceleration. By contrast, the recent debate on global warming
rates and acceleration has largely been conducted using quadratic trends,
changepoint models, and subperiod comparisons
\citep{FR2011,CahillEtAl2015,Richardson2022,BeaulieuEtAl2024,FR2026}.
These approaches are useful and interpretable, but they impose restrictive trend specifications over relatively long estimation windows. Although their fitted endpoint derivatives target the same quantities as the local methods, they can differ systematically from the true endpoint values when those trend restrictions are misspecified.
A further complication is that estimates of rates,
and especially acceleration, are sensitive not only to the trend
specification, but also to the treatment of transient covariates and serial
dependence
\citep{FR2011,Richardson2022,BeaulieuKillick2018,Mudelsee2019,BeaulieuEtAl2024,FR2026}.

Our contribution is to bring these strands together in a semiparametric
framework tailored to the joint estimation of the warming level, warming rate,
and acceleration from monthly temperature records. We incorporate observed
climate covariates, serial dependence, and time-varying irregular variance
within the same model, and use it to estimate the latent state path
\[
\{(x(t),v(t),a(t)) : t \leq T\},
\]
where \(T\) denotes the final observation time. For real-time climate
monitoring, attention naturally focuses on the recent part of this path and,
especially, on its terminal state
\[
(x(T),v(T),a(T)).
\]
These terminal values provide the current warming level, current warming rate,
and current warming acceleration. In our analysis, they are inferred jointly as
the latest values of an evolving path, rather than as isolated static
objects. The same framework also permits estimation of average acceleration over specified periods. Averaging over a recent period provides a less noisy summary than the terminal second derivative, allowing us to assess recent acceleration more precisely.

Our approach builds on the classical connection between state-space modelling
and spline smoothing \citep[e.g.,][]{Wahba1978,WeckerAnsley1983,KohnAnsley1987}.
We specify acceleration as a stochastic process that regularizes the smoothness
of the latent warming path. The dynamics of the warming rate and level follow mechanically from the identities \(v(t)=\mathrm{d}x(t)/\mathrm{d}t\) and \(a(t)=\mathrm{d}v(t)/\mathrm{d}t\). By analogy with position, velocity, and acceleration in mechanics, we refer to this as a \emph{kinematic} framework.
The specification of the acceleration dynamics serves as a reduced-form smoothness device rather than as
a structural climate model. The resulting discretized model is linear and
Gaussian, allowing the latent paths, covariate effects, and irregular-component
parameters to be estimated jointly using standard state-space methods
\citep{DK2012}. It also facilitates model verification through the
one-step-ahead prediction errors.

We compare this semiparametric estimator with a nonparametric local estimator,
as well as with parametric quadratic and changepoint benchmarks. The nonparametric
estimator imposes fewer assumptions on the warming path, but requires an
externally chosen bandwidth and estimates common covariate effects after
removing smooth time components from both temperatures and covariates. The parametric estimators, in contrast, impose strong
functional-form assumptions on the warming path: the quadratic model imposes
constant acceleration throughout the estimation period, while the changepoint
model imposes constant warming rates before and after the break. A simulation study provides further support for the proposed semiparametric state-space framework, showing substantial gains in estimation accuracy over the benchmark methods across a range of warming dynamics.

We apply the four estimators to monthly observations from five major global
temperature datasets. 
The empirical results show that the current warming level is robustly estimated
across datasets and methods. Using data through mid-2026, we estimate the
underlying warming level to be approximately
\(x(T)\approx 1.42^\circ\mathrm{C}\) above pre-industrial levels. Under this
latent-trend definition, the state-space point estimates place the current warming level
below \(1.5^\circ\mathrm{C}\), although the uncertainty intervals do not
uniformly exclude that threshold. The results for the warming rate and acceleration are more
method-sensitive. Flexible local estimators imply higher current warming rates
than the parametric benchmarks; our preferred state-space model
estimates the current warming rate at approximately
\(v(T)\approx 0.45^\circ\mathrm{C}\) per decade. For acceleration, the most
informative interpretation is pathwise as well as endpoint-based. The estimated
acceleration path is not globally positive over the full historical record, but
it becomes positive in the recent period, and the endpoint acceleration is
positive across all five temperature records. In the preferred state-space
model, the nominal \(90\%\) intervals for average acceleration over the past decade lie above zero in all five records. Although inference on \(a(T)\) is
necessarily more uncertain because it concerns a local second derivative, the
recent path behaviour and the uniformly positive endpoint estimates from the
flexible estimators support the conclusion that recent global warming has
accelerated. A similar conclusion was recently reached by  \citet{FR2026}. We obtain this finding using a more
flexible approach that jointly estimates the evolving paths of all three
kinematic quantities while explicitly accounting for covariate effects, measurement
error, serial dependence, and time-varying error variance.

The rest of the paper is organized as follows. Section~\ref{sec:data} describes the
temperature records and covariates used in the analysis. Section~\ref{sec:methods}
presents the semiparametric kinematic state-space model and the three benchmark
estimators. Section~\ref{sec:results} reports the main empirical results, including
the estimates of the current warming level, warming rate, and acceleration. Section~\ref{sec:discussion}
discusses the interpretation of the estimated recent warming dynamics. Section \ref{sec:conclusion} concludes. An appendix contains technical details, additional empirical findings, simulation studies, and various robustness analyses. Updated estimates of the warming level, rate,
and acceleration are available on the accompanying website as new observations become available.\footnote{
\url{https://sites.google.com/site/mbennedsen/research/global-warming-kinematics}}

\section{Data and controls}
\label{sec:data}

\begin{figure}[t]
\centering
\includegraphics[width=0.95\textwidth]{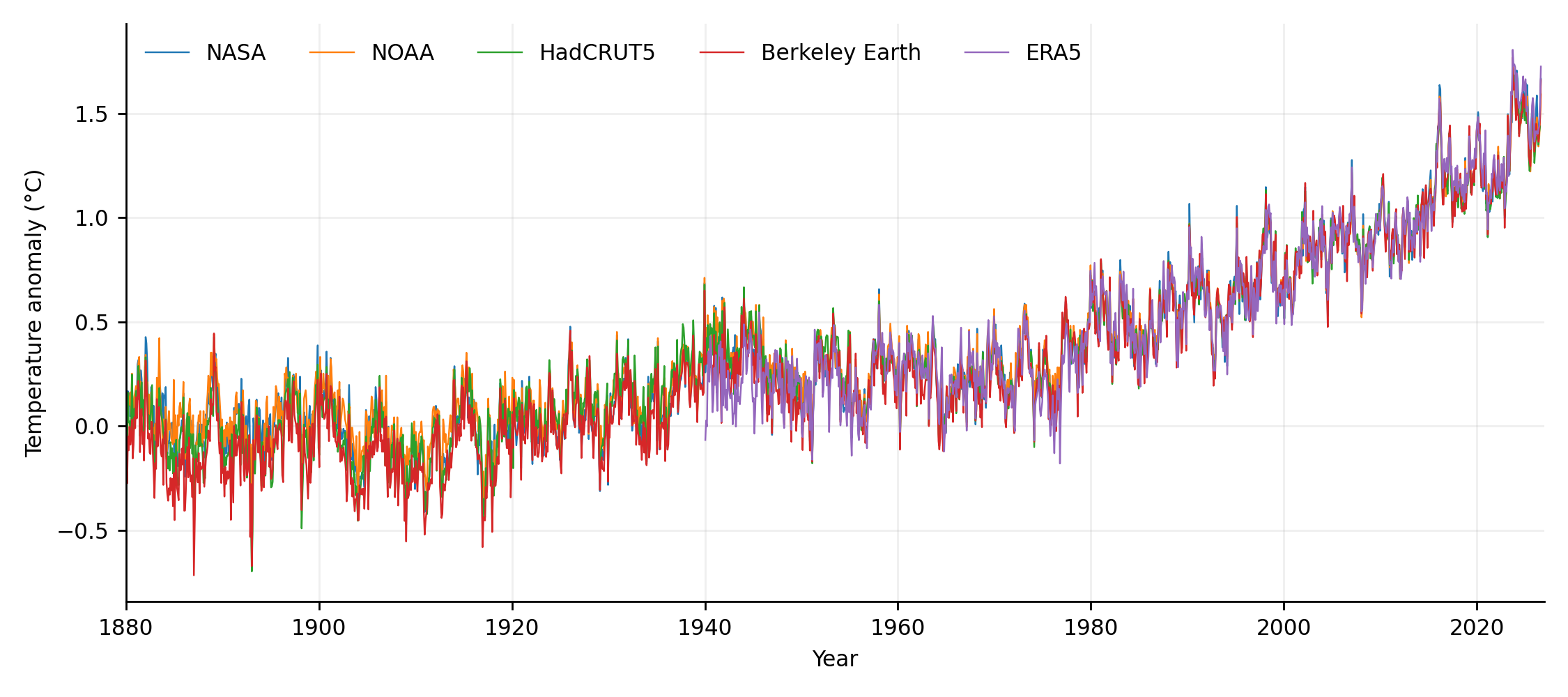}
\caption{Monthly global temperature anomaly records used in the analysis. The figure shows the five adjusted global temperature series: NASA, NOAA, HadCRUT5, Berkeley Earth, and ERA5. All series are expressed in degrees Celsius and adjusted to a common 1991--2020 reference level of \(0.88^\circ\mathrm{C}\), so that they can be interpreted as warming relative to pre-industrial conditions \citep[][]{FR2026}. To align the temperature records with the covariate series used in the baseline analysis, the  sample starts in 1880 for NASA, NOAA, HadCRUT5, and Berkeley Earth, although some of these products are available before 1880 in their current versions; ERA5 begins in 1940.}
\label{fig:global_data}
\end{figure}

We use monthly observations of global temperature anomalies. Let \(y_n = y(t_n)\) denote
the observed anomaly in month \(n=1,\ldots,N\), where \(t_n\) denotes the
corresponding calendar time measured in years and $T = t_N$ is the final available observation time. Following \citet{FR2026}, we use
five widely used global temperature anomaly records: NASA GISTEMP v4 \citep{GISTEMPTeam2024,LenssenEtAl2024}
(sample period 1880/01--2026/08), NOAA GlobalTemp \citep{VoseEtAl2021} (1880/01--2026/08), HadCRUT5
\citep{MoriceEtAl2021} (1880/01--2026/07), Berkeley Earth \citep{RohdeHausfather2020} (1880/01--2026/08), and ERA5
\citep{HersbachEtAl2017,HersbachEtAl2020} (1940/01--2026/08). Some of these records are available before 1880 in their
current versions, but we assemble the first four records from 1880 to align
with the monthly covariate series used below. After alignment with the
12-month covariate lags, the flexible estimators use observations from January
1881 for these four records and January 1940 for ERA5. The
series are reported as monthly temperature anomalies, in degrees Celsius, and
therefore have the deterministic seasonal cycle removed by construction.
To express the
series as warming relative to pre-industrial conditions, we follow \citet{FR2026}
and adjust each record to have a 1991--2020 mean of
\(0.88^\circ\mathrm{C}\). Figure~\ref{fig:global_data} displays the five
adjusted anomaly records used in the analysis.

For illustration, we also consider an ``arithmetic-mean'' series, obtained by averaging the baseline-adjusted anomalies month by month, giving equal weight to all four records before ERA5 begins in January 1940 and to all five thereafter.
The arithmetic-mean series ends in July 2026, the last month available for HadCRUT5.

\begin{figure}[tbp]
\centering
\includegraphics[width=0.95\textwidth]{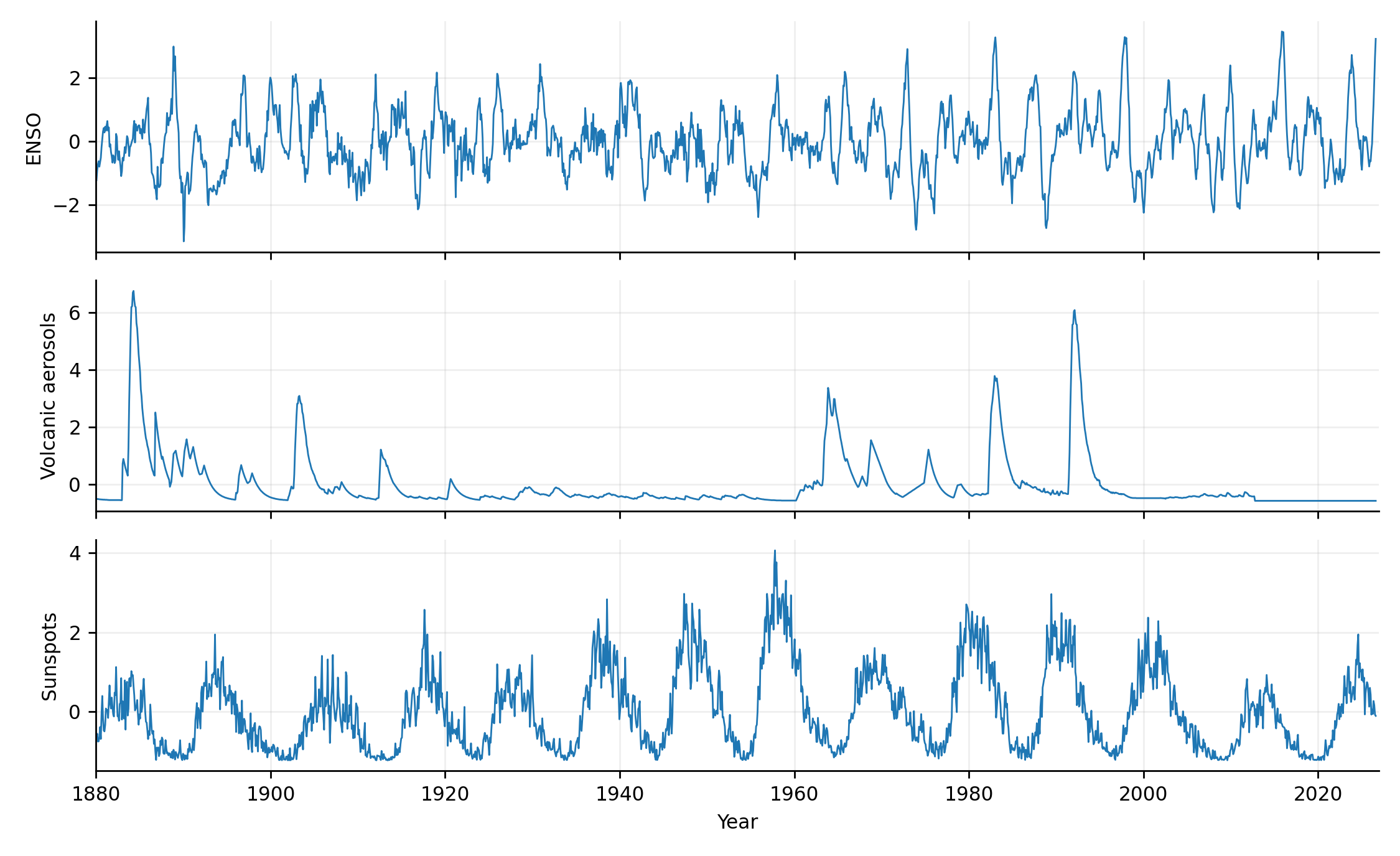}
\caption{Monthly covariates used to control for transient variability and external disturbances. The panels show standardized (z-scored) versions of the ENSO proxy, volcanic activity proxy, and sunspot series. These variables enter the empirical models through distributed lags, as described in Appendix~\ref{app:ss-model-selection}.}
\label{fig:covariates}
\end{figure}

We control for short-run internal variability and major external disturbances
using observed covariates. Specifically, the vector \(z_n=z(t_n)\) contains
distributed lags of three standardized monthly covariates: an El Ni\~no--Southern Oscillation
(ENSO) proxy, a volcanic activity proxy, and sunspot numbers, which proxy
variation in solar irradiance. These controls are included to separate
transient temperature fluctuations from the latent warming component \(x(t)\). The assembled covariate series span the period January 1880 to August 2026.\footnote{For January 1880--December 2024, all three covariate series are retained unchanged
from the replication data accompanying \citet{FR2026}
(\url{https://doi.org/10.5281/zenodo.15608929}).
The historical ENSO covariate is their HadISST-based monthly
Ni\~no 3.4 sea-surface temperature anomaly series. Values for
January 2025--August 2026 are reconstructed from HadISST sea-surface
temperature fields \citep{MetOfficeHadleyCentre2003} over
$5^\circ$S--$5^\circ$N, $170^\circ$W--$120^\circ$W. Calendar-month
additive offsets, estimated over 1981--2010, align the reconstructed index
with the reference convention of the replication series. Agreement over
the held-out 2023--2024 overlap has a maximum absolute discrepancy below
$0.00033^\circ$C; values through December 2024 are retained unchanged.
The volcanic covariate is global mean stratospheric aerosol optical depth
at 550\,nm from NASA GISS, based on \citet{SatoEtAl1993}
(\url{https://data.giss.nasa.gov/modelforce/strataer/original_GISS_data/tau.line_2012.12.txt}).
This source ends in September 2012. The replication dataset sets subsequent
values to zero through December 2024; we continue this convention through
August 2026.
The solar covariate is the monthly mean International Sunspot Number,
version 2.0, from WDC-SILSO, Royal Observatory of Belgium
(\url{https://www.sidc.be/SILSO/datafiles}). Values through December 2024
are retained from the replication dataset, and values for
January 2025--August 2026 were appended from WDC-SILSO.}
The standardized versions of the ENSO, volcanic, and sunspot covariates are
shown in Figure~\ref{fig:covariates}. In what follows, \(z_n\) denotes the full
vector of exogenous regressors at month \(n\), including the contemporaneous
covariates and their distributed lags. The lag lengths are chosen in a data-driven way, as described in
Section~\ref{app:ss-model-selection}.

Time is measured in years throughout the paper. The monthly step size is
\(\Delta=1/12\), so that consecutive observations satisfy
\(t_n-t_{n-1}=\Delta\).

\FloatBarrier

\section{Methods}
\label{sec:methods}

This section presents the methods used to estimate the kinematics of global
warming. We use the notation introduced above, where \(x(t)\) is the latent
warming path and \(v(t)\) and \(a(t)\) are its first and second derivatives. Our
main objective is to estimate the evolving kinematic state
\((x(t),v(t),a(t))\) in a unified framework. The end-of-sample values
\(x(T)\), \(v(T)\), and \(a(T)\) are the current-state summaries, obtained as the terminal values of the estimated
paths. Since endpoint derivatives, particularly acceleration, are difficult to
estimate precisely, we also consider average acceleration, i.e., the change in
warming rate per unit of time over a specified interval, as an additional estimand:
\begin{equation}
\overline{a_{T_0,T}}=\frac{v(T)-v(T_0)}{T-T_0}.
\label{eq:rate_change}
\end{equation}
For the smooth paths considered here, this equals \(\frac{1}{T-T_0}\int_{T_0}^{T}a(s)\,\mathrm{d}s\). For the changepoint model, the rate difference also captures a jump within the interval. Our primary averaging interval is the past decade, $T_0=T-10$, using exactly 120 monthly lags. This quantity complements the terminal acceleration \(a(T)\) by supplying a measure of the average acceleration over the recent interval \([T_0, T]\).

Our main empirical framework is a semiparametric kinematic state-space model that
estimates the latent path jointly for all three kinematic objects (Section~\ref{sec:ssmodel}). For comparison, we also consider
three benchmark approaches (Section~\ref{sec:benchmarks}): a nonparametric local-polynomial estimator, a
parametric quadratic trend model, and a parametric piecewise-linear changepoint
model. The state-space and local-polynomial estimators are flexible local
methods that recover an evolving path, whereas the
parametric benchmarks provide long-window summaries under more restrictive parametric trend
specifications.

We assess the accuracy of the four estimators in recovering these quantities through a simulation study, summarized in Section~\ref{sec:simstudy}, with full details and results in Appendix~\ref{app:sim_accuracy}.

\subsection{A semiparametric kinematic state-space framework}
\label{sec:ssmodel}

We specify a semiparametric model for the latent warming path \(x(t)\), the
warming rate \(v(t)=\dot{x}(t)\), and the warming acceleration
\(a(t)=\ddot{x}(t)\). The key modelling choice is to impose a stochastic law of
motion on the acceleration process. Specifically, we assume that
\begin{align}
    \mathrm{d}a(t)=\sigma\,\mathrm{d}W(t),
    \label{eq:da}
\end{align}
where \(\sigma>0\) and \(W(t)\) is a standard Brownian motion. The
continuous-time law \eqref{eq:da} is not intended as a structural physical
model of the climate system. Rather, it is a reduced-form stochastic
regularization device for the latent warming path and its derivatives.

This interpretation is closely related to the classical connection between
state-space models and spline smoothing
\citep[e.g.,][]{Wahba1978,WeckerAnsley1983,KohnAnsley1987,DK2012}.
In that literature, stochastic state-space specifications provide a
probabilistic way to regularize the smoothness of an underlying function. Our
specification plays the same role here: it regularizes changes in warming
acceleration, while allowing the warming level \(x(t)\), warming rate \(v(t)\),
and acceleration \(a(t)\) to be estimated jointly. The alternative specification
\(\mathrm{d}v(t)=\sigma\,\mathrm{d}W(t)\), examined in
Appendix~\ref{app:robust}, is the corresponding lower-order smoothing model,
which estimates the warming level and warming rate without treating
acceleration as a separate continuous latent state.

The dynamic specification \eqref{eq:da} induces dynamics for \(v(t)\) and
\(x(t)\) through the kinematic identities linking warming level, rate, and
acceleration, analogous to position, velocity, and acceleration in mechanics. That is,
\(\mathrm{d}v(t)=a(t)\mathrm{d}t\) and
\(\mathrm{d}x(t)=v(t)\mathrm{d}t\).
Let \(\Delta=1/12\) denote the monthly time step, measured in years, and write
\[
    (x_n,v_n,a_n) = (x(t_n),v(t_n),a(t_n)),
    \qquad t_n=t_1+(n-1)\Delta .
\]
The discretized state vector then evolves according to
\begin{align}
x_n &= x_{n-1} + v_{n-1}\Delta
      + \tfrac12 a_{n-1}\Delta^2 + \eta_{1,n}, \nonumber\\
v_n &= v_{n-1} + a_{n-1}\Delta + \eta_{2,n}, \label{eq:ss_state}\\
a_n &= a_{n-1} + \eta_{3,n}, \nonumber
\end{align}
where
\[
    \eta_n =
    (\eta_{1,n},\eta_{2,n},\eta_{3,n})'
    \stackrel{\mathrm{iid}}{\sim} \mathcal{N}(0,Q),
\]
with
\[
Q = \sigma^2
\begin{pmatrix}
\Delta^5/20 & \Delta^4/8 & \Delta^3/6 \\
\Delta^4/8 & \Delta^3/3 & \Delta^2/2 \\
\Delta^3/6 & \Delta^2/2 & \Delta
\end{pmatrix}.
\]
These equations constitute the state equations of the kinematic component of
the state-space model; derivation and implementation details are given in
Appendix~\ref{app:ssmodel}.

The state-space model is completed by a measurement equation for the observed
temperature anomalies. Using the representation in \eqref{eq:y}, we write
\begin{align}
    y_n = x_n + \beta' z_n + \varepsilon_n,
    \label{eq:ss_measurement}
\end{align}
where \(y_n = y(t_n)\) denotes the observed temperature anomaly in month \(n\),
\(z_n=z(t_n)\) is a vector of observed covariates and their distributed lags,
and \(\varepsilon_n = \varepsilon(t_n)\) captures remaining short-run variability, persistence,
and measurement error. This formulation allows the latent warming process and
the effects of transient covariates to be estimated jointly, rather than first
adjusting the temperature series in a separate preprocessing step.

 \subsubsection{Estimation and inference}

The state-space  model is linear and Gaussian. The
Kalman filter and smoother can therefore be used to evaluate the likelihood and to obtain
filtered and smoothed estimates of the unobserved state processes
\((x(t),v(t),a(t))\) \citep[e.g.,][]{DK2012}. The parameters are
estimated by maximum likelihood.

Uncertainty about the latent state processes is quantified using simulation smoothing and a bootstrap procedure that re-estimates the state-space model in each bootstrap sample, thereby incorporating both latent-state and parameter uncertainty. This yields pointwise bootstrap uncertainty bands for the paths
\(x(t)\), \(v(t)\), and \(a(t)\). Details of the state-space inference and bootstrap implementation are provided in Appendices~\ref{app:ssinference}--\ref{app:rate_change_inference}.

\subsubsection{Model selection and residual diagnostics}
\label{sec:model_selection}

The state-space framework requires specification of the irregular component
\(\varepsilon_n\), the time variation in its variance, and the distributed-lag
structure of the covariates \(z_n\). 
We model \(\varepsilon_n\) as an ARMA\((1,1)\) process, which captures
short-run persistence in monthly temperature anomalies not explained by the
smooth kinematic trend or the observed covariates. This specification is
selected by Bayesian information criterion (BIC)  from a set of low-order ARMA alternatives and is consistent with the treatment of
monthly temperature persistence in \citet{ClarkeRichardson2021} and
\citet{FR2026}.
Because the model is estimated on monthly data, we also
allow the variance of the ARMA innovations to vary by calendar month. We
considered additional variance breaks around 1900 and 1960, motivated by
changes in observational coverage and data construction, but these were not
supported by the BIC or by the residual
diagnostics.

The covariates enter through distributed lags, parameterized using Almon
polynomial restrictions \citep{Almon1965}. This reduces the number of free lag
coefficients while allowing flexible lag profiles. Our implementation
uses lags of 0--12 months for NINO3.4, volcanic aerosols, and sunspot numbers,
with a second-degree polynomial for NINO3.4 and first-degree polynomials for the
volcanic and sunspot variables. This specification was selected by BIC from a
grid search over lag lengths, polynomial degrees, and variance specifications,
conducted separately for each temperature record. The structure of the Almon polynomials, as well as parameter estimates from
the model, are reported in Appendix \ref{app:param_estimates}.

Model adequacy is assessed using the standardized one-step-ahead prediction
errors from the Kalman filter. Detailed
diagnostics for the five individual-dataset models are reported in
Table~\ref{tab:ss_diag} of Appendix~\ref{app:ss-model-selection}. The standardized
innovations are centred close to zero, have variances close to one, and show
little evidence of remaining conditional heteroskedasticity. Normality is not
rejected for any of the five temperature records. Ljung--Box tests at lags 1,
2 and 6 do not reject, whereas tests at lags 12 and 24 reject for all five
records. Thus, the specification captures much of the short-run behaviour but
does not remove all residual serial dependence; interval calibration is
assessed separately in Appendix~\ref{app:ssboot_mc}.

\subsection{Benchmark models}
\label{sec:benchmarks}

We compare the state-space estimates with three benchmark approaches. The first
is a nonparametric local-polynomial estimator, which estimates the latent
warming path and its first two derivatives from a covariate-adjusted temperature
series. Like the state-space model, this approach estimates the latent
kinematic paths \((x(t),v(t),a(t))\) for all $t \leq T$, but it does not model
serial dependence or time-varying variance jointly with the latent
warming path. Covariate effects are estimated after removing smooth time
components from both the temperature series and the covariates. Serial dependence
is accommodated through a moving-block bootstrap that re-estimates the covariate
effects in each replication, rather than through a jointly specified error model.

The remaining two benchmarks are parametric specifications closely related to
those used by \citet{FR2026}: a quadratic trend model and a piecewise-linear changepoint model. These models are simple and interpretable,
and provide useful reference points for recent warming rates and acceleration.
However, they impose restrictive functional forms over the full estimation
window. Following \citet{FR2026}, we estimate these models over a sample
starting in \(t_0=1970\); sensitivity to this choice is reported in
Appendix~\ref{app:parametric_t0} and summarized in Section~\ref{sec:robustness}.
The quadratic model implies constant acceleration over the full post-\(t_0\)
window, so any change in the warming rate is forced to be linear in time. The
changepoint model is restrictive in a different way: it summarizes the
post-break period by a single slope and imposes zero smooth acceleration away from
the break. We therefore
interpret these parametric benchmarks as long-window summaries of warming
dynamics. Their endpoint derivatives are conditional on these trend
restrictions, whereas the flexible models permit a more general local path.

Details on the implementation of the local-polynomial, quadratic, and
changepoint benchmarks are provided in Appendix~\ref{app:benchmark_details}.

 \subsection{Simulation evidence on estimator accuracy}
\label{sec:simstudy}
We assess the accuracy of the four estimators for $x(T)$, $v(T)$, $a(T)$,
and average acceleration over the past decade ($T_0=T-10$) using five NASA-calibrated data generating
processes (DGPs), with 1,000 replications per DGP. The designs include both
flexible warming paths and cases satisfying the parametric benchmarks'
trend restrictions. Appendix~\ref{app:sim_accuracy} provides the design and
results, including the root mean squared errors (RMSEs) in
Table~\ref{tab:simstudy}.

The results strongly favour the state-space estimator across the designs considered: it achieves the lowest RMSE in 16 of the 20 DGP--estimand combinations overall. For the endpoint warming rate, the main estimand in the empirical analysis, it has the lowest RMSE in all five designs, including those satisfying the quadratic and changepoint benchmarks' trend restrictions. The exceptions for other estimands arise under the changepoint DGP, where the warming rate undergoes a single abrupt break, and for endpoint acceleration in the zero-acceleration design, where the changepoint estimator correctly imposes \(a(T)=0\) by construction.
In the changepoint design, both flexible estimators spread the abrupt rate increase over time and therefore estimate positive endpoint and past-decade average acceleration on average, even though the post-break warming rate is constant.

Interval coverage also depends on the warming path. Under the fixed NASA-smoothed path, the nominal \(90\%\) quadratic and changepoint intervals cover the true endpoint warming rate in \(0\%\) and \(49.5\%\) of replications, respectively. Under their corresponding quadratic and broken-linear paths, coverage rises to \(79\%\) and \(84.75\%\) (Tables~\ref{tab:ss_bootstrap_coverage} and~\ref{tab:parametric_smooth_coverage}).

\section{Results}
\label{sec:results}

\begin{figure}[!htbp]
\centering
\includegraphics[width=0.95\textwidth]{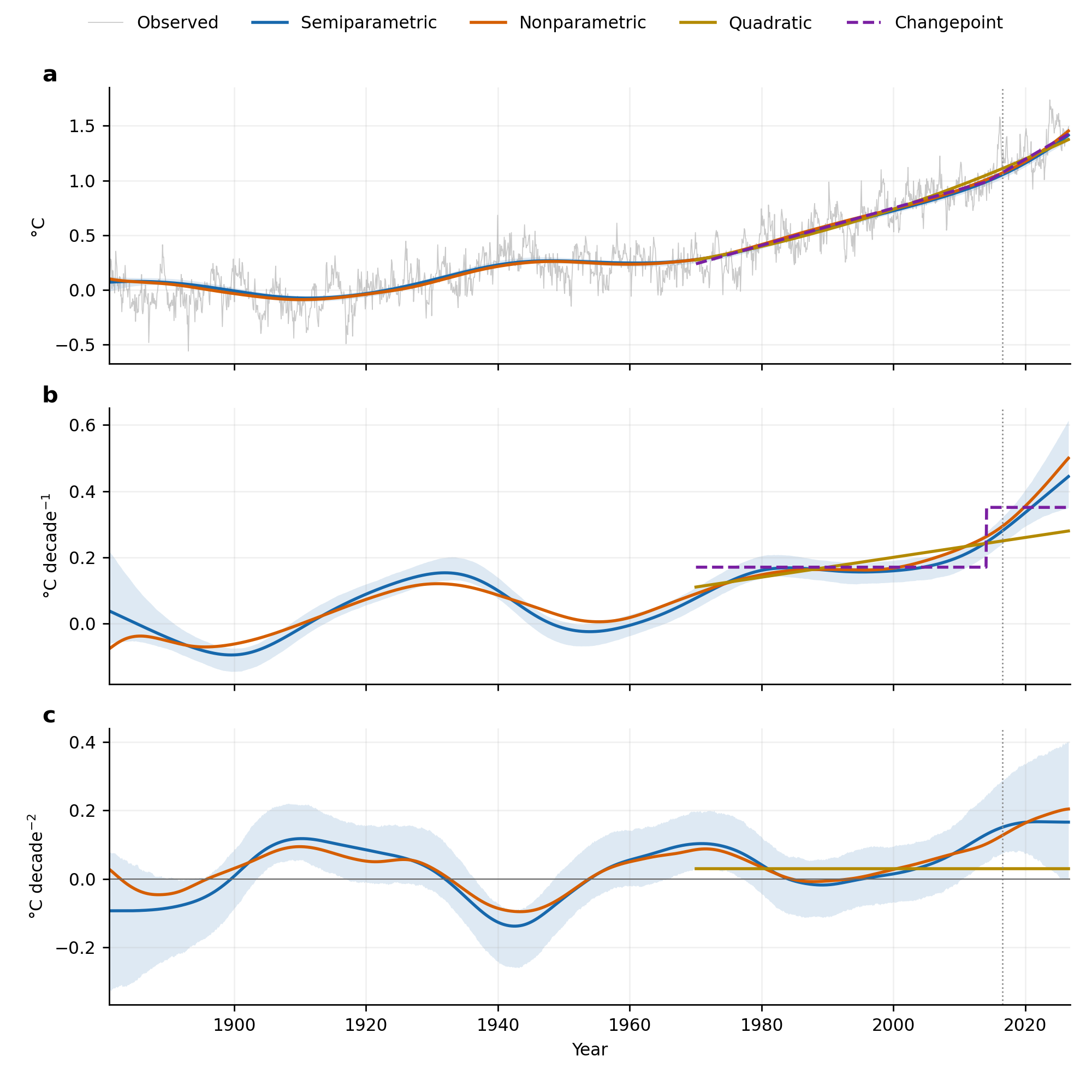}
\caption{\textbf{Illustrative decomposition for the arithmetic-mean temperature series.}
The figure compares the semiparametric state-space model, the nonparametric
local-polynomial estimator, and the parametric
quadratic and changepoint benchmarks. The panels show \textbf{a}, observed
arithmetic-mean anomalies and latent warming level \(x(t)\); \textbf{b}, warming
rate \(v(t)\); and \textbf{c}, warming acceleration \(a(t)\). The models are fitted
to the arithmetic-mean series itself, using observations from January 1881
for the flexible estimators and January 1970 for the parametric benchmarks,
through July 2026. Shaded bands are pointwise nominal 90\% state-space
bootstrap-error intervals. The vertical dotted line marks the start of the past-decade averaging
interval. All fitted paths account for the specified climate covariates.
The changepoint model imposes zero acceleration away from its break and is
omitted from \textbf{c}.}
\label{fig:global_arithmetic_xva}
\end{figure}

We apply the four estimators to the five monthly global temperature records introduced in Section~\ref{sec:data} and to their arithmetic mean. Figure~\ref{fig:global_arithmetic_xva} shows the latent warming level \(x(t)\), warming rate \(v(t)\), and warming acceleration \(a(t)\) estimated by fitting the models directly to the arithmetic-mean temperature series. Figures~\ref{fig:global_nasa_xva}--\ref{fig:global_era5_xva} in Appendix~\ref{app:additional_decompositions} show the corresponding decompositions for the five individual records.
The flexible state-space and local-polynomial estimators show broadly similar historical patterns for all three kinematic quantities. For the warming level \(x(t)\), the two
parametric benchmarks, estimated over the post-1970 sample, also agree closely
with the more flexible estimates. For the warming rate and acceleration, however, the figures show substantially higher recent estimates under the flexible estimators than under the parametric benchmarks, reflecting their different restrictions on the warming path.

To compare estimates of recent warming dynamics,  Table~\ref{tab:main}
reports estimates of \(x(T)\), \(v(T)\), \(a(T)\), and \(\overline{a_{T-10,T}}\) for all four methods and all
five individual datasets and the separately fitted arithmetic-mean series, together with nominal \(90\%\) confidence intervals.\footnote{Empirical finite-sample coverage of the nominal \(90\%\) intervals are assessed in Appendix~\ref{app:ssboot_mc}  using a Monte Carlo simulation study.} The following sections describe the results in turn.

\begin{table}[tbp]
\centering\scriptsize
\setlength{\tabcolsep}{3pt}
\caption{\textbf{Endpoint estimates and past-decade average acceleration.} Monthly results through August 2026 (July for HadCRUT5 and the arithmetic-mean series). Both flexible estimators use 1,999 bootstrap replications. Quadratic intervals use HAC standard errors. Changepoint intervals use 1,999 ARMA(1,1) bootstrap draws, repeating the full break search and covariate estimation. Panel D uses exactly 120 monthly lags, $T_0=T-10$. Units are $^{\circ}$C for level, $^{\circ}$C decade$^{-1}$ for rate, and $^{\circ}$C decade$^{-2}$ for both acceleration estimands. $\dagger$ denotes a zero under the original fitted changepoint path. Endpoint acceleration is identically zero across bootstrap fits and shown without an interval; average-acceleration intervals incorporate breakpoint selection. The arithmetic-mean series is fitted separately to the mean temperature record; it is not an average of the individual estimates. Cross-dataset means average the five separately fitted estimates and have no pooled interval.}\label{tab:main}
\begin{adjustbox}{max width=\linewidth}
\begin{tabular}{lcccccccc}
\toprule
 & \multicolumn{2}{c}{Semiparametric} & \multicolumn{2}{c}{Nonparametric} & \multicolumn{2}{c}{Quadratic} & \multicolumn{2}{c}{Changepoint} \\
Dataset & Est. & 90\% CI & Est. & 90\% CI & Est. & 90\% CI & Est. & 90\% CI \\\midrule
\multicolumn{9}{l}{\textbf{Panel A: Endpoint warming level $x(T)$}}\\
NASA & 1.437 & [1.390, 1.520] & 1.482 & [1.391, 1.546] & 1.392 & [1.347, 1.437] & 1.443 & [1.390, 1.504] \\
NOAA & 1.410 & [1.362, 1.477] & 1.441 & [1.369, 1.509] & 1.376 & [1.337, 1.415] & 1.424 & [1.376, 1.476] \\
HadCRUT5 & 1.381 & [1.325, 1.460] & 1.420 & [1.339, 1.486] & 1.347 & [1.305, 1.388] & 1.390 & [1.340, 1.453] \\
Berkeley Earth & 1.414 & [1.364, 1.496] & 1.442 & [1.355, 1.513] & 1.360 & [1.318, 1.402] & 1.412 & [1.363, 1.479] \\
ERA5 & 1.457 & [1.416, 1.547] & 1.502 & [1.418, 1.573] & 1.414 & [1.370, 1.459] & 1.473 & [1.414, 1.546] \\
Arithmetic mean & 1.415 & [1.364, 1.489] & 1.453 & [1.375, 1.521] & 1.374 & [1.333, 1.416] & 1.424 & [1.373, 1.486] \\
\addlinespace[2pt]Cross-dataset mean & 1.420 &  & 1.457 &  & 1.378 &  & 1.428 &  \\\addlinespace[5pt]
\multicolumn{9}{l}{\textbf{Panel B: Endpoint warming rate $v(T)$}}\\
NASA & 0.504 & [0.410, 0.714] & 0.556 & [0.324, 0.668] & 0.292 & [0.260, 0.324] & 0.366 & [0.295, 0.484] \\
NOAA & 0.433 & [0.320, 0.589] & 0.453 & [0.292, 0.602] & 0.286 & [0.259, 0.313] & 0.349 & [0.290, 0.441] \\
HadCRUT5 & 0.394 & [0.292, 0.567] & 0.476 & [0.259, 0.608] & 0.259 & [0.230, 0.288] & 0.322 & [0.250, 0.478] \\
Berkeley Earth & 0.473 & [0.374, 0.664] & 0.502 & [0.282, 0.652] & 0.265 & [0.236, 0.295] & 0.340 & [0.270, 0.492] \\
ERA5 & 0.427 & [0.360, 0.615] & 0.513 & [0.300, 0.652] & 0.302 & [0.272, 0.333] & 0.387 & [0.304, 0.542] \\
Arithmetic mean & 0.444 & [0.350, 0.614] & 0.500 & [0.290, 0.623] & 0.280 & [0.251, 0.308] & 0.351 & [0.278, 0.484] \\
\addlinespace[2pt]Cross-dataset mean & 0.446 &  & 0.500 &  & 0.281 &  & 0.353 &  \\\addlinespace[5pt]
\multicolumn{9}{l}{\textbf{Panel C: Endpoint acceleration $a(T)$}}\\
NASA & 0.229 & [0.046, 0.548] & 0.255 & [0.033, 0.356] & 0.035 & [0.024, 0.045] & $0^{\dagger}$ & -- \\
NOAA & 0.151 & [-0.056, 0.395] & 0.152 & [0.006, 0.301] & 0.035 & [0.026, 0.044] & $0^{\dagger}$ & -- \\
HadCRUT5 & 0.137 & [-0.028, 0.373] & 0.202 & [-0.006, 0.324] & 0.023 & [0.014, 0.033] & $0^{\dagger}$ & -- \\
Berkeley Earth & 0.197 & [0.027, 0.461] & 0.212 & [0.003, 0.352] & 0.024 & [0.014, 0.033] & $0^{\dagger}$ & -- \\
ERA5 & 0.132 & [0.034, 0.332] & 0.193 & [-0.012, 0.329] & 0.035 & [0.024, 0.045] & $0^{\dagger}$ & -- \\
Arithmetic mean & 0.166 & [0.001, 0.403] & 0.204 & [0.008, 0.322] & 0.030 & [0.021, 0.039] & $0^{\dagger}$ & -- \\
\addlinespace[2pt]Cross-dataset mean & 0.169 &  & 0.203 &  & 0.030 &  & 0.000 &  \\\addlinespace[5pt]
\multicolumn{9}{l}{\textbf{Panel D: Past-decade average acceleration $\overline{a}_{T-10,T}$}}\\
NASA & 0.221 & [0.133, 0.438] & 0.254 & [0.037, 0.346] & 0.035 & [0.024, 0.045] & $0^{\dagger}$ & [0.000, 0.312] \\
NOAA & 0.152 & [0.038, 0.312] & 0.157 & [0.009, 0.292] & 0.035 & [0.026, 0.044] & $0^{\dagger}$ & [0.000, 0.237] \\
HadCRUT5 & 0.135 & [0.039, 0.302] & 0.202 & [-0.004, 0.315] & 0.023 & [0.014, 0.033] & $0^{\dagger}$ & [0.000, 0.300] \\
Berkeley Earth & 0.193 & [0.101, 0.380] & 0.213 & [0.009, 0.344] & 0.024 & [0.014, 0.033] & $0^{\dagger}$ & [0.000, 0.312] \\
ERA5 & 0.130 & [0.065, 0.293] & 0.198 & [-0.004, 0.322] & 0.035 & [0.024, 0.045] & $0^{\dagger}$ & [0.000, 0.358] \\
Arithmetic mean & 0.164 & [0.070, 0.329] & 0.206 & [0.012, 0.312] & 0.030 & [0.021, 0.039] & $0^{\dagger}$ & [0.000, 0.307] \\
\addlinespace[2pt]Cross-dataset mean & 0.166 &  & 0.205 &  & 0.030 &  & 0.000 &  \\\addlinespace[5pt]
\bottomrule
\end{tabular}
\end{adjustbox}
\end{table}

\subsection{Recent warming level}
Panel~A of Table~\ref{tab:main} shows that the estimated current warming level
is highly robust across datasets and methods. All four estimators place the
endpoint warming level close to
\(1.42\,^\circ\mathrm{C}\) above the pre-industrial reference level, with
cross-dataset means ranging from \(1.378\,^\circ\mathrm{C}\) for the quadratic
benchmark to \(1.457\,^\circ\mathrm{C}\) for the local-polynomial benchmark. The
state-space estimates are particularly stable across records, ranging from
\(1.381\,^\circ\mathrm{C}\) for HadCRUT5 to \(1.457\,^\circ\mathrm{C}\) for ERA5,
with a cross-dataset mean of \(1.420\,^\circ\mathrm{C}\).

Under the latent endpoint definition used here, the estimates place the
current underlying warming level close to, but generally below,
\(1.5\,^\circ\mathrm{C}\). For the state-space model, the \(90\%\)
confidence intervals have upper endpoints below \(1.5\,^\circ\mathrm{C}\)
for NOAA, HadCRUT5 and Berkeley Earth. They include that threshold for NASA and
ERA5. In simulation experiments, the upper endpoint of the state-space interval for the warming level falls below the true level in \(11.5\%\) of replications (Appendix~\ref{app:ssboot_mc}). Thus, the state-space
results do not uniformly rule out a latent warming level of
\(1.5\,^\circ\mathrm{C}\).

\subsection{Recent warming rate}
Panel~B of Table~\ref{tab:main} shows greater disagreement across methods for
the endpoint warming rate \(v(T)\) than for the warming level \(x(T)\). All
methods imply positive current warming rates, but the magnitude depends strongly
on whether the estimator targets a local endpoint quantity or a long-window
summary. The state-space estimates range from
\(0.394\,^\circ\mathrm{C}\,\mathrm{decade}^{-1}\) for HadCRUT5 to
\(0.504\,^\circ\mathrm{C}\,\mathrm{decade}^{-1}\) for NASA, with a
cross-dataset mean of
\(0.446\,^\circ\mathrm{C}\,\mathrm{decade}^{-1}\). The local-polynomial
estimates are higher on average, with a cross-dataset mean of
\(0.500\,^\circ\mathrm{C}\,\mathrm{decade}^{-1}\). 

The parametric benchmarks imply substantially lower warming-rate estimates.
The quadratic model gives a cross-dataset mean of
\(0.281\,^\circ\mathrm{C}\,\mathrm{decade}^{-1}\), while the changepoint model
gives a post-break slope of
\(0.353\,^\circ\mathrm{C}\,\mathrm{decade}^{-1}\). For every record, both flexible
estimators give higher point estimates than either parametric benchmark. 
This is consistent with continued increases in the warming
rate over the most recent period, rather than a single constant post-break slope.

\subsection{Recent acceleration}
Panel~C of Table~\ref{tab:main} shows the estimates of endpoint acceleration \(a(T)\). Uncertainty varies across records and estimators,
and the confidence bands widen towards the sample endpoint in the lower panels of
Figure~\ref{fig:global_arithmetic_xva} and
Figures~\ref{fig:global_nasa_xva}--\ref{fig:global_era5_xva} in
Appendix~\ref{app:additional_decompositions}. Such endpoint widening is natural: without observations beyond \(T\), the terminal acceleration \(a(T)\) must be inferred using only observations on one side of the endpoint. Nevertheless, the endpoint estimates are consistently positive. The
state-space estimates range from
\(0.132\,^\circ\mathrm{C}\,\mathrm{decade}^{-2}\) for ERA5 to
\(0.229\,^\circ\mathrm{C}\,\mathrm{decade}^{-2}\) for NASA, with a
cross-dataset mean of
\(0.169\,^\circ\mathrm{C}\,\mathrm{decade}^{-2}\). The local-polynomial
estimates are also positive for all five records and are generally larger,
with a
cross-dataset mean of
\(0.203\,^\circ\mathrm{C}\,\mathrm{decade}^{-2}\).

For the state-space estimator, the lower endpoints of the reported \(90\%\)
intervals can be read as nominal one-sided \(95\%\) lower confidence bounds.
They exceed zero for NASA, Berkeley Earth and ERA5, while the intervals for
NOAA and HadCRUT5 include zero. In simulation experiments, the state-space interval covers true endpoint acceleration in \(95.75\%\) of replications. Its lower bound exceeds the truth in \(1\%\) of replications (Appendix~\ref{app:ssboot_mc}).
The lower-bound exclusions therefore provide statistical evidence of positive
endpoint acceleration, subject to the modelling assumptions.

For the local-polynomial estimator, nominal endpoint intervals exclude zero
for NASA, NOAA and Berkeley Earth. In the same simulation experiment, the local-polynomial lower bound exceeds the true acceleration in none of the 400 replications for either acceleration measure. The upper bound falls below the truth in \(17.25\%\) of replications for endpoint acceleration and \(16\%\) for past-decade average acceleration (Table~\ref{tab:ss_bootstrap_coverage}). The record-specific evidence for endpoint
acceleration thus remains less uniform than the evidence for positive current
warming rates. This motivates complementing endpoint estimates with average
acceleration over a recent period.

To this end, Panel~D of Table~\ref{tab:main} reports average acceleration
\(\overline{a_{T_0,T}}\), defined in \eqref{eq:rate_change}, with \(T_0=T-10\).
The state-space estimates range from
\(0.130\,^\circ\mathrm{C}\,\mathrm{decade}^{-2}\) for ERA5 to
\(0.221\,^\circ\mathrm{C}\,\mathrm{decade}^{-2}\) for NASA, with a
cross-dataset mean of \(0.166\,^\circ\mathrm{C}\,\mathrm{decade}^{-2}\).
These estimates are close to the corresponding endpoint acceleration
estimates, indicating that the
estimated magnitude of recent acceleration is not confined to the
sample endpoint. Further,
the nominal \(90\%\) intervals lie above zero in all five records. In the simulation experiments, these intervals cover the true past-decade average acceleration in \(81.5\%\) of replications. The lower bound exceeds the truth in \(2.25\%\) of replications, while the upper bound falls below it in \(16.25\%\).
The local-polynomial average-acceleration estimates are also positive across
records, with a cross-dataset mean of
\(0.205\,^\circ\mathrm{C}\,\mathrm{decade}^{-2}\); their nominal intervals
exclude zero for NASA, NOAA and Berkeley Earth, while the HadCRUT5 and ERA5
intervals include zero. Both acceleration intervals also exclude zero for the separately fitted arithmetic-mean series under each flexible estimator.
Alternative starting years $T_0$ yield similar findings (Table~\ref{tab:rate_change_alt_t0} of Appendix~\ref{app:robust}).

The quadratic benchmark also yields positive acceleration estimates for all five
datasets, and the associated confidence intervals exclude zero. These estimates
are much smaller, with a cross-dataset mean of
\(0.030\,^\circ\mathrm{C}\,\mathrm{decade}^{-2}\), because the quadratic model
imposes a constant acceleration over the full post-1970 estimation window. The
narrow confidence intervals should therefore be interpreted conditional on this
strong parametric restriction. The changepoint benchmark, by contrast, does not
provide a directly comparable estimate of \(a(T)\): the fitted path is
piecewise linear, so local acceleration is zero away from the changepoint and
undefined at the changepoint itself. Its past-decade average-acceleration point estimates are also zero because the selected breaks precede the averaging interval. Their bootstrap intervals are not degenerate, since bootstrap breaks can fall within that interval.

\subsection{Sensitivity to modelling choices}
\label{sec:robustness}

The preceding results are based on monthly data, a state-space specification in
which stochastic variation enters through the acceleration process, baseline
smoothing choices for the state-space and local-polynomial estimators, a
post-1970 estimation window for the parametric benchmarks, a starting date \(T_0=T-10\) for the averaging interval of the average acceleration estimand $\overline{a_{T_0,T}}$, and adjustment for climate covariates. We assess the
sensitivity of the conclusions to these choices in six ways. First, we repeat
the analysis using annual rather than monthly data.
For the annual data, we fix each state-space smoothing parameter
\(\sigma\) at its corresponding monthly estimate and re-estimate the
remaining parameters. This avoids the tendency of maximum-likelihood
estimates of state innovation variances to pile up at zero in short
samples \citep{StockWatson1998}; see Appendix~\ref{app:robust_yearly}
for details.
For the annual state-space
specification, the irregular component is selected at annual frequency: the BIC
selects an AR\((1)\) irregular for NASA, NOAA, HadCRUT5, and Berkeley Earth, and
an IID irregular for ERA5. This reflects that annual aggregation removes much of
the short-run persistence that motivates the monthly ARMA\((1,1)\) irregular.
Second, we estimate an alternative state-space specification in which stochastic
variation enters through the warming rate,
\(\mathrm{d}v(t)=\sigma\,\mathrm{d}W(t)\), rather than through the warming
acceleration, \(\mathrm{d}a(t)=\sigma\,\mathrm{d}W(t)\). Third, we vary the
smoothing parameters in the state-space and local-polynomial estimators. Fourth,
we vary the starting year \(t_0\) of the estimation window for the quadratic and
changepoint benchmarks. Fifth, we vary the starting year \(T_0\) of the averaging interval for the average acceleration estimand $\overline{a_{T_0,T}}$.
Sixth, we repeat the state-space analysis without the climate covariates. Details are reported in Appendix~\ref{app:robust}.

Figure~\ref{fig:robust_summary} summarizes the sensitivity to modelling
choices by showing the range of cross-dataset mean estimates across
the alternative specifications. The warming level is comparatively stable, whereas the magnitudes of
the warming rate and acceleration are more sensitive to temporal aggregation
and smoothing. The annual state-space mean warming
rate remains above the changepoint mean, although the gap is narrower than in
the monthly results (Table~\ref{tab:annual_robustness_endpoint}). The state-space  endpoint and average acceleration estimates remain positive across
the smoothing values considered (Table~\ref{tab:ss_q_robustness}).
The local-polynomial endpoint acceleration estimates are sensitive to bandwidth
choice. They remain positive at bandwidths of 15--35 years except for ERA5 at
15 years, where endpoint acceleration is slightly negative. Average
acceleration is positive for every record throughout this bandwidth range
(Table~\ref{tab:kernel_bandwidth_robustness}).
Changing the estimation window
also affects the parametric benchmarks, especially when a single changepoint
captures an earlier transition in the warming record (Tables~\ref{tab:quadratic_startdate_robustness} and~\ref{tab:changepoint_startdate_robustness}). These results support
the baseline findings while showing that their magnitudes and inferential
strength are not uniform across specifications. For average acceleration, the
state-space intervals exclude zero in all five records for every alternative
starting year considered (Table~\ref{tab:rate_change_alt_t0}). Removing climate
covariates lowers mean past-decade average acceleration from \(0.166\) to
\(0.107\,^\circ\mathrm{C}\,\mathrm{decade}^{-2}\), but preserves positive point
estimates for all records and intervals excluding zero for four of the five
records and the arithmetic-mean series (Appendix~\ref{app:no_covariates}).
These results support positive recent average acceleration, while showing that
its estimated magnitude and formal significance depend on the specification.

\begin{figure}[tbp]
\centering
\includegraphics[width=0.95\textwidth]{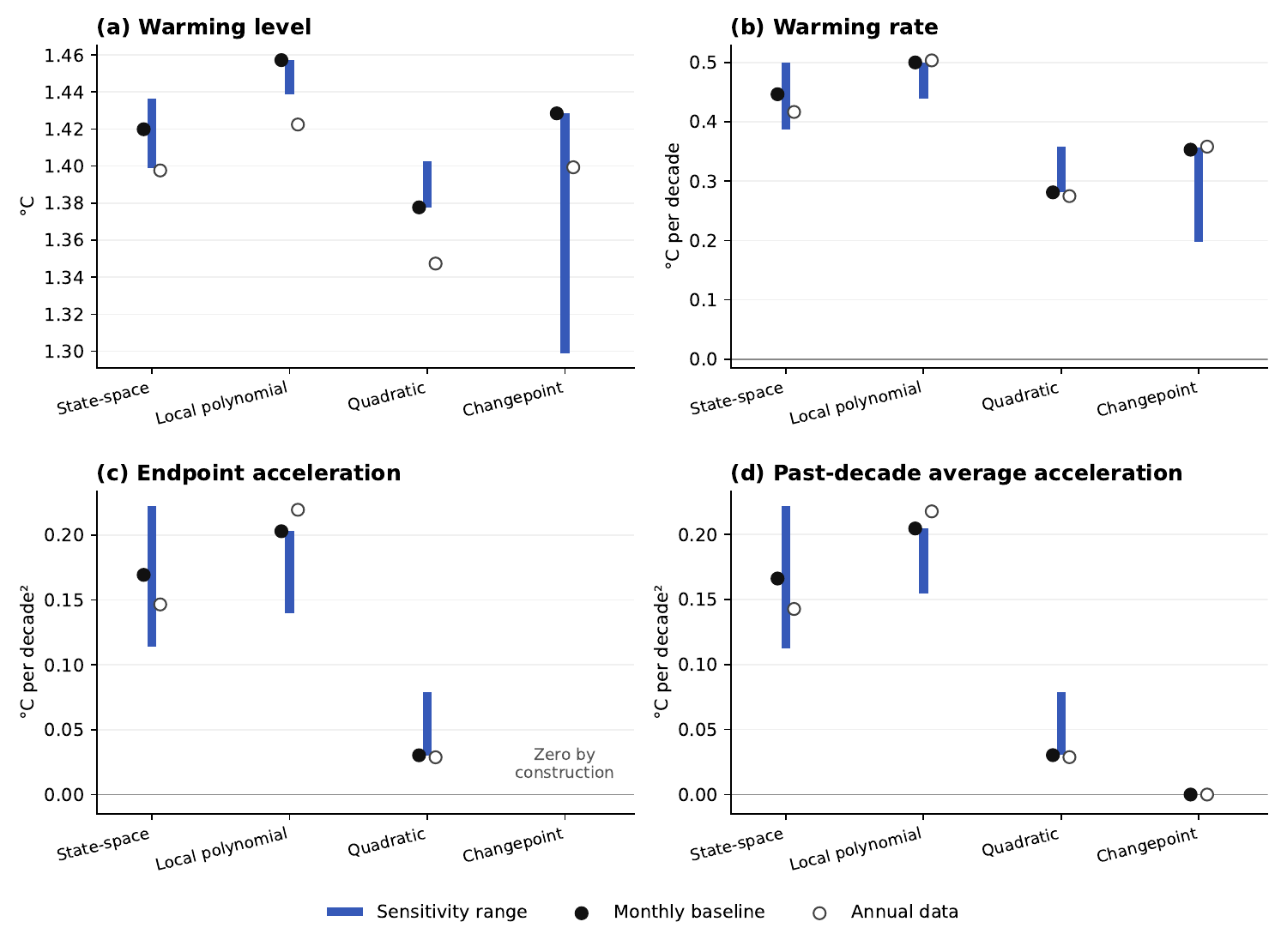}
\caption{\textbf{Sensitivity of endpoint and average acceleration estimates to modelling choices.}
The figure summarizes how the estimates of (a) the endpoint warming level
\(x(T)\), (b) the endpoint warming rate \(v(T)\), (c) the endpoint warming acceleration
\(a(T)\), and (d) the average warming acceleration \(\overline{a}_{T-10,T}\) vary across robustness analyses. Circles denote cross-dataset arithmetic
means across the five temperature records. Black circles show the baseline
monthly estimates reported in Table~\ref{tab:main}. Open circles show the
corresponding estimates obtained using annual data; for the annual state-space
model, each \(\sigma\) is fixed at its corresponding monthly estimate and the
remaining parameters are re-estimated. Blue vertical ranges show
the range of point estimates obtained under estimator-specific sensitivity
analyses. For the state-space model, this includes variation in the smoothing
parameter \(\sigma\) and the alternative specification in which stochastic variation
enters through \(\mathrm{d}v(t)=\sigma\,\mathrm{d}W(t)\) rather than through
\(\mathrm{d}a(t)=\sigma\,\mathrm{d}W(t)\). For both acceleration panels, the state-space range uses only the baseline trend order; the alternative model is included only in the level and rate panels. For the local-polynomial estimator, it reflects variation in the bandwidth \(h\). For the quadratic and
changepoint benchmarks, it reflects variation in the starting year \(t_0\) of
the estimation window. The changepoint model does not provide a directly
comparable smooth estimate of \(a(T)\), since the fitted path is piecewise linear.}
\label{fig:robust_summary}
\end{figure}

 \FloatBarrier

\section{Discussion}
\label{sec:discussion}

We focus the discussion on the state-space estimates, which combine
a flexible warming path with joint estimation of covariate effects,
serial dependence, and time-varying error variance. This focus is supported
by the simulation study in Appendix~\ref{app:sim_accuracy}, which
shows strong comparative accuracy for the state-space estimator across the
selected designs. The benchmark comparisons help distinguish findings that are stable across
specifications from those that depend more strongly on trend restrictions
on the warming path.

\subsection{Recent warming level}

The most robust result concerns the current warming level. Across datasets and
methods, the endpoint estimate \(x(T)\) is close to
\(1.42\,^\circ\mathrm{C}\) above the pre-industrial reference level. This
agreement is notable because the four estimators differ substantially in their
treatment of smoothness, covariates, serial dependence, and endpoint behaviour.
The estimated warming level is therefore much less sensitive to the choice of
statistical specification than either the warming rate or the warming
acceleration.

Our endpoint level estimate is consistent with recent independent assessments.
\citet{ThorneEtAl2026} report long-term realised warming of
\(1.40\,^\circ\mathrm{C}\) with 90\% uncertainty interval of \(\,[1.23\text{--}1.58]\,^\circ\mathrm{C}\) and underlying
human-induced warming of \(1.34\,[1.18\text{--}1.50]\,^\circ\mathrm{C}\) as of
end-2024, while \citet{KirchengastPichler2025} place the current traceable
warming level at \(1.39\,[1.29\text{--}1.49]\,^\circ\mathrm{C}\). Our
cross-dataset endpoint estimate of
\(x(T)\approx 1.42\,^\circ\mathrm{C}\) in mid-2026 is therefore closely
aligned with these estimates. 

Under the latent endpoint definition used in this paper, the state-space point estimates
place the underlying warming level below \(1.5\,^\circ\mathrm{C}\) in mid-
2026, although the uncertainty intervals do not uniformly exclude that
threshold.
By separating the evolving warming path from transient
fluctuations, our framework provides an estimate of underlying warming
at the current endpoint without waiting for subsequent observations
to complete a centred averaging window. This addresses the need for timely indicators of current warming discussed by \citet{Betts2023}, including for assessing proximity to the Paris temperature thresholds.\footnote{To this end, the accompanying website \href{https://sites.google.com/site/mbennedsen/research/global-warming-kinematics}{(\texttt{LINK})} shows state-space estimates of the warming level, rate, and acceleration updated using the latest available temperature data.} 

\subsection{Recent warming rate}

The current warming rate is more sensitive to the choice of estimator than the warming
level. Across the five datasets, the preferred state-space model estimates an
endpoint warming rate of approximately
\(v(T)\approx 0.45\,^\circ\mathrm{C}\,\mathrm{decade}^{-1}\). This is
substantially higher than the rates implied by the parametric benchmarks, which
estimate the endpoint rate under restrictions on the warming path over the
post-1970 estimation window. The local-polynomial estimates are
higher still, while interval widths vary across records and estimators.

These differences reflect both the temporal summaries and the restrictions
imposed on the warming path. The quadratic and
changepoint benchmarks provide useful long-window summaries of recent warming
dynamics. The quadratic model imposes a linear path for \(v(t)\), while the changepoint model summarizes the post-break
period by a single slope. The state-space and local-polynomial estimators,
which target \(v(T)\) directly, indicate that the current warming rate is higher
than these long-window summaries suggest. Across the five records, the mean state-space rate rose from \(0.28\,^\circ\mathrm{C}\,\mathrm{decade}^{-1}\) in mid-2016 to \(0.45\,^\circ\mathrm{C}\,\mathrm{decade}^{-1}\) in mid-2026. The mean post-break slope of \(0.35\,^\circ\mathrm{C}\,\mathrm{decade}^{-1}\) lies between them.

Studies that summarize warming over fixed windows or through
assessment-style attribution methods tend to obtain rates close to our
parametric benchmarks. \citet{ForsterEtAl2026} report an observed warming trend of \(0.30\,^\circ\mathrm{C}\,\mathrm{decade}^{-1}\) over 2016--2025 and assess the human-induced rate at \(0.27\,^\circ\mathrm{C}\,\mathrm{decade}^{-1}\) (likely range \(0.2\)--\(0.4\)). \citet{FR2026} obtain post-1970 quadratic endpoint rates of \(0.25\)--\(0.29\,^\circ\mathrm{C}\,\mathrm{decade}^{-1}\) and post-changepoint slopes of \(0.34\)--\(0.42\,^\circ\mathrm{C}\,\mathrm{decade}^{-1}\). Similarly,
\citet{KirchengastPichler2025} report recent trend rates rising from about
\(0.19\) to \(0.25\,^\circ\mathrm{C}\,\mathrm{decade}^{-1}\), and
\citet{SamsetEtAl2023} find filtered warming rates that increase as the estimation
window is shortened. These estimates concern different periods and target quantities; in particular, the attributed rate is distinct from our smooth observational component. Their magnitudes are broadly consistent with our quadratic and changepoint
summaries, but remain below the flexible local estimates from the
state-space and local-polynomial methods.

Fitting linear trends to the annual means of our five state-space warming paths over 2016--2025 gives a mean trend of \(0.35\,^\circ\mathrm{C}\,\mathrm{decade}^{-1}\). Those same paths give a mean endpoint rate of \(0.45\,^\circ\mathrm{C}\,\mathrm{decade}^{-1}\) in mid-2026. Part of the difference between published decade-average rates and our endpoint estimate thus reflects the period over which warming is measured.

\subsection{Recent acceleration}

Our acceleration results speak directly to the recent finding of
\citet{FR2026} that global warming has accelerated significantly. Their study
tests the null hypothesis of no change in the warming rate since 1970 using two
parametric alternatives: a quadratic trend model and a continuous
piecewise-linear changepoint model. Both tests are applied after adjusting the
temperature records for the estimated effects of ENSO, volcanic activity, and
solar variability. For the adjusted data, \citet{FR2026} find statistically
significant evidence of acceleration using both approaches, with the
changepoint model locating a slope change around 2013--2014. These results
provide an important benchmark for our analysis, but they are conditional on the
chosen parametric trend specifications and on a two-step adjustment procedure in
which climate covariates are partialled out before the trend models are
estimated. 

Our results corroborate the main conclusion of \citet{FR2026}, but from a
different statistical perspective. The semiparametric state-space framework avoids strong parametric restrictions and
 jointly estimates the evolving warming path and covariate effects in a flexible way,
while simultaneously accounting for serial dependence and time-varying error variance.  In this sense, the paper extends the flexible local and
adaptive monitoring perspective, which has previously been used primarily
to estimate the current warming level or climate mean
\citep{ClarkeRichardson2021,WuEtAl2011,RigalEtAl2019,
ScherrerEtAl2024,NicklasEtAl2025}, to the pathwise estimation of warming
rates and accelerations.

Differences in estimated acceleration across studies partly reflect the trend restrictions and time horizons used in the analysis.
Our quadratic benchmark yields acceleration estimates close
to those reported by \citet{Richardson2022} and \citet{FR2026}, whereas the
larger state-space and local-polynomial estimates are obtained without imposing
constant curvature over the post-1970 window. Recent-window evidence points in the same direction. For
example, \citet{JenkinsEtAl2022} report observed GMST trends increasing from
about \(0.18\) to \(0.35\,^\circ\mathrm{C}\,\mathrm{decade}^{-1}\) between
2000--2009 and 2010--2019, while \citet{SamsetEtAl2023} find increasing
filtered warming rates over shorter windows.

The concerns raised by \citet{BeaulieuEtAl2024} highlight the
importance of distinguishing changes in the underlying warming
rate from persistent variability. Our framework addresses this
issue by jointly estimating the warming path and transient
covariate effects, explicitly modelling serial dependence and
time-varying error variance, and incorporating latent-state and
parameter uncertainty through bootstrap inference. It allows
acceleration to evolve flexibly without requiring a discrete
breakpoint. Our analysis does not directly assess Richardson's
(\citeyear{Richardson2022}) concern that uncertainty estimated
from the observed record may understate the contribution of
internal climate variability. Assessing this concern would
require evaluating the method under alternative representations
of internal climate variability, for example using
climate-model ensembles.

Further, the physical interpretation of the estimated acceleration is a
separate question. Earth's energy imbalance more than doubled between 1976--1995 and 2006--2025 \citep{ForsterEtAl2026}, and \citet{MerchantEtAl2025} link the growing imbalance to accelerating sea-surface warming in the satellite record. Changes in aerosol forcing may contribute to
the recent increase in warming rates \citep{JenkinsEtAl2022}.
Our kinematic model captures such contributions through curvature
in \(x(t)\), but does not separately identify their physical causes.

\newpage
\subsection{Conditional extrapolations}

\begin{figure}[!htbp]
\centering
\includegraphics[width=\textwidth]{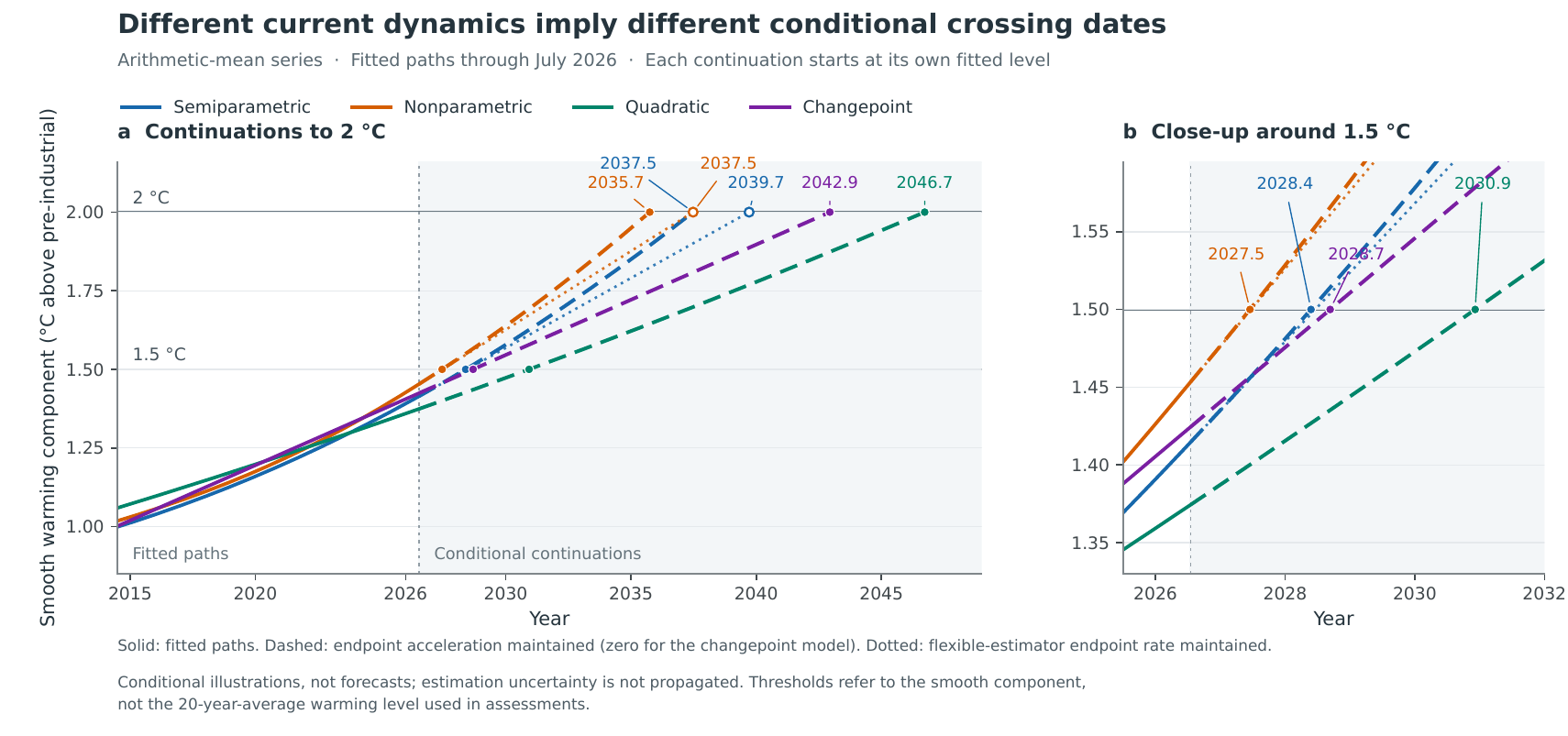}
\caption{\textbf{Conditional continuations of the estimated warming paths.}
Solid curves show the smooth warming paths fitted to the arithmetic-mean series through July 2026. Dashed curves continue each path at its estimated endpoint acceleration. Dotted curves hold the semiparametric and nonparametric endpoint rates constant. Panel~a shows the \(2\,^\circ\mathrm{C}\) crossings, with open markers for the constant-rate cases. Panel~b enlarges the region around \(1.5\,^\circ\mathrm{C}\).}
\label{fig:continuations_constant_rate}
\end{figure}

Figure~\ref{fig:continuations_constant_rate} shows when the arithmetic-mean warming path would reach \(1.5\,^\circ\mathrm{C}\) and \(2\,^\circ\mathrm{C}\) if the estimated endpoint dynamics persist. Maintaining the semiparametric and nonparametric
endpoint accelerations brings the smooth component to \(2\,^\circ\mathrm{C}\)
around 2037 and 2036, respectively, compared with around 2043 under the
changepoint continuation and 2047 under the quadratic continuation. Holding
the semiparametric and nonparametric rates constant gives crossings around 2040 and 2037, respectively, so much of the difference from the changepoint continuation arises from the higher current rates. These calculations are
conditional illustrations, not forecasts: they do not propagate estimation
uncertainty and concern the smooth observational component rather than the
20-year-average warming level used in assessments. Appendix~\ref{app:extrapolation} gives the calculation and crossing dates for both thresholds (Tables~\ref{tab:15c_continuations} and~\ref{tab:2c_continuations}).

\FloatBarrier

\section{Conclusion}
\label{sec:conclusion}

This paper has developed a semiparametric kinematic framework for estimating the
warming level, warming rate, and warming acceleration jointly from monthly
global temperature records. The framework estimates the latent kinematic path
\((x(t),v(t),a(t))\), with the current endpoint values
\(x(T)\), \(v(T)\), and \(a(T)\) serving as main empirical summaries. The
central distinction is between flexible methods that estimate an
evolving local kinematic path and parametric benchmarks that provide
long-window summaries. This distinction matters empirically: methods can agree
closely on the warming level while giving materially different answers about the
current warming rate and recent acceleration.

By analyzing five major temperature records, our study finds three key results concerning warming level, rate, and acceleration. Firstly,  the current latent warming level is estimated robustly at approximately \(1.42\,^\circ\mathrm{C}\) above pre-industrial conditions. Under the latent endpoint definition used here, the state-space point estimates place the underlying warming level below \(1.5\,^\circ\mathrm{C}\) in mid-2026, although the uncertainty intervals do not uniformly exclude that threshold. Secondly, the current warming rate is more estimator-sensitive. Flexible local estimators imply substantially higher current warming rates than the long-window benchmarks, with the preferred state-space model estimating \(v(T)\) at approximately \(0.45\,^\circ\mathrm{C}\) per decade. Thirdly, we find evidence of positive recent warming acceleration. In the preferred state-space model, nominal \(90\%\) intervals for average acceleration over the past decade lie above zero in all five temperature records, while endpoint acceleration estimates are also positive but less precisely determined. Together, these results suggest that recent warming has proceeded
faster, and accelerated more strongly, than conventional long-window
parametric estimators imply.

\subsection*{Use of AI tools}
We used ChatGPT to assist with coding, manuscript drafting, language editing and proofreading. We used Claude to obtain feedback on earlier drafts. The authors evaluated the suggestions and take responsibility for the final manuscript.

\section*{Data availability}
Historical covariate inputs through 2024 were obtained from the replication data accompanying \citet{FR2026}, available at \url{https://doi.org/10.5281/zenodo.15608929}. Temperature records are available from NASA GISTEMP (\url{https://data.giss.nasa.gov/gistemp/}), NOAA GlobalTemp (\url{https://www.ncei.noaa.gov/products/land-based-station/noaa-global-temp}), the Met Office Hadley Centre (\url{https://www.metoffice.gov.uk/hadobs/hadcrut5/}), Berkeley Earth (\url{https://berkeleyearth.org/data/}) and the Copernicus Climate Data Store (\url{https://cds.climate.copernicus.eu/}). Covariate sources and extension conventions are documented in Section~\ref{sec:data}. The Zenodo replication package \citep{BennedsenVeraValdes2026Replication} provides a fixed snapshot of the analysis inputs and numerical results.
Interactive visualizations of the semiparametric estimates are available at \url{https://sites.google.com/site/mbennedsen/research/global-warming-kinematics}. The website is updated as new observations become available and may therefore display estimates based on a later data vintage than those reported here.

\section*{Code availability}
The MATLAB replication package for the empirical analysis and simulation studies is archived on Zenodo \citep[version v1.0.1]{BennedsenVeraValdes2026Replication}. The authors' code and software documentation are released under the MIT License. Third-party data retain their providers' terms.

\section*{Acknowledgements}
We thank Siem Jan Koopman for valuable input and helpful comments. We are grateful for financial support from the Center for Research in Energy:
Economics and Markets (CoRE). M.B. further acknowledges financial support from the Aarhus Center for Econometrics (ACE) funded by the Danish National Research
Foundation grant number DNRF186.



 \singlespacing
\begin{spacing}{0.65}
	\phantomsection
	\addcontentsline{toc}{section}{References}
	\bibliography{arxiv_v60_support/references}

@article{Rodriguez2012,
	author = {Rodr\'{\i}guez, Alejandro and Ruiz, Esther},
	journal = {Computational Statistics \& Data Analysis},
	number = {1},
	pages = {62--74},
	title = {{Bootstrap prediction mean squared errors of unobserved states based on the Kalman filter with estimated parameters}},
	volume = {56},
	year = {2012}}

@article{Betts2023,
	author = {Richard A. Betts and Stephen E. Belcher and Leon Hermanson and Albert Klein Tank and Jason A. Lowe and Chris D. Jones and Colin P. Morice and Nick A. Rayner and Adam A. Scaife and Peter A. Stott},
	journal = {Nature},
	pages = {33-35},
	title = {Approaching $1.5^\circ${C}: how will we know we've reached this crucial warming mark?},
	volume = {624},
	year = {2023}}

@article{wahba1978,
	author = {Wahba, Grace},
	journal = {Journal of the Royal Statistical Society, Series B},
	number = {3},
	pages = {364--372},
	title = {Improper Priors, Spline Smoothing and the Problem of Guarding Against Model Errors in Regression},
	volume = {40},
	year = {1978}}

@article{weckeransley1983,
	author = {Wecker, William E. and Ansley, Craig F.},
	journal = {Journal of the American Statistical Association},
	number = {381},
	pages = {81--89},
	title = {The Signal Extraction Approach to Nonlinear Regression and Spline Smoothing},
	volume = {78},
	year = {1983}}

@article{kohnansley1987,
	author = {Kohn, Robert and Ansley, Craig F.},
	journal = {SIAM Journal on Scientific and Statistical Computing},
	number = {1},
	pages = {33--48},
	title = {A New Algorithm for Spline Smoothing Based on Smoothing a Stochastic Process},
	volume = {8},
	year = {1987}}

@article{BHK2024,
	author = {Bennedsen, Mikkel and Hillebrand, Eric and Koopman, Siem Jan},
	journal = {Nature Communications},
	number = {1},
	pages = {8507},
	title = {A regression-based approach to the {CO}2 airborne fraction},
	volume = {15},
	year = {2024}}

@article{BeaulieuKillick2018,
	author = {Beaulieu, Claudie and Killick, Rebecca},
	journal = {Journal of Climate},
	number = {23},
	pages = {9519--9543},
	title = {Distinguishing Trends and Shifts from Memory in Climate Data},
	volume = {31},
	year = {2018}}

@article{ScherrerEtAl2024,
	author = {Scherrer, Simon C. and de Valk, Cees and Begert, Michael and Gubler, Stefanie and Kotlarski, Sven and Croci-Maspoli, Mischa},
	date = {2024/01/01/},
	doi = {https://doi.org/10.1016/j.cliser.2023.100428},
	isbn = {2405-8807},
	journal = {Climate Services},
	pages = {100428},
	title = {Estimating trends and the current climate mean in a changing climate},
	url = {https://www.sciencedirect.com/science/article/pii/S2405880723000900},
	volume = {33},
	year = {2024}}

@article{RigalEtAl2019,
	author = {Rigal, Alix and Aza{\"\i}s, Jean-Marc and Ribes, Aur{\'e}lien},
	date = {2019/07/01},
	doi = {10.1007/s00382-018-4584-6},
	id = {Rigal2019},
	isbn = {1432-0894},
	journal = {Climate Dynamics},
	number = {1},
	pages = {275--286},
	title = {Estimating daily climatological normals in a changing climate},
	url = {https://doi.org/10.1007/s00382-018-4584-6},
	volume = {53},
	year = {2019}}

@article{Almon1965,
	author = {Shirley Almon},
	journal = {Econometrica},
	number = {1},
	pages = {178--196},
	title = {The Distributed Lag Between Capital Appropriations and Expenditures},
	volume = {33},
	year = {1965}}

@article{Fruhwirth-Schnatter1994,
	author = {Fr{\"u}hwirth-Schnatter, Sylvia},
	journal = {Journal of Time Series Analysis},
	number = {2},
	pages = {183--202},
	title = {DATA AUGMENTATION AND DYNAMIC LINEAR MODELS},
	volume = {15},
	year = {1994}}

@article{CarterKohn1994,
	author = {C. K. Carter and R. Kohn},
	journal = {Biometrika},
	number = {3},
	pages = {541--553},
	title = {On {Gibbs} Sampling for State Space Models},
	volume = {81},
	year = {1994}}

@article{FR2026,
	author = {G. Foster and S. Rahmstorf},
	doi = {https://doi.org/10.1029/2025GL118804},
	journal = {Geophysical Research Letters},
	number = {5},
	pages = {e2025GL118804},
	title = {Global Warming Has Accelerated Significantly},
	volume = {53},
	year = {2026}}

@article{FR2011,
	author = {G. Foster and S. Rahmstorf},
	doi = {https://doi.org/10.1088/1748-9326/6/4/044022},
	journal = {Environmental Research Letters},
	number = {4},
	pages = {044022},
	title = {Global temperature evolution 1979--2010},
	volume = {6},
	year = {2011}}

@article{NW1987,
	author = {W. K. Newey and K. D. West},
	doi = {https://doi.org/10.2307/1913610},
	journal = {Econometrica},
	number = {3},
	pages = {703--708},
	title = {A Simple, Positive Semi-Definite, Heteroskedasticity and Autocorrelation Consistent Covariance Matrix},
	volume = {55},
	year = {1987}}

@book{DK2012,
	author = {J. Durbin and S. J. Koopman},
	publisher = {Oxford University Press},
	title = {Time series analysis by state space methods},
	year = {2012}}

@article{ClarkeRichardson2021,
	author = {D. C. Clarke and M. Richardson},
	doi = {https://doi.org/10.1029/2020EA001082},
	journal = {Earth and Space Science},
	number = {5},
	pages = {e2020EA001082},
	title = {The Benefits of Continuous Local Regression for Quantifying Global Warming},
	volume = {8},
	year = {2021}}

@article{NicklasEtAl2025,
	author = {J. Matthew Nicklas and Baylor Fox-Kemper and Charles Lawrence},
	doi = {https://doi.org/10.1175/JCLI-D-23-0580.1},
	journal = {Journal of Climate},
	number = {5},
	pages = {1235--1270},
	title = {Efficient Estimation of Climate State and Its Uncertainty Using {K}alman Filtering with Application to Policy Thresholds and Volcanism},
	volume = {38},
	year = {2025}}

@article{ThorneEtAl2026,
	author = {Peter W. Thorne and John M. Nicklas and John J. Kennedy and Bruce Calvert and Baylor Fox-Kemper and Mark T. Richardson and Adrian Simmons and Ed Hawkins and Robert Rohde and Kathryn Cowtan and others},
	doi = {https://doi.org/10.5194/essd-2025-825},
	journal = {Earth System Science Data Discussions},
	note = {Preprint under review},
	title = {How well can we quantify when 1.5 $^\circ${C} of global warming has been exceeded?},
	year = {2026}}

@article{KirchengastPichler2025,
	author = {Gottfried Kirchengast and Moritz Pichler},
	doi = {https://doi.org/10.1038/s43247-025-02368-0},
	journal = {Communications Earth \& Environment},
	number = {1},
	pages = {402},
	title = {A traceable global warming record and clarity for the 1.5$^\circ${C} and well-below-2$^\circ${C} goals},
	volume = {6},
	year = {2025}}

@article{WuEtAl2011,
	author = {Zhaohua Wu and Norden E. Huang and John M. Wallace and Brian V. Smoliak and Xianyao Chen},
	doi = {https://doi.org/10.1007/s00382-011-1128-8},
	journal = {Climate Dynamics},
	number = {3--4},
	pages = {759--773},
	title = {On the time-varying trend in global-mean surface temperature},
	volume = {37},
	year = {2011}}

@article{SamsetEtAl2023,
	author = {B. H. Samset and C. Zhou and J. S. Fuglestvedt and M. T. Lund and J. Marotzke and M. D. Zelinka},
	doi = {https://doi.org/10.1038/s43247-023-01061-4},
	journal = {Communications Earth \& Environment},
	number = {1},
	pages = {400},
	title = {Steady global surface warming from 1973 to 2022 but increased warming rate after 1990},
	volume = {4},
	year = {2023}}

@article{CahillEtAl2015,
	author = {Niamh Cahill and Stefan Rahmstorf and Andrew C. Parnell},
	doi = {https://doi.org/10.1088/1748-9326/10/8/084002},
	journal = {Environmental Research Letters},
	number = {8},
	pages = {084002},
	title = {Change points of global temperature},
	volume = {10},
	year = {2015}}

@article{BeaulieuEtAl2024,
	author = {Claudie Beaulieu and Colm Gallagher and Rebecca Killick and Robert Lund and Xiaogang Shi},
	doi = {https://doi.org/10.1038/s43247-024-01711-1},
	journal = {Communications Earth \& Environment},
	number = {1},
	pages = {576},
	title = {A recent surge in global warming is not detectable yet},
	volume = {5},
	year = {2024}}

@article{Richardson2022,
	author = {Mark T. Richardson},
	doi = {https://doi.org/10.1029/2021GL095782},
	journal = {Geophysical Research Letters},
	number = {2},
	pages = {e2021GL095782},
	title = {Prospects for Detecting Accelerated Global Warming},
	volume = {49},
	year = {2022}}

@article{JenkinsEtAl2022,
	author = {Stuart Jenkins and Adam Povey and Andrew Gettelman and Roy Grainger and Philip Stier and Myles Allen},
	doi = {https://doi.org/10.1175/JCLI-D-22-0081.1},
	journal = {Journal of Climate},
	number = {24},
	pages = {7873--7890},
	title = {Is Anthropogenic Global Warming Accelerating?},
	volume = {35},
	year = {2022}}

@article{Mudelsee2019,
	author = {Manfred Mudelsee},
	doi = {https://doi.org/10.1016/j.earscirev.2018.12.005},
	journal = {Earth-Science Reviews},
	pages = {310--322},
	title = {Trend analysis of climate time series: A review of methods},
	volume = {190},
	year = {2019}}

@misc{GISTEMPTeam2024,
	author = {{GISTEMP Team}},
	note = {[Dataset]. {NASA} Goddard Institute for Space Studies},
	title = {{GISS} Surface Temperature Analysis ({GISTEMP}), version 4},
	url = {https://data.giss.nasa.gov/gistemp/},
	year = {2024}}

@article{LenssenEtAl2024,
	author = {Nathan Lenssen and Gavin A. Schmidt and Michael Hendrickson and Peter Jacobs and Matthew J. Menne and Reto Ruedy},
	doi = {10.1029/2023JD040179},
	journal = {Journal of Geophysical Research: Atmospheres},
	number = {17},
	pages = {e2023JD040179},
	title = {A {NASA} {GISTEMPv4} observational uncertainty ensemble},
	volume = {129},
	year = {2024}}

@article{VoseEtAl2021,
	author = {Russell S. Vose and Boyin Huang and Xungang Yin and Derek Arndt and David R. Easterling and Jay H. Lawrimore and others},
	doi = {10.1029/2020GL090873},
	journal = {Geophysical Research Letters},
	number = {4},
	pages = {e2020GL090873},
	title = {Implementing Full Spatial Coverage in {NOAA}'s Global Temperature Analysis},
	volume = {48},
	year = {2021}}

@article{MoriceEtAl2021,
	author = {Colin P. Morice and John J. Kennedy and Nick A. Rayner and Jason P. Winn and Emma Hogan and Rebecca E. Killick and others},
	doi = {10.1029/2019JD032361},
	journal = {Journal of Geophysical Research: Atmospheres},
	number = {3},
	pages = {e2019JD032361},
	title = {An updated assessment of near-surface temperature change from 1850: The {HadCRUT5} dataset},
	volume = {126},
	year = {2021}}

@article{RohdeHausfather2020,
	author = {Robert A. Rohde and Zeke Hausfather},
	doi = {10.5194/essd-12-3469-2020},
	journal = {Earth System Science Data},
	number = {4},
	pages = {3469--3479},
	title = {The {Berkeley Earth} Land/Ocean Temperature Record},
	volume = {12},
	year = {2020}}

@misc{HersbachEtAl2017,
	author = {Hans Hersbach and Bill Bell and Paul Berrisford and Shoji Hirahara and Andr{\'a}s Hor{\'a}nyi and Joaqu{\'\i}n Mu{\~n}oz-Sabater and others},
	doi = {10.24381/cds.143582cf},
	note = {[Dataset]. Copernicus Climate Change Service ({C3S}) Climate Data Store ({CDS})},
	title = {Complete {ERA5} global atmospheric reanalysis},
	year = {2017}}

@article{HersbachEtAl2020,
	author = {Hans Hersbach and Bill Bell and Paul Berrisford and Shoji Hirahara and Andr{\'a}s Hor{\'a}nyi and Joaqu{\'\i}n Mu{\~n}oz-Sabater and others},
	doi = {10.1002/qj.3803},
	journal = {Quarterly Journal of the Royal Meteorological Society},
	number = {730},
	pages = {1999--2049},
	title = {The {ERA5} global reanalysis},
	volume = {146},
	year = {2020}}

@misc{MetOfficeHadleyCentre2003,
	author = {{Met Office Hadley Centre}},
	note = {[Dataset]},
	title = {Hadley Centre Sea Ice and Sea Surface Temperature data set ({HadISST})},
	url = {https://www.metoffice.gov.uk/hadobs/hadisst},
	year = {2003}}

@article{SatoEtAl1993,
	author = {Makiko Sato and James E. Hansen and M. Patrick McCormick and John B. Pollack},
	doi = {10.1029/93JD02553},
	journal = {Journal of Geophysical Research},
	number = {D12},
	pages = {22987--22994},
	title = {Stratospheric aerosol optical depths, 1850--1990},
	volume = {98},
	year = {1993}}

@article{StockWatson1998,
	author = {Stock, James H. and Watson, Mark W.},
	doi = {10.1080/01621459.1998.10474116},
	journal = {Journal of the American Statistical Association},
	number = {441},
	pages = {349--358},
	title = {Median Unbiased Estimation of Coefficient Variance in a Time-Varying Parameter Model},
	volume = {93},
	year = {1998}}

@article{ForsterEtAl2026,
  author = {Forster, Piers M. and Walsh, Tristram and Smith, Chris and Lamb, William F. and Lamboll, Robin and others},
  title = {Indicators of Global Climate Change 2025: annual update of key indicators of the state of the climate system and human influence},
  journal = {Earth System Science Data},
  year = {2026},
  volume = {18},
  pages = {3889--3933},
  doi = {10.5194/essd-18-3889-2026}
}

@article{MerchantEtAl2025,
  author = {Merchant, Christopher J. and Allan, Richard P. and Embury, Owen},
  title = {Quantifying the acceleration of multidecadal global sea surface warming driven by {E}arth's energy imbalance},
  journal = {Environmental Research Letters},
  year = {2025},
  volume = {20},
  number = {2},
  pages = {024037},
  doi = {10.1088/1748-9326/adaa8a}
}

@misc{BennedsenVeraValdes2026Replication,
  author = {Bennedsen, Mikkel and Vera-Vald{\'e}s, J. Eduardo},
  title = {Replication package for ``{C}onventional trend models may understate the current rate of global warming''},
  howpublished = {Version v1.0.1. Zenodo \url{https://doi.org/10.5281/zenodo.24570241}},
  year = {2026},
  doi = {10.5281/zenodo.24570241}
}
\end{spacing}

\clearpage
\newpage

\appendix

\section{Details on the semiparametric state-space model}
\label{app:ssmodel}

\subsection{Model specification}

\subsubsection{Latent kinematic states}

Let \(x(t)\) denote the latent warming level, with derivatives
\[
    \dot{x}(t) = v(t),
    \qquad
    \ddot{x}(t) = \dot{v}(t) = a(t),
\]
where \(v(t)\) is the warming rate and \(a(t)\) is the warming acceleration. At
the monthly observation grid, let
\[
    (x_n,v_n,a_n)
    =
    (x(t_n),v(t_n),a(t_n)),
    \qquad
    t_n=t_1 + (n-1)\Delta,
    \qquad
    n=1,\ldots,N,
\]
where \(\Delta=1/12\) is one month measured in years.

\subsubsection{Measurement equation}

Observed temperature anomalies are noisy measurements of the latent warming
level after controlling for observed covariates. The measurement equation is
\begin{equation}
    y_n
    =
    x_n + z_n'\beta + \varepsilon_n,
    \label{eq:meas}
\end{equation}
where \(z_n\) contains the observed controls and their distributed lags, and
\(\varepsilon_n\) is a persistent irregular component capturing remaining
short-run variability and measurement error.

\subsubsection{Discrete-time kinematics}

We specify the continuous-time law of motion for the warming acceleration as
\[
    \mathrm{d}a(t)
    =
    \sigma\,\mathrm{d}W(t),
\]
where \(\sigma>0\) and \(W(t)\) is a standard Brownian motion. Integrating the
continuous-time system over the interval \([t_{n-1},t_n]\) gives the exact
discrete-time kinematic equations
\begin{align}
    a_n
    &=
    a_{n-1} + \eta_{3,n},
    \label{eq:an}\\
    v_n
    &=
    v_{n-1} + a_{n-1}\Delta + \eta_{2,n},
    \label{eq:vt}\\
    x_n
    &=
    x_{n-1} + v_{n-1}\Delta
    + \tfrac{1}{2}a_{n-1}\Delta^2 + \eta_{1,n}.
    \label{eq:xt}
\end{align}
The state innovations are
\begin{align*}
    \eta_{3,n}
    &=
    \sigma
    \int_{t_{n-1}}^{t_n}
    \mathrm{d}W(s), \\
    \eta_{2,n}
    &=
    \sigma
    \int_{t_{n-1}}^{t_n}
    (t_n-s)\,\mathrm{d}W(s), \\
    \eta_{1,n}
    &=
    \frac{\sigma}{2}
    \int_{t_{n-1}}^{t_n}
    (t_n-s)^2\,\mathrm{d}W(s).
\end{align*}
Let
\[
    \eta_n
    =
    (\eta_{1,n},\eta_{2,n},\eta_{3,n})' .
\]
Then \(\eta_n\sim\mathcal{N}(0,Q_{xva})\), where
\begin{align}
Q_{xva}
=
\sigma^2
\begin{pmatrix}
 \Delta^5 / 20 & \Delta^4 / 8 & \Delta^3 / 6 \\
 \Delta^4 / 8 & \Delta^3 / 3 & \Delta^2 / 2 \\
 \Delta^3 / 6 & \Delta^2 / 2 & \Delta
\end{pmatrix}.
\label{eq:Qxva}
\end{align}
In particular,
\[
    \operatorname{Var}(a_n-a_{n-1})
    =
    \sigma^2\Delta.
\]
The diffusion scale \(\sigma\) governs the smoothness of the latent path. Over one monthly interval, the acceleration innovation has standard deviation $\sqrt{\operatorname{Var}(a_n-a_{n-1})} =  \sigma\sqrt{\Delta}$. Smaller values of \(\sigma\) impose a smoother acceleration path, and hence smoother
implied paths for the warming rate and warming level, whereas larger values of
\(\sigma\) allow the latent kinematic states to adapt more rapidly to recent changes.

\subsubsection{Irregular component}

Monthly temperature anomalies exhibit short-run persistence beyond what is
captured by the latent kinematic state \((x_n,v_n,a_n)\) and the observed
covariates \(z_n\). We model this remaining component as an ARMA\((1,1)\)
process:
\begin{equation}
    \varepsilon_n
    =
    \rho_\varepsilon \varepsilon_{n-1}
    + u_n
    + \theta_\varepsilon u_{n-1},
    \qquad
    u_n \stackrel{independent}{\sim} \mathcal{N}(0,R_n),
    \label{eq:irregular}
\end{equation}
where \(R_n\) is allowed to vary over time.

\subsubsection{Time-varying irregular variance}

Diagnostics indicate that the variance of the irregular innovations differs by
calendar month. We therefore allow a multiplicative month-specific variance:
\begin{equation}
    R_n
    =
    \sigma_\varepsilon^2
    \exp\{\gamma_{m(n)}\}
    \exp\{\delta_{\operatorname{era}(n)}\},
    \label{eq:Rt}
\end{equation}
where \(m(n)\in\{1,\ldots,12\}\) indexes the month of the year and
\(\operatorname{era}(n)\) indexes possible historical variance regimes. The
month effects are normalized by imposing
\[
    \sum_{m=1}^{12}\gamma_m = 0,
\]
and the first era effect is normalized by setting \(\delta_1=0\). In the final
specification used in the main analysis, the month effects are retained, while
era-specific variance breaks are set to zero. This choice follows the
model-selection exercise described in Section~\ref{app:ss-model-selection}.

\subsection{State-space representation}

For the individual semiparametric specification, let the latent state be
\begin{equation}
    \alpha_n = (x_n, v_n, a_n, e_n, u_n)' ,
\end{equation}
where \(x_n\) is the latent warming level, \(v_n\) is the warming rate,
\(a_n\) is the warming acceleration, and \((e_n,u_n)\) provides a
state-space representation of the ARMA\((1,1)\) irregular component, with
\(e_n=\varepsilon_n\). The transition equation is
\begin{equation}
    \alpha_n = F\alpha_{n-1} + w_n,
    \qquad
    w_n \sim \mathcal{N}(0,Q_n),
    \label{eq:state}
\end{equation}
with
\begin{equation}
F =
\begin{pmatrix}
1 & \Delta & \tfrac{1}{2}\Delta^2 & 0 & 0\\
0 & 1 & \Delta & 0 & 0\\
0 & 0 & 1 & 0 & 0\\
0 & 0 & 0 & \rho_\varepsilon & \theta_\varepsilon\\
0 & 0 & 0 & 0 & 0
\end{pmatrix}.
\label{eq:F_individual}
\end{equation}
The upper-left \(3\times 3\) block is the kinematic trend system. The
lower-right block gives the ARMA\((1,1)\) irregular component
\begin{equation}
    e_n
    =
    \rho_\varepsilon e_{n-1}
    + u_n
    + \theta_\varepsilon u_{n-1},
\end{equation}
where the second irregular state stores the current innovation \(u_n\).

The state innovation covariance is block diagonal,
\begin{equation}
    Q_n =
    \begin{pmatrix}
        Q_{xva} & 0\\
        0 & Q_{e,n}
    \end{pmatrix},
    \label{eq:Qn_individual}
\end{equation}
where $Q_{xva}$ is given in \eqref{eq:Qxva} and the ARMA innovation covariance is
\begin{equation}
Q_{e,n}
=
\sigma_{u,n}^2
\begin{pmatrix}
1 & 1\\
1 & 1
\end{pmatrix},
\end{equation}
because the same innovation \(u_n\) enters both the irregular component \(e_n\)
and the innovation state. Here \(\sigma_{u,n}^2=R_n\), with
\(\sigma_u^2=\sigma_\varepsilon^2\) and \(r(n)=\operatorname{era}(n)\)
in the variance specification above. Thus,
\begin{equation}
    \sigma_{u,n}^2
    =
    \sigma_u^2
    \exp\!\left(\gamma_{m(n)}+\delta_{r(n)}\right),
    \label{eq:sigmaun}
\end{equation}
where \(m(n)\) indexes calendar month and \(r(n)\) indexes variance regimes if
variance breaks are included. In the specification used in the main analysis,
monthly variance effects are allowed, while variance breaks are set to zero.

The measurement equation is
\begin{equation}
    y_n = H\alpha_n + z_n'\beta,
    \qquad
    H = (1,0,0,1,0),
    \label{eq:meas2}
\end{equation}
so that the observed temperature anomaly is the sum of the latent warming level
\(x_n\), the ARMA irregular component \(e_n\), and the contribution of observed
covariates \(z_n\). There is therefore no additional white-noise measurement
disturbance in the baseline individual specification beyond the irregular
component itself.

\subsection{Estimation and inference}
\label{sec:estimation}

\subsubsection{Maximum likelihood estimation}

Let
\begin{equation}
    \theta
    =
    (\beta,\rho_\varepsilon,\theta_\varepsilon,\sigma^2,\sigma_u^2,\gamma,\delta)
\end{equation}
collect the static parameters, where \(\rho_\varepsilon\) and
\(\theta_\varepsilon\) are the autoregressive and moving-average parameters of
the irregular component, \(\sigma\) is the standard deviation parameter of the acceleration dynamics, \(\gamma\) collects the month-specific log-variance
effects, and \(\delta\) collects any additional variance-shift parameters. Given
\(\theta\), the model is linear and Gaussian, with time-varying state innovation
covariance \(Q_n\). The log-likelihood is evaluated by the Kalman filter and
maximized numerically.

\subsubsection{Smoothed estimates of the main estimands}
Given the maximum-likelihood estimate \(\hat{\theta}\), we compute smoothed
state estimates \(\hat{\alpha}_{n\mid N}\) using the
Rauch--Tung--Striebel smoother. The main quantities of interest are the
components of the state vector representing the latent warming level \(x_n\),
the warming rate \(v_n\), and the warming acceleration \(a_n\).

\subsubsection{Inference}\label{app:ssinference}

Conditional on \(\hat{\theta}\), inference for the unobserved states
\((x_n,v_n,a_n)\), \(n=1,\ldots,N\), can be based on the Kalman smoother.
To also reflect the finite-sample effect of estimating the static parameters $\theta$,
we propose a ``local'' bootstrap-based approach. A conventional parametric bootstrap
would simulate \((x_n,v_n,a_n)\) forward from the fitted model. With the
 random-walk specification for acceleration, however, such long
forward simulations can generate latent warming histories far from the
observed record. The local procedure instead conditions its latent warming
trajectory on the observed data and regenerates only the irregular component. Related conditional bootstrap approaches for state-space models are discussed by, e.g., \cite{Rodriguez2012}.
Our approach is described in Appendix~\ref{app:ssbootstrap}. Its finite-sample
calibration is assessed in Appendix~\ref{app:ssboot_mc}.

\subsubsection{Confidence interval via bootstrap}
\label{app:ssbootstrap}

Fix \(n\in\{1,\ldots,N\}\), and let \(\psi_n\) denote one scalar component
of \((x_n,v_n,a_n)\). Let \(B\) be the number of bootstrap replications. For
replication \(b=1,\ldots,B\), proceed as follows.

\begin{enumerate}
	\item \textbf{Draw a latent warming path.} Using the backward-sampling
    simulation smoother of \citet{CarterKohn1994} and \citet{Fruhwirth-Schnatter1994},
    draw a complete state path
    \[
        \tilde{\alpha}_{1:N}^{(b)}
        \sim p(\alpha_{1:N}\mid y_{1:N},\hat{\theta}).
    \]
    Write
    \[
        \tilde{\alpha}_t^{(b)}
        =
        \left(
            \tilde{x}_t^{(b)},\tilde{v}_t^{(b)},\tilde{a}_t^{(b)},
            \tilde{r}_t^{(b)\prime}
        \right)^\prime,
    \]
    where \(r_t\) collects the ARMA irregular-state components. In what follows, we retain the
    \(x\)-\(v\)-\(a\) path $( \tilde{x}_t^{(b)},\tilde{v}_t^{(b)},\tilde{a}_t^{(b)})$, which also defines the known bootstrap truth
    \(\psi_n^{*(b)}\). This conditions the experiment on a latent warming
    history compatible with the observed record, rather than simulating an
    unrestricted long random-walk path. The remaining ARMA states are replaced by a newly generated irregular path,
initialized at \(r_1^{*(b)}=\tilde r_1^{(b)}\), as described in the next step.

\item \textbf{Regenerate the irregular component and observations.}
    Retain \((\tilde{x}_t^{(b)},\tilde{v}_t^{(b)},\tilde{a}_t^{(b)})\) for every
    $t = 1, \ldots, N$. The ARMA state is regenerated, retaining only its initial value from
    the simulation-smoother draw,
    \[
        r_1^{*(b)}=\tilde{r}_1^{(b)},
    \]
    and, for \(t=2,\ldots,N\), simulating
    \[
        r_t^{*(b)}=F_r(\hat{\theta})r_{t-1}^{*(b)}+u_t^{*(b)},
    \]
    where \(F_r(\hat{\theta})\) is the fitted ARMA transition matrix and the
    innovations \(u_t^{*(b)}\) use the fitted, calendar-time-specific
    innovation variances. The bootstrap temperature record is constructed on
    the original observation calendar as
    \[
        y_t^{*(b)}
        =
        H_{xva}
        \begin{pmatrix}
            \tilde{x}_t^{(b)}\\
            \tilde{v}_t^{(b)}\\
            \tilde{a}_t^{(b)}
        \end{pmatrix}
        +H_r r_t^{*(b)}+z_t^\prime\hat{\beta}.
    \]
    There is no separate observation-noise draw: the irregular component is
    represented in the state vector.

 \item \textbf{Re-estimate and smooth.} Re-estimate the model using
    \(y_{1:N}^{*(b)}\), obtaining \(\hat{\theta}^{*(b)}\), and apply the
    smoother to the bootstrap record. Let \(\hat{\psi}_n^{*(b)}\) denote the
    relevant component of the resulting smoothed state estimate.

 \item \textbf{Form the bootstrap error.} Record
    \[
        e_n^{*(b)}=\hat{\psi}_n^{*(b)}-\psi_n^{*(b)}.
    \]

\end{enumerate}

Writing \(q_p(e_n^*)\) for the empirical \(p\)-quantile of the bootstrap
errors, a two-sided \(100(1-\alpha)\%\) basic bootstrap interval is
\[
    \left[
        \hat{\psi}_n-q_{1-\alpha/2}(e_n^*),\,
        \hat{\psi}_n-q_{\alpha/2}(e_n^*)
    \right].
\]
We use \(B=1{,}999\) bootstrap replications for the monthly endpoint intervals, path bands, average-acceleration intervals, and annual-data intervals. Monthly path bands and endpoint and average-acceleration intervals are derived from the same retained bootstrap errors. At
\(n=N\), this construction gives intervals for the terminal estimands
\(x(T)\), \(v(T)\), and \(a(T)\). The same construction may be used to
obtain pointwise intervals at every date $n = 1, \ldots, N$ by retaining the bootstrap errors at
each date. Such intervals are pointwise, rather than simultaneous confidence
bands.

\subsubsection{Inference for average acceleration}
\label{app:rate_change_inference}
Both rates in \eqref{eq:rate_change} are estimated from the same full-sample
smoothed path. Average acceleration therefore describes retrospective evolution,
rather than a difference between two real-time estimates, and does not
require acceleration to be positive throughout the interval. Time is measured
in years, so the rate difference is divided by the elapsed time in years and
multiplied by 100 for reporting in
\(^{\circ}\mathrm{C}\,\mathrm{decade}^{-2}\).

For each retained bootstrap replication in Section~\ref{app:ssbootstrap},
let \(e_v^{*(b)}(t)=\hat v^{*(b)}(t)-v^{*(b)}(t)\) be the rate error.
The error for average acceleration in \eqref{eq:rate_change} is
\[
e_{\bar a}^{*(b)}=
\frac{e_v^{*(b)}(T)-e_v^{*(b)}(T_0)}{T-T_0}.
\]
Both terms use the same simulated path and refitted model. This pairing
preserves dependence between the two rate estimates; subtracting marginal
interval endpoints would not provide the corresponding interval for their
difference. After expressing estimates and errors in
\(^{\circ}\mathrm{C}\,\mathrm{decade}^{-2}\), the nominal \(90\%\) interval is
\[
\left[\widehat{\overline{a_{T_0,T}}}-q_{0.95}(e_{\bar a}^*),\,
      \widehat{\overline{a_{T_0,T}}}-q_{0.05}(e_{\bar a}^*)\right].
\]
For fixed dates, this is a rescaling of the paired rate-change interval and
leaves inference against zero unchanged. The analysis is conducted separately
for each temperature record, with the baseline averaging interval starting exactly ten years before the endpoint and alternative starting years reported in
Table~\ref{tab:rate_change_alt_t0}.

\subsection{Parameter estimates}
\label{app:param_estimates}

This section reports the parameter estimates from the baseline state-space
models. In the state-space specification, the regression coefficients, dynamic
parameters, irregular-component parameters, and variance parameters are estimated
jointly by maximum likelihood. The covariates are therefore not partialled out in
a first step in the state-space analysis; the observation equation is estimated
jointly with the latent warming path and the irregular component.

Table~\ref{tab:ss_param_dynamic} reports the main dynamic and variance
parameters. The parameter \(\sigma\) controls the variability of the stochastic
acceleration component and therefore governs the smoothness of the latent
warming path. Smaller values of \(\sigma\) imply smoother paths for \(a(t)\), and,
through integration, even smoother paths for \(v(t)\) and \(x(t)\). Since the
monthly time step is small, \(\Delta=1/12\), shocks to acceleration enter the
level and rate equations only through the small powers of \(\Delta\) in the
state covariance matrix. This is the discrete-time counterpart of the smoothing
interpretation discussed in Section~\ref{sec:ssmodel}. The table also reports
\(100\sqrt{\sigma^2\Delta}\), the implied monthly innovation standard deviation for
the acceleration state when acceleration is expressed in
\({}^\circ\mathrm{C}\,\mathrm{decade}^{-2}\). The AR and MA coefficients
describe the ARMA\((1,1)\) irregular component, while \(\sigma_u^2\) is the
baseline innovation variance before applying the month-specific variance
scales.

Tables~\ref{tab:ss_param_monthly_variance} and
\ref{tab:ss_param_monthly_residual_variance} report the estimated monthly
variance pattern of the irregular component. The temperature series are monthly
anomalies and therefore have the deterministic seasonal mean removed by
construction. The month-specific variance parameters are not seasonal mean
effects; they allow the remaining short-run innovation variance to differ across
calendar months. Specifically, if \(m(n)\) denotes the calendar month of
observation \(n\), the innovation variance of the irregular component is
\[
    \sigma_u^2 \exp\{\gamma_{m(n)}\}.
\]
Table~\ref{tab:ss_param_monthly_variance} reports the multiplicative factors
\(\exp(\gamma_m)\), while Table~\ref{tab:ss_param_monthly_residual_variance}
reports the corresponding month-specific innovation variances
\(\sigma_u^2\exp(\gamma_m)\).

Table~\ref{tab:ss_param_beta} reports the estimated coefficients for the Almon
distributed-lag basis used for the covariates. Let \(c_{k,n}\) denote covariate
\(k\), standardized before constructing the lag basis, and let \(L=12\) denote
the maximum lag length. The implied coefficient on lag \(\ell=0,\ldots,L\) is
parameterized as
\begin{align}
    \theta_{k,\ell}
    =
    \sum_{r=0}^{p_k} b_{k,r}
    \left(\frac{\ell}{L}\right)^r ,  \label{eq:AlmontCoeff}
\end{align}
where \(p_k\) is the polynomial degree for covariate \(k\). In the baseline
specification, \(p_k=2\) for NINO3.4 and \(p_k=1\) for volcanic aerosols and
sunspot numbers. The covariate contribution in the observation equation is
therefore
\[
    \sum_k \sum_{\ell=0}^{L} \theta_{k,\ell} c_{k,n-\ell}.
\]
Table~\ref{tab:ss_param_beta} reports the basis coefficients \(b_{k,r}\), while
Table~\ref{tab:ss_param_lag_weights} reports the implied lag weights
\(\theta_{k,\ell}\). The estimated NINO3.4 lag profiles are positive across the
lag window and decline over the first several months, the volcanic coefficients
are negative across the lag window, and the sunspot profiles are small in
magnitude. The broad similarity of these lag profiles across temperature records
indicates that the covariate adjustment is stable across datasets.

\begin{table}[t]
\centering
\scriptsize
\setlength{\tabcolsep}{4pt}
\caption{\textbf{State-space dynamic and variance parameter estimates.} The parameter $\sigma^2$ scales the continuous-time integrated-Brownian process covariance for acceleration; $100\sqrt{\sigma^2\Delta t}$ reports the monthly acceleration shock standard deviation in $^\circ$C decade$^{-2}$. $\sigma_u^2$ is the baseline residual innovation variance before monthly variance scaling.}\label{tab:ss_param_dynamic}
\begin{adjustbox}{max width=\linewidth}
\begin{tabular}{lcccccc}
\toprule
Dataset & $\sigma^2$ & $100\sqrt{\sigma^2\Delta t}$ & AR(1) & MA(1) & $\sigma_u^2$ & $\sigma_u$ \\
\midrule
NASA & $2.534\times 10^{-7}$ & 0.0145 & 0.712 & -0.275 & $9.223\times 10^{-3}$ & 0.0960 \\
NOAA & $2.080\times 10^{-7}$ & 0.0132 & 0.708 & -0.285 & $7.499\times 10^{-3}$ & 0.0866 \\
HadCRUT5 & $1.401\times 10^{-7}$ & 0.0108 & 0.743 & -0.303 & $8.934\times 10^{-3}$ & 0.0945 \\
Berkeley Earth & $1.652\times 10^{-7}$ & 0.0117 & 0.680 & -0.252 & $1.108\times 10^{-2}$ & 0.1053 \\
ERA5 & $4.828\times 10^{-8}$ & 0.0063 & 0.719 & -0.254 & $9.109\times 10^{-3}$ & 0.0954 \\
\bottomrule
\end{tabular}
\end{adjustbox}
\end{table}

\begin{table}[t]
\centering
\scriptsize
\setlength{\tabcolsep}{4pt}
\caption{\textbf{Covariate coefficient estimates.} Coefficients are for the Almon distributed-lag basis used in the state-space observation equation. Covariates are standardized before constructing lag bases.}\label{tab:ss_param_beta}
\begin{adjustbox}{max width=\linewidth}
\begin{tabular}{lccccc}
\toprule
Coefficient & NASA & NOAA & HadCRUT5 & Berkeley Earth & ERA5 \\
\midrule
NINO34 Almon degree 0 & 0.0202 & 0.0212 & 0.0208 & 0.0209 & 0.0219 \\
NINO34 Almon degree 1 & -0.0585 & -0.0659 & -0.0602 & -0.0600 & -0.0609 \\
NINO34 Almon degree 2 & 0.0472 & 0.0554 & 0.0488 & 0.0479 & 0.0493 \\
volc2 Almon degree 0 & -0.0041 & -0.0045 & -0.0064 & -0.0063 & -0.0058 \\
volc2 Almon degree 1 & $-1.343\times 10^{-5}$ & 0.0019 & 0.0036 & 0.0030 & 0.0039 \\
ssn Almon degree 0 & 0.0047 & 0.0038 & 0.0047 & 0.0021 & 0.0071 \\
ssn Almon degree 1 & -0.0090 & -0.0076 & -0.0089 & -0.0045 & -0.0112 \\
\bottomrule
\end{tabular}
\end{adjustbox}
\end{table}

\begin{table}[t]
\centering
\tiny
\setlength{\tabcolsep}{3pt}
\caption{\textbf{Monthly residual variance scales.} Entries report $\exp(\gamma_m)$, the multiplicative variance scale for month $m$ relative to the baseline residual innovation variance.}\label{tab:ss_param_monthly_variance}
\begin{adjustbox}{max width=\linewidth}
\begin{tabular}{lccccc}
\toprule
Month & NASA & NOAA & HadCRUT5 & Berkeley Earth & ERA5 \\
\midrule
Jan & 2.356 & 1.893 & 1.969 & 1.913 & 1.508 \\
Feb & 1.777 & 1.785 & 1.673 & 1.490 & 1.596 \\
Mar & 1.886 & 1.751 & 1.988 & 1.573 & 1.555 \\
Apr & 0.816 & 0.928 & 0.724 & 0.786 & 0.695 \\
May & 0.643 & 0.688 & 0.717 & 0.694 & 0.992 \\
Jun & 0.675 & 0.658 & 0.693 & 0.654 & 0.636 \\
Jul & 0.516 & 0.588 & 0.578 & 0.634 & 0.635 \\
Aug & 0.672 & 0.669 & 0.690 & 0.718 & 0.776 \\
Sep & 0.608 & 0.584 & 0.568 & 0.630 & 0.925 \\
Oct & 0.883 & 0.948 & 0.809 & 1.050 & 1.185 \\
Nov & 1.077 & 1.021 & 1.173 & 1.310 & 0.985 \\
Dec & 1.781 & 1.813 & 1.976 & 1.585 & 1.145 \\
\bottomrule
\end{tabular}
\end{adjustbox}
\end{table}

\begin{table}[t]
\centering
\tiny
\setlength{\tabcolsep}{3pt}
\caption{\textbf{Monthly residual innovation variances.} Entries report $\sigma_u^2\exp(\gamma_m)$ for each month.}\label{tab:ss_param_monthly_residual_variance}
\begin{adjustbox}{max width=\linewidth}
\begin{tabular}{lccccc}
\toprule
Month & NASA & NOAA & HadCRUT5 & Berkeley Earth & ERA5 \\
\midrule
Jan & 0.0217 & 0.0142 & 0.0176 & 0.0212 & 0.0137 \\
Feb & 0.0164 & 0.0134 & 0.0149 & 0.0165 & 0.0145 \\
Mar & 0.0174 & 0.0131 & 0.0178 & 0.0174 & 0.0142 \\
Apr & 0.0075 & 0.0070 & 0.0065 & 0.0087 & 0.0063 \\
May & 0.0059 & 0.0052 & 0.0064 & 0.0077 & 0.0090 \\
Jun & 0.0062 & 0.0049 & 0.0062 & 0.0072 & 0.0058 \\
Jul & 0.0048 & 0.0044 & 0.0052 & 0.0070 & 0.0058 \\
Aug & 0.0062 & 0.0050 & 0.0062 & 0.0080 & 0.0071 \\
Sep & 0.0056 & 0.0044 & 0.0051 & 0.0070 & 0.0084 \\
Oct & 0.0081 & 0.0071 & 0.0072 & 0.0116 & 0.0108 \\
Nov & 0.0099 & 0.0077 & 0.0105 & 0.0145 & 0.0090 \\
Dec & 0.0164 & 0.0136 & 0.0177 & 0.0176 & 0.0104 \\
\bottomrule
\end{tabular}
\end{adjustbox}
\end{table}

\FloatBarrier
\begin{longtable}{lccccc}
\caption{\textbf{Implied distributed-lag covariate weights.} Entries are the implied coefficient at each monthly lag from the estimated Almon basis coefficients.}\label{tab:ss_param_lag_weights}\\
\toprule
Covariate/lag & NASA & NOAA & HadCRUT5 & Berkeley Earth & ERA5 \\
\midrule
\endfirsthead
\toprule
Covariate/lag & NASA & NOAA & HadCRUT5 & Berkeley Earth & ERA5 \\
\midrule
\endhead
\multicolumn{6}{l}{\textbf{NINO34}}\\
0 & 0.0202 & 0.0212 & 0.0208 & 0.0209 & 0.0219 \\
1 & 0.0156 & 0.0161 & 0.0161 & 0.0162 & 0.0172 \\
2 & 0.0117 & 0.0117 & 0.0121 & 0.0122 & 0.0131 \\
3 & 0.0085 & 0.0082 & 0.0088 & 0.0088 & 0.0098 \\
4 & 0.0059 & 0.0053 & 0.0061 & 0.0062 & 0.0071 \\
5 & 0.0040 & 0.0033 & 0.0042 & 0.0042 & 0.0051 \\
6 & 0.0027 & 0.0021 & 0.0029 & 0.0028 & 0.0038 \\
7 & 0.0021 & 0.0016 & 0.0023 & 0.0021 & 0.0032 \\
8 & 0.0021 & 0.0018 & 0.0023 & 0.0021 & 0.0032 \\
9 & 0.0028 & 0.0029 & 0.0031 & 0.0028 & 0.0040 \\
10 & 0.0041 & 0.0047 & 0.0045 & 0.0041 & 0.0054 \\
11 & 0.0062 & 0.0073 & 0.0066 & 0.0061 & 0.0076 \\
12 & 0.0088 & 0.0106 & 0.0094 & 0.0087 & 0.0104 \\
\addlinespace[2pt]
\multicolumn{6}{l}{\textbf{volc2}}\\
0 & -0.0041 & -0.0045 & -0.0064 & -0.0063 & -0.0058 \\
1 & -0.0041 & -0.0044 & -0.0061 & -0.0060 & -0.0055 \\
2 & -0.0041 & -0.0042 & -0.0058 & -0.0058 & -0.0051 \\
3 & -0.0041 & -0.0040 & -0.0055 & -0.0055 & -0.0048 \\
4 & -0.0041 & -0.0039 & -0.0052 & -0.0053 & -0.0045 \\
5 & -0.0041 & -0.0037 & -0.0049 & -0.0050 & -0.0042 \\
6 & -0.0041 & -0.0036 & -0.0046 & -0.0048 & -0.0038 \\
7 & -0.0041 & -0.0034 & -0.0043 & -0.0045 & -0.0035 \\
8 & -0.0041 & -0.0033 & -0.0040 & -0.0043 & -0.0032 \\
9 & -0.0041 & -0.0031 & -0.0037 & -0.0040 & -0.0029 \\
10 & -0.0041 & -0.0029 & -0.0034 & -0.0038 & -0.0025 \\
11 & -0.0041 & -0.0028 & -0.0031 & -0.0035 & -0.0022 \\
12 & -0.0041 & -0.0026 & -0.0028 & -0.0033 & -0.0019 \\
\addlinespace[2pt]
\multicolumn{6}{l}{\textbf{ssn}}\\
0 & 0.0047 & 0.0038 & 0.0047 & 0.0021 & 0.0071 \\
1 & 0.0039 & 0.0032 & 0.0039 & 0.0017 & 0.0062 \\
2 & 0.0032 & 0.0025 & 0.0032 & 0.0014 & 0.0053 \\
3 & 0.0024 & 0.0019 & 0.0024 & 0.0010 & 0.0043 \\
4 & 0.0016 & 0.0013 & 0.0017 & 0.0006 & 0.0034 \\
5 & 0.0009 & 0.0006 & 0.0009 & 0.0002 & 0.0025 \\
6 & 0.0001 & $-1.121\times 10^{-5}$ & 0.0002 & -0.0001 & 0.0016 \\
7 & -0.0006 & -0.0006 & -0.0005 & -0.0005 & 0.0006 \\
8 & -0.0014 & -0.0013 & -0.0013 & -0.0009 & -0.0003 \\
9 & -0.0021 & -0.0019 & -0.0020 & -0.0013 & -0.0012 \\
10 & -0.0029 & -0.0025 & -0.0028 & -0.0017 & -0.0022 \\
11 & -0.0036 & -0.0032 & -0.0035 & -0.0020 & -0.0031 \\
12 & -0.0044 & -0.0038 & -0.0042 & -0.0024 & -0.0040 \\
\bottomrule
\end{longtable}

\newpage

\subsection{Model selection and residual diagnostics}
\label{app:ss-model-selection}

The main
specification choices concern the irregular component \(\varepsilon_n\), the
time variation in its innovation variance, and the distributed-lag structure of
the covariates \(z_n\). In a model-selection exercise, we considered
for each temperature record a grid of
candidate models varying the lag lengths and Almon-polynomial restrictions for
the control variables, the inclusion of calendar-month-specific innovation
variances, and the inclusion of longer-run variance breaks. Candidate
specifications were compared using the maximized log-likelihood, BIC, and
standard residual diagnostics based on one-step-ahead prediction errors from the
Kalman filter.

The preferred individual specification used in the main analysis models the
irregular component as an ARMA\((1,1)\) process and allows the variance of its
innovations to vary by calendar month. This specification was selected by BIC in that exercise
from a set of low-order irregular-component alternatives, including IID,
AR\((1)\), ARMA\((1,1)\), ARMA\((2,1)\), and ARMA\((2,2)\) specifications, with
and without calendar-month-specific innovation variances. Additional variance breaks around 1900
and 1960 were considered, motivated by historical changes in observational
coverage and data construction, but were not supported by the BIC or by the
diagnostic checks. The covariates enter through Almon-polynomial distributed
lags with coefficients given by \eqref{eq:AlmontCoeff}. The baseline specification uses lags of 0--12 months for NINO3.4,
volcanic aerosols, and sunspot numbers, with a second-degree polynomial for
NINO3.4 and first-degree polynomials for the volcanic and sunspot variables.
The main empirical conclusions are robust to moderate changes in this lag
structure.

Table~\ref{tab:ss_diag} reports residual diagnostics for the final individual
state-space models. The diagnostics are based on the standardized
one-step-ahead innovations
\[
    \xi_n = \frac{\nu_n}{\sqrt{F_n}},
\]
where
\[
    \nu_n = y_n - \operatorname{E}(y_n \mid y_{1:n-1})
\]
is the Kalman-filter prediction error and
\[
    F_n = \operatorname{Var}(y_n \mid y_{1:n-1})
\]
is its conditional variance. Under the maintained Gaussian state-space
specification, the standardized innovations should be approximately independent
and standard normal. The table reports sample moments, Jarque--Bera tests for
normality, auxiliary AR\((1)\) regressions, Durbin--Watson statistics,
Ljung--Box tests for serial correlation at several horizons, Ljung--Box tests
for serial correlation in squared innovations, and ARCH LM tests.

The standardized innovations have means close to zero and standard
deviations close to one for all five temperature records, consistent with
their assumed centring and variance. These moments alone do not establish
independence or interval calibration. Skewness and excess kurtosis are modest, and
the Jarque--Bera test does not reject normality for any of the five records at the 5\% significance level.
The estimated AR\((1)\) coefficients are close to zero, the Durbin--Watson
statistics are close to two, and the Ljung--Box tests at lags 1, 2 and 6 do not
reject at the 5\% level for any record.

The Ljung--Box tests applied to squared standardized innovations and the ARCH LM
tests do not reject at the 5\% level for any record. These diagnostics show
little evidence of remaining conditional heteroskedasticity under the
calendar-month-specific variance specification.

However, the Ljung--Box tests at lags 12 and 24 reject the null hypothesis of
no serial correlation for all five records at the 5\% level. Thus, the
ARMA\((1,1)\) irregular component and observed covariates do not remove all
serial dependence. These joint tests do not identify its source or isolate
correlation at a particular lag.

Overall, the diagnostics are broadly consistent with the assumed marginal
innovation distribution and show little evidence of remaining conditional
heteroskedasticity, but reveal residual serial dependence at longer lags. The state-space
specification therefore remains an approximation; interval calibration is
assessed separately in Appendix~\ref{app:ssboot_mc}.

\begin{landscape}
\begin{table}[t]
\centering
\tiny
\setlength{\tabcolsep}{3.5pt}
\caption{Residual diagnostics for the individual state-space models. Diagnostics
are based on standardized one-step-ahead Kalman-filter innovations
\(\xi_n=\nu_n/\sqrt{F_n}\), where \(\nu_n\) is the prediction error and \(F_n\)
is its conditional variance. The table reports sample moments, Jarque--Bera
\(p\)-values, auxiliary AR\((1)\) estimates and the associated $p$-values, Durbin--Watson statistics,
Ljung--Box \(p\)-values for innovations and squared innovations, and ARCH LM
\(p\)-values. Temperature records are adjusted to the common reference level
described in Section~\ref{sec:data}.}
\label{tab:ss_diag}
\begin{adjustbox}{max width=\linewidth}
\begin{tabular}{lcccccccccccccccccccc}
\toprule
Dataset & Mean & Std & Skew & Kurt & JB \(p\) & AR(1) & AR(1) \(p\) & DW & LB(1) & LB(2) & LB(6) & LB(12) & LB(24) & LB\(^{2}\)(1) & LB\(^{2}\)(2) & LB\(^{2}\)(6) & LB\(^{2}\)(12) & LB\(^{2}\)(24) & ARCH \(p\) \\
\midrule
\addlinespace[2pt]
NASA & 0.013 & 0.999 & -0.089 & 3.107 & 0.207 & 0.002 & 0.930 & 1.995 & 0.930 & 0.810 & 0.347 & 0.018 & $<10^{-4}$ & 0.556 & 0.261 & 0.117 & 0.387 & 0.629 & 0.371 \\
NOAA & 0.010 & 0.999 & -0.007 & 3.044 & 0.926 & -0.003 & 0.914 & 2.004 & 0.914 & 0.254 & 0.468 & 0.013 & $<10^{-4}$ & 0.599 & 0.869 & 0.931 & 0.884 & 0.997 & 0.894 \\
HadCRUT5 & -0.016 & 0.999 & -0.018 & 3.064 & 0.824 & 0.003 & 0.899 & 1.993 & 0.899 & 0.780 & 0.780 & 0.019 & $<10^{-4}$ & 0.410 & 0.554 & 0.145 & 0.273 & 0.674 & 0.257 \\
Berkeley Earth & 0.018 & 0.999 & -0.039 & 3.043 & 0.752 & -0.009 & 0.702 & 2.017 & 0.701 & 0.737 & 0.087 & 0.001 & $<10^{-4}$ & 0.517 & 0.763 & 0.392 & 0.779 & 0.843 & 0.843 \\
ERA5 & 0.006 & 0.998 & -0.140 & 3.212 & 0.069 & -0.007 & 0.825 & 2.011 & 0.825 & 0.142 & 0.307 & $<10^{-4}$ & $<10^{-4}$ & 0.815 & 0.398 & 0.649 & 0.885 & 0.930 & 0.899 \\
\bottomrule
\end{tabular}
\end{adjustbox}
\end{table}
\end{landscape}

 \section{Details of the benchmark estimators}\label{app:benchmark_details}

\subsection{Details of the nonparametric local-polynomial estimator}
\label{app:kernel_details}

This section provides additional details on the nonparametric
local-polynomial benchmark. The purpose
of this estimator is to recover the latent warming path \(x(t)\), together with
its first and second derivatives, while imposing as little parametric structure
as possible on the time path of the trend itself. Relative to the
semiparametric state-space model, the local-polynomial estimator is more
agnostic about the dynamics of \(x(t)\). It incorporates common covariate
effects through a partially linear adjustment, but does not specify a joint
model for the trend, serial dependence, and time-varying error variance.
This motivates its use as a flexible benchmark rather than as our preferred
specification.

\subsubsection{Partialling out observed covariates}

The starting point is the measurement equation
\begin{align}
    y_n = x(t_n) + \beta' z_n + \varepsilon_n,
\end{align}
where \(z_n\) contains the same observed controls and distributed lags as in
the main state-space specification. Let \(S_h\) be the local-quadratic level
smoother evaluated at all sample dates, with the kernel and bandwidth defined
below. Writing \(y\) for the temperature vector and \(Z\) for the covariate
matrix, we first remove smooth time components from both and then estimate $\beta$:
\begin{align}
    \widetilde y &= (I-S_h)y, & \widetilde Z &= (I-S_h)Z, \\
    \widehat\beta &= \arg\min_\beta
        \|\widetilde y-\widetilde Z\beta\|^2 .
    \label{eq:kernel_beta_aux}
\end{align}
Thus, the covariate regression accounts for the unknown climate trend instead
of attributing its projection onto the covariates to their effects. The same
bandwidth is used to smooth the temperature series and every covariate. One
common coefficient vector is estimated over the full sample; no additional
intercept is included in \(Z\), since it is part of the smooth trend.

We then construct \(r=y-Z\widehat\beta\) and apply the local-polynomial
estimator to \(r\) to obtain the climate path and its derivatives. The
bootstrap repeats both the covariate adjustment and the climate fit in every
replication, propagating uncertainty from estimating \(\beta\).

\subsubsection{Local-polynomial estimator}

Let \(\tau\) denote an evaluation point in continuous time, measured in years.
Around \(\tau\), we approximate the latent warming path by a second-order Taylor
expansion,
\begin{align}
    x(t) \approx b_0(\tau)
        + b_1(\tau)(t-\tau)
        + b_2(\tau)(t-\tau)^2 .
\end{align}
Given a bandwidth \(h>0\), the local-polynomial estimator chooses
\((b_0(\tau),b_1(\tau),b_2(\tau))\) to minimize
\begin{align}
\sum_{n=1}^{N}
K\!\left(\frac{t_n-\tau}{h}\right)
\left[
r_n - b_0(\tau) - b_1(\tau)(t_n-\tau)
      - b_2(\tau)(t_n-\tau)^2
\right]^2 ,
\label{eq:kernel_localpoly_objective}
\end{align}
where \(N\) is the number of monthly observations and \(K(\cdot)\) is the kernel
function.

In the implementation used in the paper, we employ the tricube kernel,
\begin{align}
    K(u) = \frac{70}{81}\left(1-\lvert u\rvert^3\right)^3
    \mathbf{1}\{\lvert u\rvert<1\}.
\end{align}
Thus, only observations within the window \(\lvert t_n-\tau\rvert<h\) receive
positive weight, and observations closer to \(\tau\) receive larger weight.

Given the solution
\((\hat{b}_0(\tau),\hat{b}_1(\tau),\hat{b}_2(\tau))\), the estimated warming
level, warming rate, and warming acceleration are
\begin{align}
    \hat{x}(\tau) &= \hat{b}_0(\tau), \\
    \hat{v}(\tau) &= \hat{b}_1(\tau), \\
    \hat{a}(\tau) &= 2\hat{b}_2(\tau).
\end{align}
Since time is measured in years, \(\hat{v}(\tau)\) is initially measured in
\({}^\circ\mathrm{C}\) per year and \(\hat{a}(\tau)\) in
\({}^\circ\mathrm{C}\) per year squared. In the tables reported in this paper, we rescale these quantities to \({}^\circ\mathrm{C}\) per decade and
\({}^\circ\mathrm{C}\) per decade squared by multiplying by \(10\) and \(100\),
respectively.

We evaluate the local-polynomial estimator at every observed time point in the
sample. The endpoint estimands are then constructed analogously to the
state-space case:
\begin{align}
    \hat{x}(T), \qquad \hat{v}(T), \qquad \hat{a}(T).
\end{align}

\subsubsection{Bandwidth choice}

The bandwidth \(h\) governs the smoothness of the estimated local-polynomial
trend. A smaller bandwidth allows the fitted path to respond more strongly to
short-run fluctuations, whereas a larger bandwidth imposes greater smoothness.
In the baseline empirical implementation, we use
\begin{align}
    h = 25 \text{ years}.
\end{align}
This value provides a useful compromise between flexibility and smoothness for
the monthly global temperature records considered here. The same bandwidth is
used for all datasets in order to facilitate comparisons across temperature
records. Sensitivity to the bandwidth choice is assessed in
Section~\ref{app:robust}.

\subsubsection{Bootstrap inference}\label{app:kernel_bootstrap}

To quantify uncertainty for the local-polynomial estimator, we use a
trend-preserving moving-block bootstrap with fixed observed covariates. Let \(\hat{x}(t_n)\) denote the fitted local-polynomial trend evaluated
at month \(n\), and define the residuals
\begin{align}
    \hat{e}_n = r_n - \hat{x}(t_n).
\end{align}
The bootstrap proceeds as follows.

\begin{enumerate}
    \item Estimate \(\hat{\beta}\) from the residualized regression in
    \eqref{eq:kernel_beta_aux}, and
    construct the covariate-adjusted series
    \begin{align}
        r_n = y_n - \hat{\beta}'z_n .
    \end{align}

    \item Apply the local-polynomial estimator to \(r_n\) to obtain
    \(\hat{x}(t_n)\), \(\hat{v}(t_n)\), and \(\hat{a}(t_n)\).

    \item Form the detrended residuals
    \begin{align}
        \hat{e}_n = r_n - \hat{x}(t_n).
    \end{align}

    Center these residuals by subtracting their sample mean.

    \item Generate a bootstrap residual series \(\hat{e}^{\ast}_n\) using a
    moving-block bootstrap. Blocks of consecutive residuals of length
    \(L\) are sampled with replacement and concatenated until a bootstrap
    residual series of length \(N\) is obtained. The implementation uses the
    baseline block length
    \begin{align}
        L = 24 \text{ months}.
    \end{align}

    \item Construct the bootstrap pseudo-series
    \begin{align}
        y^{\ast}_n = \hat{x}(t_n) + \widehat\beta' z_n + \hat{e}^{\ast}_n .
    \end{align}

    \item Apply \eqref{eq:kernel_beta_aux} to \(y^{\ast}\) and the fixed
    covariate matrix \(Z\) to obtain \(\widehat\beta^{\ast}\). Re-estimate
    the local-polynomial model on \(r^{\ast}=y^{\ast}-Z\widehat\beta^{\ast}\), obtaining
    bootstrap estimates \(\hat{x}^{\ast}(t_n)\), \(\hat{v}^{\ast}(t_n)\), and
    \(\hat{a}^{\ast}(t_n)\). The bootstrap endpoint estimates are then
    \begin{align}
        \hat{x}^{\ast}(T), \qquad
        \hat{v}^{\ast}(T), \qquad
        \hat{a}^{\ast}(T).
    \end{align}

    \item Repeat Steps 4--6 \(B\) times. In the empirical analysis, we use
    \(B=1{,}999\) bootstrap replications for monthly fits and
    \(B=1{,}999\) for annual fits. Annual fits use two-year blocks and
    contemporaneous annual covariates.
\end{enumerate}

Confidence intervals are constructed as percentile intervals from the empirical
bootstrap distribution. Thus, a two-sided \(100(1-\alpha)\%\) interval is given
by the empirical \(\alpha/2\) and \(1-\alpha/2\) quantiles of the bootstrap
estimates. The reported nominal \(90\%\) intervals use the 5th and 95th
percentiles directly. The finite-sample
coverage of these confidence intervals is assessed in Appendix~\ref{app:ssboot_mc}.

The bandwidth and observed covariates are held fixed, but the covariate
coefficients are re-estimated in every bootstrap sample. For average
acceleration, both endpoint velocities are evaluated within the same draw
and their difference is divided by elapsed time, preserving their dependence.
The primary starting date is exactly ten years before the endpoint. For sensitivity analyses at specified starting years, it is the first observed date at or after the requested year.
The percentile procedure does not explicitly correct smoothing bias.

Unlike the
state-space model, the local-polynomial estimator does not jointly model
covariates, persistence, time-varying variance, and latent states in a single
system.

\subsection{Parametric quadratic estimator}
\label{sec:regression}

As a parametric benchmark, we fit a quadratic trend over a recent estimation
window. Specifically, over the window starting in year \(t_0\), with baseline
choice \(t_0=1970\) as in Foster and Rahmstorf\cite{FR2026}, we estimate
\begin{equation}
    y_n
    =
    \alpha
    + \gamma_1 \tilde{t}_n
    + \gamma_2 \tilde{t}_n^{\,2}
    + \beta' z_n
    + u_n,
    \label{eq:quad}
\end{equation}
where \(\tilde{t}_n=t_n-\bar{t}\) is time, measured in years, centered at the
midpoint of the estimation window. The covariates \(z_n\) enter with the same
Almon-polynomial distributed-lag structure used in the state-space model.

Under the quadratic specification, the implied latent warming path is
\[
    x(t_n)
    =
    \alpha
    + \gamma_1 \tilde{t}_n
    + \gamma_2 \tilde{t}_n^{\,2}.
\]
Thus, \(\gamma_1\) is the warming rate at the midpoint of the estimation window,
and \(2\gamma_2\) is the constant acceleration imposed throughout the window.
The fitted endpoint quantities are obtained by evaluating the fitted quadratic
and its derivatives at the end of the sample:
\begin{align}
    \hat{x}(T)
    &=
    \hat{\alpha}
    + \hat{\gamma}_1 \tilde{t}_T
    + \hat{\gamma}_2 \tilde{t}_T^{\,2},
    \nonumber\\
    \hat{v}(T)
    &=
    \hat{\gamma}_1
    + 2\hat{\gamma}_2 \tilde{t}_T,
    \label{eq:quad_estimands}\\
    \hat{a}(T)
    &=
    2\hat{\gamma}_2 .
    \nonumber
\end{align}

Inference is based on the heteroskedasticity- and autocorrelation-consistent
covariance estimator of \citet{NW1987}, using a Bartlett kernel and a baseline
lag truncation of \(L=24\) months. Because the model imposes a quadratic trend,
the resulting estimates should be interpreted as parametric long-window
summaries of warming level, rate, and acceleration over the post-\(t_0\) period,
rather than as flexible local estimates.

\subsection{Parametric changepoint estimator}
\label{sec:changepoint}

As a second parametric benchmark, we fit a piecewise-linear changepoint model
over the post-1970 subsample. Specifically, we estimate
\begin{align}
    y_n
    =
    \alpha
    + \gamma_1 t_n
    + \gamma_2 (t_n-\tau)^+
    + \beta' z_n
    + u_n,
    \label{eq:changepoint}
\end{align}
where \((t_n-\tau)^+=\max\{t_n-\tau,0\}\). For a given changepoint date
\(\tau\), the warming rate before the changepoint is \(\gamma_1\), while the
warming rate after the changepoint is \(\gamma_1+\gamma_2\). The changepoint
date is chosen by minimizing the residual sum of squares over admissible
monthly dates, requiring at least 60 observations on each side of the break. The
estimated changepoint is consistently close to 2014 across datasets, matching
the break location reported by \citet{FR2026}.

As in the quadratic benchmark, the covariates \(z_n\) enter with the same
Almon-polynomial distributed-lag structure used in the state-space model.
Inference uses the full-procedure parametric bootstrap described below, with ARMA(1,1) errors for monthly data and AR(1) errors for annual data. The fitted endpoint warming level is obtained
by evaluating the piecewise-linear trend at \(T\), with the covariate
contribution set to zero, and the endpoint warming-rate summary is the
post-changepoint slope \(\hat{\gamma}_1+\hat{\gamma}_2\). Because the fitted
path is piecewise linear, it imposes zero local acceleration away from the
changepoint and has an undefined second derivative at the changepoint itself.
The zero endpoint acceleration reported in Table~\ref{tab:main} is therefore
imposed by the fitted specification rather than estimated freely.

\subsubsection{Bootstrap inference with breakpoint selection}\label{app:changepoint_bootstrap}
The breakpoint search is part of the estimator and is repeated in every bootstrap replication. This follows the principle behind the Monte Carlo treatment of breakpoint selection by \citet{FR2026}, although their changepoint experiment concerns a test for a break under a linear-trend null. Here, the bootstrap is generated under the fitted single-break model to quantify uncertainty in its estimated warming quantities; it is not a test that a break exists.

For monthly data, fit a zero-mean Gaussian ARMA(1,1) model to the residuals of the original changepoint regression,
\[
 u_n=\phi u_{n-1}+e_n+\theta e_{n-1},\qquad e_n\overset{\mathrm{iid}}{\sim}N(0,\sigma_e^2).
\]
For annual data, use an AR(1) model, setting $\theta=0$. Noise parameters are estimated by exact stationary Gaussian maximum likelihood, with constant innovation variance. These are bootstrap noise specifications; the point estimator remains least squares with a grid search for the breakpoint. The covariates and adopted baseline adjustment are held fixed.

For each of $B=1{,}999$ independent replications:
\begin{enumerate}
\item Draw a stationary error path from the fitted noise model. For ARMA(1,1), initialize $(u_0,e_0)$ jointly with covariance
\[
 \sigma_e^2\begin{pmatrix}
 (1+\theta^2+2\phi\theta)/(1-\phi^2)&1\\1&1
 \end{pmatrix},
\]
then propagate the recursion with fresh Gaussian innovations. The AR(1) case uses the same construction with $\theta=0$.
\item Generate $y_n^*=\hat x(t_n)+z_n'\hat\beta+u_n^*$.
\item Refit the complete changepoint regression, searching the original admissible dates and re-estimating the trend and all covariate coefficients at every candidate date. Select the date minimizing residual sum of squares.
\item Evaluate the endpoint level, rate, slope change, and paired rate contrasts from the selected bootstrap fit. Record its breakpoint as well.
\end{enumerate}
The empirical 5th and 95th percentiles give nominal 90\% percentile intervals. The noise parameters fitted to the original residuals define the bootstrap data-generating process. Re-estimating noise parameters inside a draw is unnecessary for these unstudentized intervals because the least-squares point estimator does not use them. The procedure propagates covariate-coefficient and breakpoint-selection uncertainty, conditional on the specified noise process and single-break model.

For rate contrasts, define the rate at a break by its right-hand slope,
\[
 \hat v(t)=\hat\gamma_1+\hat\gamma_2\mathbf{1}\{t\geq\hat\tau\}.
\]
A selected break strictly inside $(T_0,T)$ can therefore contribute to average acceleration, even when the original fitted break precedes $T_0$. A break exactly at $T_0$ is excluded from the contrast. All admissible breaks precede $T$, so endpoint acceleration remains zero under every bootstrap fit and its interval is omitted. In contrast, monthly past-decade average-acceleration intervals are not constrained to zero. The annual search requires ten observations on each side of the break. Its latest admissible date is 2015, so the 2015--2025 annual contrast is zero under every fitted path; the resulting $[0,0]$ interval reflects the search restriction, not independent evidence against acceleration.

Table~\ref{tab:changepoint_bootstrap_diagnostics} reports fitted noise parameters and breakpoint summaries. For the six adjusted monthly fits, repeating the search increases the widths of the nominal \(90\%\) endpoint warming-rate intervals by a factor of 1.9 to 2.4 compared with a bootstrap using the same synthetic samples but holding the original break fixed. This paired comparison isolates the effect of breakpoint selection.

\begin{table}[tbp]\centering\scriptsize
\setlength{\tabcolsep}{4pt}
\caption{\textbf{Changepoint-bootstrap noise and breakpoint summaries.} Adjusted monthly fits use ARMA(1,1); annual fits use AR(1), with $\theta=0$. The 90\% breakpoint range gives the 5th and 95th percentiles of the reselected dates (decimal years). The final column reports the percentage of draws with a break strictly inside the past-decade interval. All fits use 1,999 draws.}\label{tab:changepoint_bootstrap_diagnostics}
\begin{adjustbox}{max width=\linewidth}
\begin{tabular}{lrrrrr}\toprule
Dataset & $\phi$ & $\theta$ & $\sigma_e$ & Breakpoint range & Inside decade (\%)\\\midrule
\multicolumn{6}{l}{\textbf{Monthly: ARMA(1,1)}}\\
NASA & 0.628 & -0.263 & 0.099 & [2009.38, 2017.71] & 11.0\\
NOAA & 0.638 & -0.241 & 0.085 & [2009.38, 2016.84] & 5.5\\
HadCRUT5 & 0.666 & -0.292 & 0.090 & [2006.38, 2019.42] & 22.1\\
Berkeley Earth & 0.575 & -0.177 & 0.101 & [2007.88, 2019.12] & 20.3\\
ERA5 & 0.713 & -0.289 & 0.092 & [2008.83, 2018.46] & 15.5\\
Arithmetic mean & 0.661 & -0.271 & 0.089 & [2008.54, 2018.29] & 13.6\\
\addlinespace[4pt]
\multicolumn{6}{l}{\textbf{Annual: AR(1)}}\\
NASA & 0.023 & 0.000 & 0.050 & [2009.00, 2015.00] & 0.0\\
NOAA & -0.038 & 0.000 & 0.045 & [2010.00, 2015.00] & 0.0\\
HadCRUT5 & -0.118 & 0.000 & 0.049 & [2008.00, 2015.00] & 0.0\\
Berkeley Earth & -0.139 & 0.000 & 0.050 & [2009.00, 2015.00] & 0.0\\
ERA5 & 0.205 & 0.000 & 0.058 & [2006.00, 2015.00] & 0.0\\
\addlinespace[4pt]
\bottomrule\end{tabular}
\end{adjustbox}\end{table}

\clearpage
\section{Additional full-sample decompositions}
\label{app:additional_decompositions}
\begin{figure}[!htbp]
\centering
\includegraphics[width=0.95\textwidth]{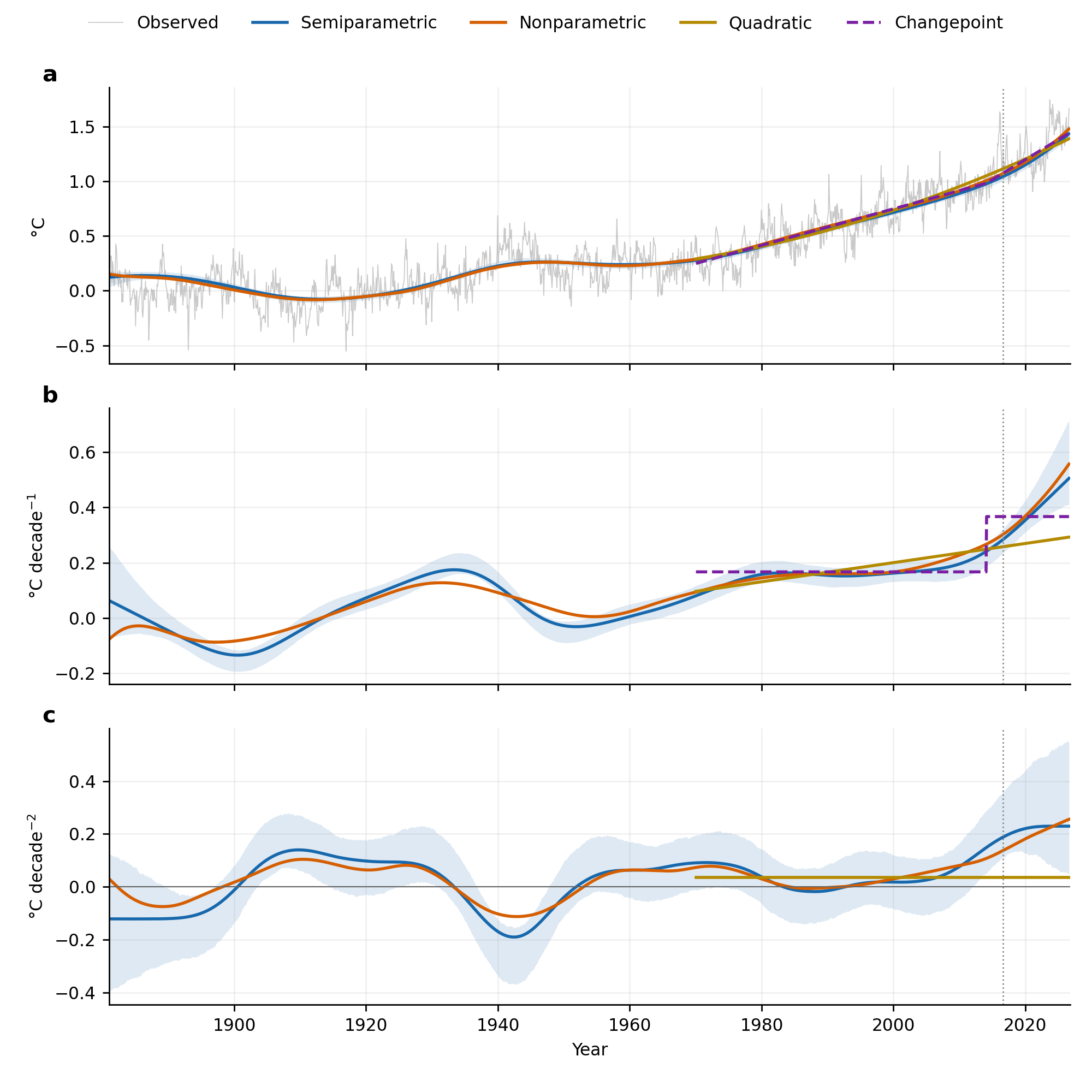}
\caption{\textbf{Illustrative decomposition for NASA global temperature anomalies.}
The figure compares the state-space (SS) model, the nonparametric
local-polynomial estimator, and the parametric
quadratic and changepoint benchmark models. The panels show (i) the observed
anomalies \(y(t)\) and the estimated latent warming level \(x(t)\), (ii) the
estimated warming rate \(v(t)\), and (iii) the estimated warming acceleration
\(a(t)\). Shaded bands are pointwise nominal 90\% bootstrap uncertainty bands for
the state-space model from the bootstrap procedure described in Appendix~\ref{app:ssbootstrap}. The changepoint estimator does not provide a smooth estimate
of acceleration and is therefore omitted from panel (iii). The vertical dotted line marks the start of the past-decade averaging interval.}
\label{fig:global_nasa_xva}
\end{figure}

\begin{figure}[t]
\centering
\includegraphics[width=0.95\textwidth]{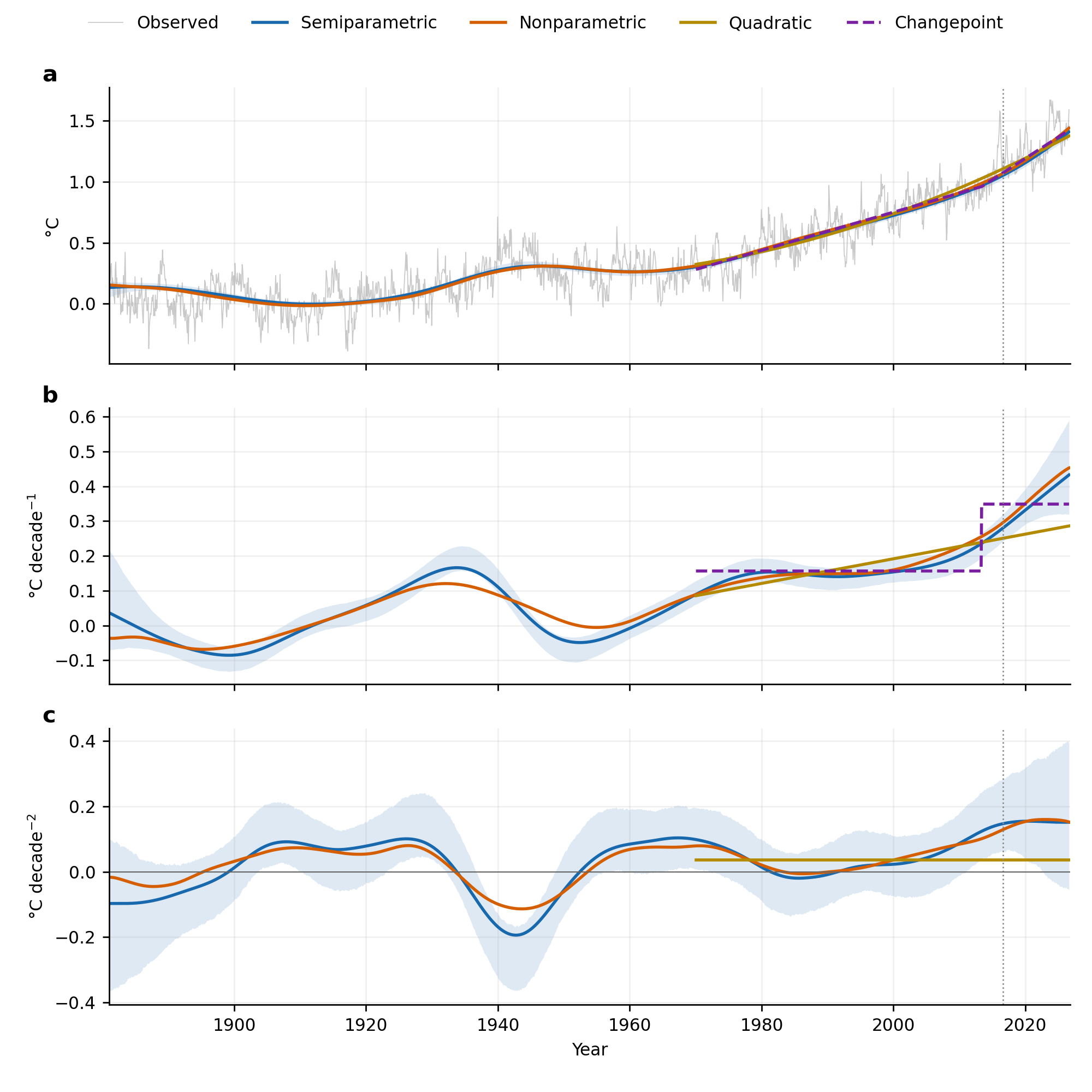}
\caption{\textbf{Illustrative decomposition for NOAA global temperature
anomalies.} The figure compares the semiparametric state-space model, the nonparametric
local-polynomial estimator, and the parametric
quadratic and changepoint benchmarks. The panels show \textbf{a}, observed anomalies and latent warming level \(x(t)\); \textbf{b}, warming rate \(v(t)\); and \textbf{c}, warming acceleration \(a(t)\). Shaded bands are pointwise nominal 90\% state-space
bootstrap-error intervals. The vertical dotted line marks the start of the past-decade averaging
interval. All fitted paths account for the specified climate covariates.
The changepoint model imposes zero acceleration away from its break and is
omitted from \textbf{c}.}
\label{fig:global_noaa_xva}
\end{figure}

\begin{figure}[t]
\centering
\includegraphics[width=0.95\textwidth]{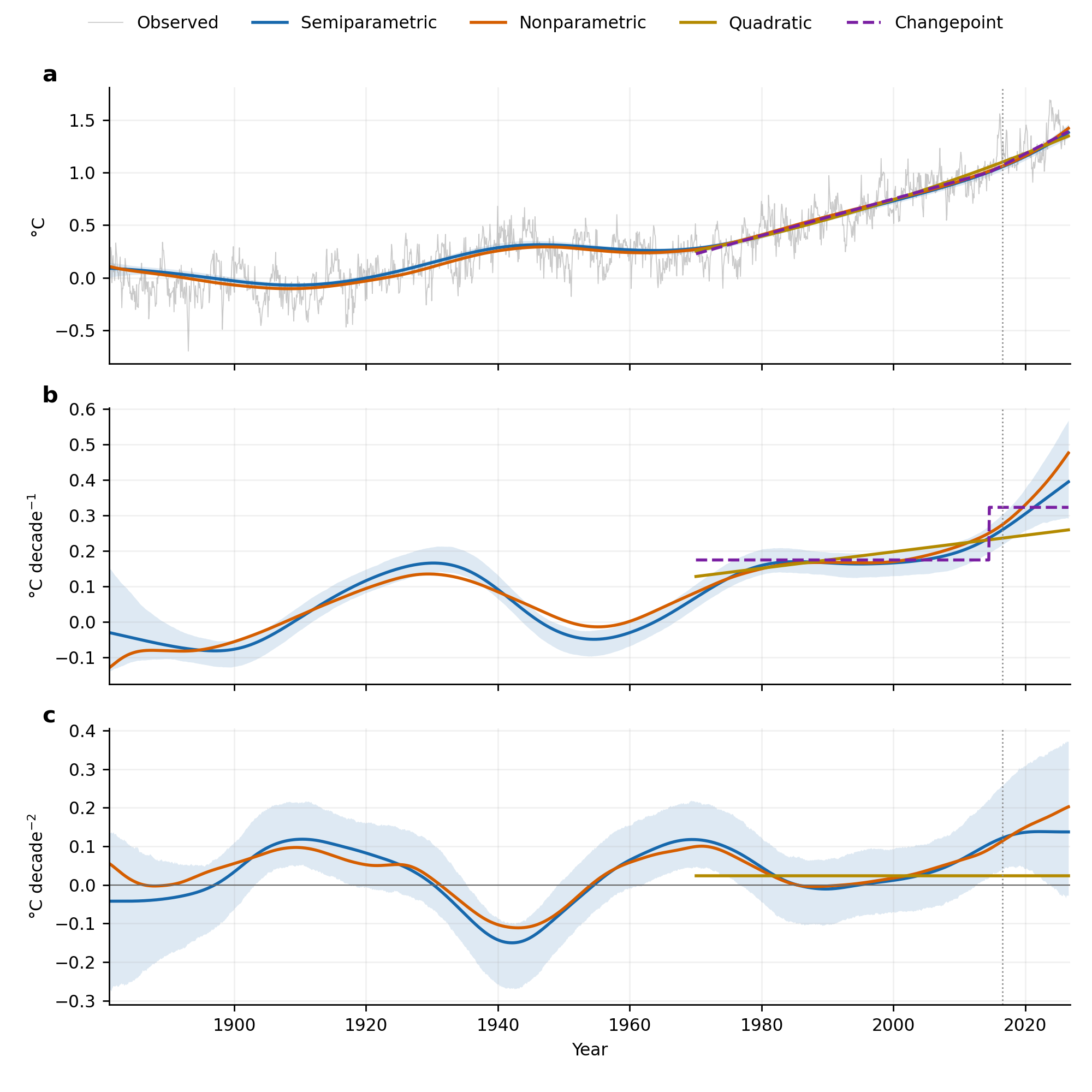}
\caption{\textbf{Illustrative decomposition for HadCRUT5 global temperature
anomalies.} The figure compares the semiparametric state-space model, the nonparametric
local-polynomial estimator, and the parametric
quadratic and changepoint benchmarks. The panels show \textbf{a}, observed anomalies and latent warming level \(x(t)\); \textbf{b}, warming rate \(v(t)\); and \textbf{c}, warming acceleration \(a(t)\). Shaded bands are pointwise nominal 90\% state-space
bootstrap-error intervals. The vertical dotted line marks the start of the past-decade averaging
interval. All fitted paths account for the specified climate covariates.
The changepoint model imposes zero acceleration away from its break and is
omitted from \textbf{c}.}
\label{fig:global_hadcru_xva}
\end{figure}

\begin{figure}[t]
\centering
\includegraphics[width=0.95\textwidth]{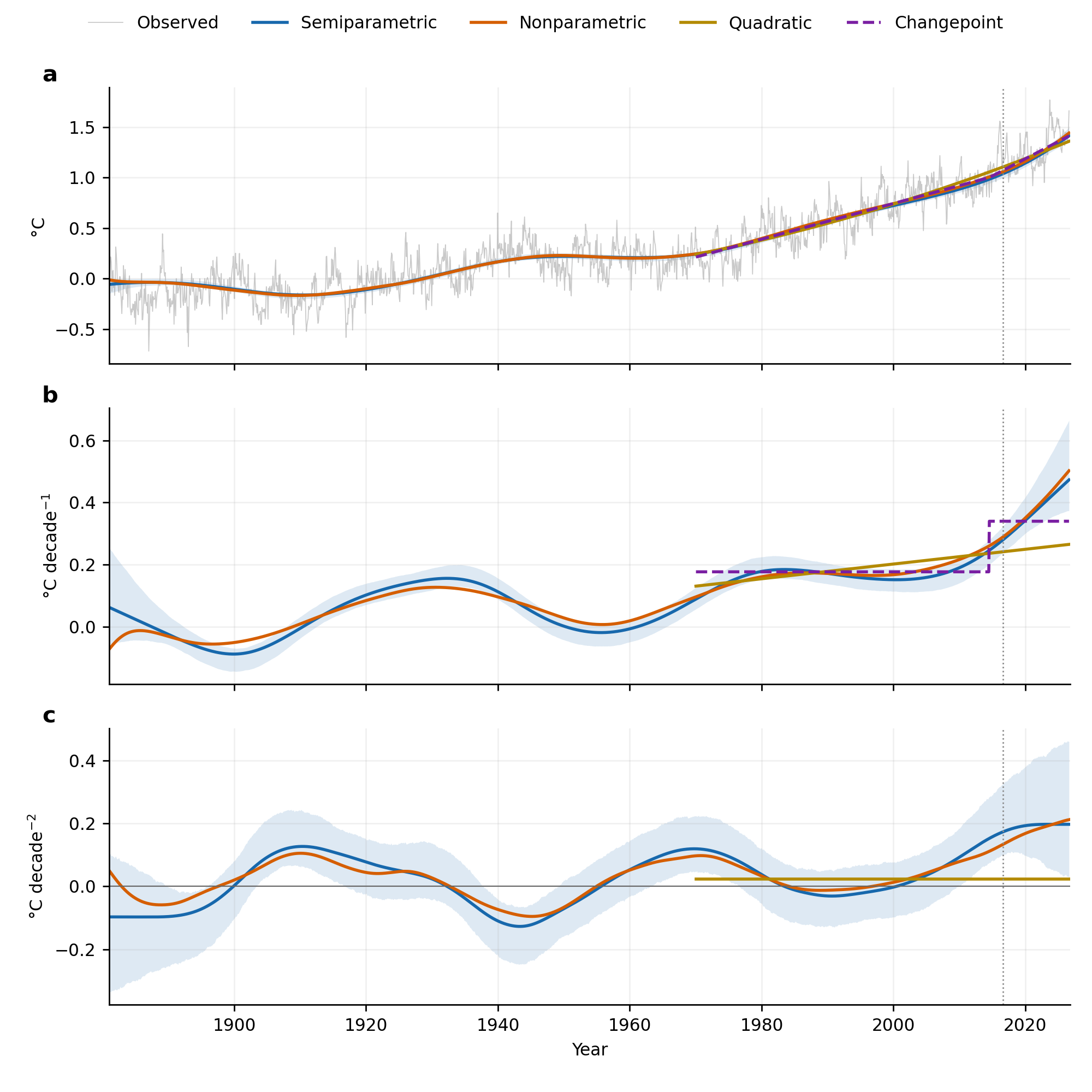}
\caption{\textbf{Illustrative decomposition for Berkeley Earth global temperature
anomalies.} The figure compares the semiparametric state-space model, the nonparametric
local-polynomial estimator, and the parametric
quadratic and changepoint benchmarks. The panels show \textbf{a}, observed anomalies and latent warming level \(x(t)\); \textbf{b}, warming rate \(v(t)\); and \textbf{c}, warming acceleration \(a(t)\). Shaded bands are pointwise nominal 90\% state-space
bootstrap-error intervals. The vertical dotted line marks the start of the past-decade averaging
interval. All fitted paths account for the specified climate covariates.
The changepoint model imposes zero acceleration away from its break and is
omitted from \textbf{c}.}
\label{fig:global_berk_xva}
\end{figure}

\begin{figure}[t]
\centering
\includegraphics[width=0.95\textwidth]{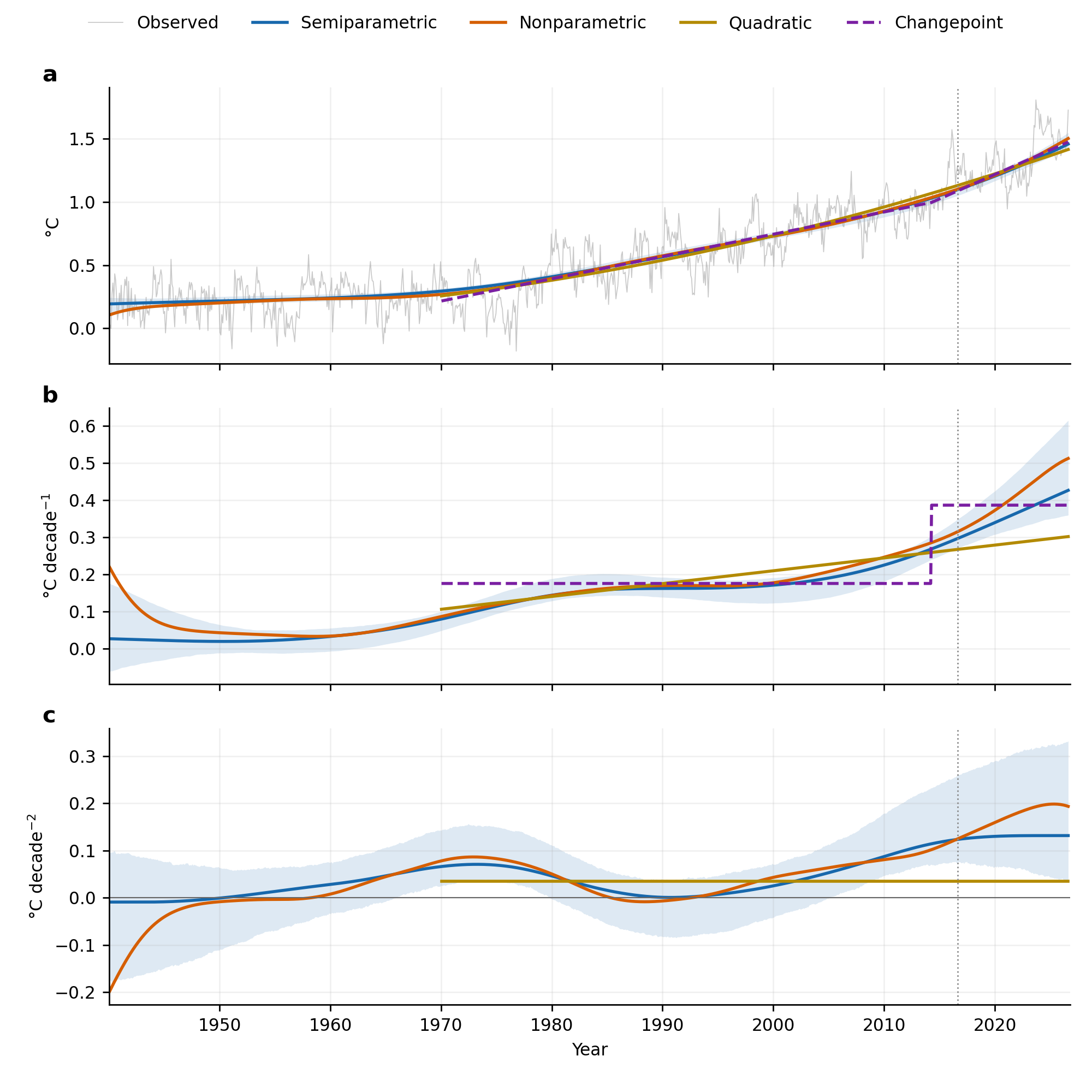}
\caption{\textbf{Illustrative decomposition for ERA5 global temperature
anomalies.} The figure compares the semiparametric state-space model, the nonparametric
local-polynomial estimator, and the parametric
quadratic and changepoint benchmarks. The panels show \textbf{a}, observed anomalies and latent warming level \(x(t)\); \textbf{b}, warming rate \(v(t)\); and \textbf{c}, warming acceleration \(a(t)\). Shaded bands are pointwise nominal 90\% state-space
bootstrap-error intervals. The vertical dotted line marks the start of the past-decade averaging
interval. All fitted paths account for the specified climate covariates.
The changepoint model imposes zero acceleration away from its break and is
omitted from \textbf{c}.}
\label{fig:global_era5_xva}
\end{figure}

Figures~\ref{fig:global_nasa_xva}--\ref{fig:global_era5_xva} show the full-sample decompositions for NASA, NOAA, HadCRUT5, Berkeley Earth, and ERA5, respectively. Figure~\ref{fig:global_arithmetic_xva} in the main text shows the separately fitted arithmetic-mean series.

\clearpage
\section{Simulation studies}\label{app:simstudy}
Section~\ref{app:sim_accuracy} compares estimator bias and root mean squared error across five warming paths. Section~\ref{app:coverage} assesses empirical coverage properties of the confidence intervals used in the paper, by examining how often the nominal confidence intervals contain the true warming quantities.

\subsection{Estimator accuracy}\label{app:sim_accuracy}

\subsubsection{Calibration and observation equation}
The experiment compares the four empirical estimation procedures on a common
monthly grid of 1,748 observations, from January 1881 to August 2026.
Dates are monthly midpoints, so $T=2026.625$ and the past-decade averaging
interval starts at $T_0=T-10=2016.625$ (August 2016). All calibration parameters and covariate paths are
held fixed across replications. The calibration uses the NASA record on the
same temperature scale as the empirical analysis; simulated datasets are not
individually rebased. The calibration uses the covariate series described in Section~\ref{sec:data} and the fitted NASA model, as in the coverage experiment in Section~\ref{app:ssboot_mc}.

For DGP $d$ and replication $r$, observations are generated as
\begin{equation}
 y_{i}^{(d,r)}=x_d(t_i)+Z_i'\beta_d+e_i^{(r)},\qquad
 e_i^{(r)}=\rho e_{i-1}^{(r)}+u_i^{(r)}+\theta u_{i-1}^{(r)},
 \label{eq:sim_observation}
\end{equation}
where $Z_i$ contains the observed NASA covariates and their distributed-lag
terms, with the same construction as in the empirical analysis. The Gaussian
innovations are independent over time and use the innovation-variance profile
estimated from the NASA state-space model, including its monthly variation.
The calibrated AR and MA coefficients are $\rho=0.71197$ and
$\theta=-0.27505$, and the baseline innovation variance is $0.009223$.
The same noise calibration is used for all five DGPs, with a burn-in of 6,000
months using the initial variance regime and recurring monthly variance
pattern. Within a replication, the same observation-noise realization is
used across DGPs; different replications have independent noise draws.
All estimators are applied to the same simulated dataset within a DGP and
replication.

For DGPs 1--3, $\beta_d$ is the NASA state-space estimate. For DGPs 4 and 5,
the covariate contribution is taken from the corresponding NASA quadratic
or changepoint fit. The covariates are thus treated as fixed design, while
their coefficients are re-estimated in every simulated dataset.

\subsubsection{Five latent-path designs}\label{app:sim_designs}
The five designs span zero acceleration, linearly increasing acceleration,
the NASA state-space smoothed acceleration path, constant positive
acceleration calibrated to the NASA quadratic fit, and a continuous
changepoint trend calibrated to the NASA changepoint fit. The first two
provide simple reference cases, the third supplies an empirically calibrated
flexible path, and the final two satisfy the respective parametric
benchmarks' trend restrictions.

\begin{figure}[p]
\centering
\includegraphics[width=\textwidth]{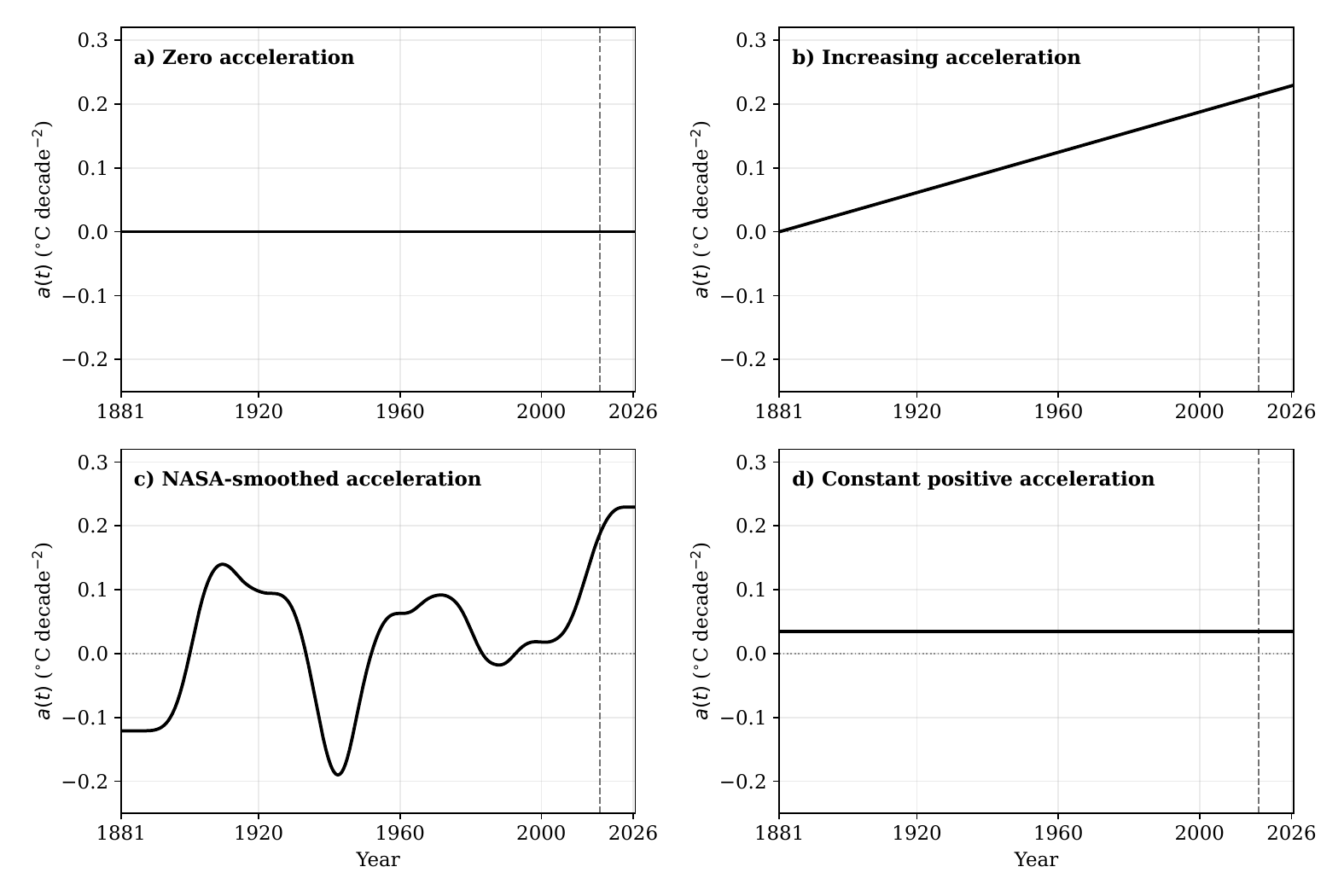}
\caption{\textbf{Acceleration paths in the simulation designs.}
Panels (a)--(d) show zero acceleration, increasing acceleration,
NASA-smoothed acceleration, and constant positive acceleration, respectively,
over the full monthly sample, January 1881 to August 2026. All panels use
a common vertical scale in $^{\circ}$C/decade$^2$. Dashed vertical lines mark
August 2016, the start of the average-acceleration interval.
The fifth design, the changepoint trend, is omitted from the figure:
acceleration is zero away from the February 2014 break and undefined at
the break, where the warming rate increases by
$0.200$ $^{\circ}$C/decade. The same latent path is used in all 1,000
replications of each DGP.}
\label{fig:simulation_acceleration_dgps}
\end{figure}

Write $t_s$ for the first sample date, $u=t-t_s$, and $L=T-t_s$, with time
measured in years. Let $\widetilde a_i$ be the NASA state-space smoothed
acceleration in $^{\circ}$C/year$^2$. The first three designs share the
NASA-smoothed initial level and velocity. Their level and velocity paths
are obtained by integrating the specified acceleration twice and once,
respectively; no additional state innovations are added.
\begin{enumerate}
\item \textit{Zero acceleration:} $a_1(t)=0$. This gives a linear latent
warming path and checks performance when acceleration is absent.
\item \textit{Increasing acceleration:} $a_2(t)=A u/L$, where
$A=\max_i|\widetilde a_i|=0.00229455$ $^{\circ}$C/year$^2$.
Acceleration rises linearly from zero to approximately $0.229$
$^{\circ}$C/decade$^2$ over the sample.
\item \textit{NASA-smoothed acceleration:} $a_3(t_i)=\widetilde a_i$,
with linear interpolation between monthly dates. Integration is exact
for this piecewise-linear acceleration path. The smoothed acceleration
is held fixed in every replication; the latent level and velocity are
constructed by integration from the common initial conditions.
\item \textit{Constant positive acceleration:} the latent level is the
quadratic trend fitted jointly with covariates to the NASA observations
from 1970 onwards. Its acceleration is $0.034593$
$^{\circ}$C/decade$^2$. The fitted polynomial is evaluated over the full
simulation grid, including dates before 1970. This design satisfies the
quadratic benchmark's trend restriction.
\item \textit{Changepoint:} the latent level is the continuous broken-linear
trend fitted jointly with covariates to the NASA observations from 1970
onwards and evaluated over the full simulation grid. The break is in
February 2014 ($\tau=2014.125$); the warming rate increases from $0.166712$
to $0.366395$ $^{\circ}$C/decade. Instantaneous acceleration is zero away
from the break and undefined at the break. Since $\tau<T_0<T$, both
acceleration estimands considered here are zero. This design satisfies the
changepoint benchmark's trend restriction.
\end{enumerate}
Figure~\ref{fig:simulation_acceleration_dgps} shows the acceleration paths
for the first four designs on common axes, with August 2016 marked to
locate the start of the average-acceleration interval. The changepoint
path is omitted; it has acceleration zero away from the February 2014 break and an undefined acceleration at
the break, where the warming rate increases by
$0.200$ $^{\circ}$C/decade.

The designs vary the shape of acceleration rather than requiring all
simulated level paths to resemble the observed historical record. In
particular, integrating the increasing acceleration path over the full
sample produces a substantially larger endpoint level than in the NASA
record. Table~\ref{tab:sim_truth} reports the fixed estimand values.

\begin{table}[htbp]
\centering
\caption{True estimand values in the five simulation designs. Levels are in
$^{\circ}$C, velocities in $^{\circ}$C/decade, and both acceleration quantities
in $^{\circ}$C/decade$^2$.}
\label{tab:sim_truth}
\begingroup\small
\begin{adjustbox}{max width=\linewidth}
\begin{tabular}{lrrrr}\toprule
DGP & $x(T)$ & $v(T)$ & $a(T)$ & $\overline{a_{T_0,T}}$ \\\midrule
Zero acceleration & 1.040 & 0.063 & 0.000 & 0.000 \\
Increasing acceleration & 9.146 & 1.733 & 0.229 & 0.222 \\
NASA-smoothed acceleration & 1.437 & 0.504 & 0.229 & 0.221 \\
Constant positive acceleration & 1.392 & 0.292 & 0.035 & 0.035 \\
Changepoint & 1.443 & 0.366 & 0.000 & 0.000 \\
\bottomrule\end{tabular}
\end{adjustbox}
\endgroup
\end{table}

\FloatBarrier

\subsubsection{Estimation and Monte Carlo summaries}
For each DGP, we generate $R=1{,}000$ simulated datasets. The state-space estimator is fitted
to the full sample, re-estimating the state-innovation variance, covariate
coefficients, and observation-noise parameters before smoothing the states.
The kernel estimator also uses the full sample and the partially linear
covariate adjustment described in Section~\ref{app:kernel_details}. The tricube bandwidth is fixed at 25 years.

The quadratic and changepoint estimators use observations from 1970 onwards
and estimate the trend and covariate coefficients jointly by least squares.
The changepoint is re-estimated by searching the admissible monthly dates,
requiring at least 60 observations on each side. None of the estimators is
given the true latent path, covariate coefficients, or breakpoint. These
settings reproduce the empirical comparison.

For each fitted path, we evaluate the three endpoint quantities and compute
average acceleration,
\begin{equation}
 \overline{a_{T_0,T}}=\frac{v(T)-v(T_0)}{T-T_0}.
 \label{eq:sim_average}
\end{equation}
This formula uses velocity in annual units; velocities and accelerations
are multiplied by 10 and 100, respectively, for reporting in decadal units.
For a quadratic trend, average and endpoint acceleration coincide. For a
changepoint trend, average acceleration includes the velocity jump only
when the estimated break falls strictly between $T_0$ and $T$. Endpoint acceleration for the changepoint model
is zero by construction because the admissible break dates precede the endpoint.

Let $E_r=\widehat\psi_r-\psi_d$ be the estimation error in the $r$'th replication for estimand $\psi_d \in \{x(T),v(T),a(T), \overline{a_{T_0,T}}\}$.  We report the Monte Carlo bias and root mean squared error (RMSE), given by, respectively,
\begin{equation}
 \widehat{\mathrm{Bias}}=\frac{1}{R}\sum_{r=1}^R E_r,
 \qquad
 \widehat{\mathrm{RMSE}}=\left(\frac{1}{R}\sum_{r=1}^R E_r^2\right)^{1/2}.
\end{equation}

\subsubsection{Simulation results}
Table~\ref{tab:simstudy} reports the RMSEs across the 1,000 replications per
DGP. For the endpoint warming rate, the main estimand in the empirical analysis, the state-space estimator has the lowest RMSE in all five designs, including those satisfying the quadratic and changepoint benchmarks' trend restrictions. The state-space estimator has the lowest RMSE in 16 of the 20
DGP--estimand combinations, including all four estimands under increasing
acceleration, the NASA-smoothed path, and constant positive acceleration.
The gains over the parametric benchmarks are substantial for the warming rate and both acceleration estimands under increasing acceleration, the NASA-smoothed path, and constant positive acceleration. The exceptions primarily concern designs with zero acceleration throughout the sample or away from a changepoint. In the zero-acceleration design, the changepoint estimator correctly imposes zero endpoint acceleration, although the state-space estimator also achieves a very small RMSE. In the changepoint design, the state-space estimator spreads the abrupt velocity shift over time, producing positive acceleration estimates after the break, when true acceleration is zero. This illustrates a limitation of smoothing abrupt changes. Nevertheless, the state-space estimator has the lowest endpoint-velocity RMSE in both designs and the lowest average-acceleration RMSE in the zero-acceleration design; the quadratic estimator performs best for average acceleration under the changepoint design.
Overall, these results provide strong support for the state-space estimator
as a method for estimating the warming level, rate, and acceleration within
the designs considered. Table~\ref{tab:sim_bias} complements the RMSE
comparison with the mean estimation errors, i.e. biases.

\begin{table}[p]
\centering
\caption{\textbf{Root mean squared errors across DGPs and estimators.} Results use 1,000 replications per DGP. Units are \(^{\circ}\mathrm{C}\) for level, \(^{\circ}\mathrm{C}\,\mathrm{decade}^{-1}\) for rate, and \(^{\circ}\mathrm{C}\,\mathrm{decade}^{-2}\) for acceleration. Each simulated dataset is fitted by all four estimators. The kernel estimator uses the covariate-adjusted local quadratic specification with a 25-year bandwidth. State-space and kernel estimation use the full sample; the parametric benchmarks use observations from 1970 onwards. $T_0$ denotes August 2016. The changepoint estimator imposes zero instantaneous acceleration away from the break; its zero RMSE in Panel C for the zero-acceleration and changepoint DGPs follows from this restriction. The lowest RMSE in each row is shown in bold, based on unrounded values. Values are rounded to three decimal places.}
\label{tab:simstudy}
\begingroup\scriptsize
\setlength{\tabcolsep}{4pt}
\begin{adjustbox}{max width=\linewidth}
\begin{tabular}{lcccc}\toprule
DGP & State-space (SS) & Kernel & Quadratic & Changepoint \\\midrule
\multicolumn{5}{l}{\textbf{Panel A: Estimand $x(T)$}}\\
\addlinespace[2pt]
Zero acceleration & \textbf{0.020} & 0.050 & 0.033 & 0.053 \\
Increasing acceleration & \textbf{0.032} & 0.050 & 0.039 & 0.127 \\
NASA-smoothed acceleration & \textbf{0.042} & 0.050 & 0.086 & 0.060 \\
Constant positive acceleration & \textbf{0.020} & 0.050 & 0.033 & 0.043 \\
Changepoint & 0.058 & 0.052 & 0.067 & \textbf{0.049} \\
\addlinespace[4pt]
\multicolumn{5}{l}{\textbf{Panel B: Estimand $v(T)$}}\\
\addlinespace[2pt]
Zero acceleration & \textbf{0.010} & 0.113 & 0.025 & 0.091 \\
Increasing acceleration & \textbf{0.041} & 0.113 & 0.055 & 0.281 \\
NASA-smoothed acceleration & \textbf{0.093} & 0.117 & 0.213 & 0.167 \\
Constant positive acceleration & \textbf{0.010} & 0.113 & 0.025 & 0.075 \\
Changepoint & \textbf{0.069} & 0.142 & 0.085 & 0.105 \\
\addlinespace[4pt]
\multicolumn{5}{l}{\textbf{Panel C: Estimand $a(T)$}}\\
\addlinespace[2pt]
Zero acceleration & 0.003 & 0.107 & 0.008 & \textbf{0.000} \\
Increasing acceleration & \textbf{0.033} & 0.108 & 0.045 & 0.229 \\
NASA-smoothed acceleration & \textbf{0.097} & 0.119 & 0.192 & 0.229 \\
Constant positive acceleration & \textbf{0.003} & 0.107 & 0.008 & 0.035 \\
Changepoint & 0.081 & 0.181 & 0.034 & \textbf{0.000} \\
\addlinespace[4pt]
\multicolumn{5}{l}{\textbf{Panel D: Estimand $\overline{a_{T_0,T}}$}}\\
\addlinespace[2pt]
Zero acceleration & \textbf{0.003} & 0.102 & 0.008 & 0.090 \\
Increasing acceleration & \textbf{0.027} & 0.103 & 0.037 & 0.222 \\
NASA-smoothed acceleration & \textbf{0.090} & 0.112 & 0.184 & 0.221 \\
Constant positive acceleration & \textbf{0.003} & 0.102 & 0.008 & 0.077 \\
Changepoint & 0.080 & 0.182 & \textbf{0.034} & 0.171 \\
\bottomrule\end{tabular}
\end{adjustbox}
\endgroup
\end{table}

The zero- and constant-acceleration designs show the benefit of pooling
information when curvature is simple. The kernel estimator has little bias
in these cases but appreciably larger dispersion in the derivative
estimates. Under the NASA-smoothed design, the kernel estimator has less
negative bias than the state-space estimator for both acceleration
quantities, but higher RMSE. Thus the state-space advantage in this case
reflects a bias--variance trade-off rather than uniformly smaller bias.

The changepoint design highlights a different difficulty. Smooth estimators
spread a velocity jump over time and consequently have positive acceleration
bias after the true break. The changepoint estimator is exact for endpoint
acceleration by construction, but estimating the breakpoint can place it on
the wrong side of the August 2016 cutoff. This occurs in 20.5\% of the replications,
producing a nonzero estimated average acceleration even though its true
value is zero. The quadratic estimator has the lowest RMSE for this average
in this particular design, despite having the wrong trend specification.
This result depends on the break's proximity to the chosen cutoff and is
not a general ranking for all changepoint locations.

The increasing-acceleration and NASA-smoothed designs illustrate why the
parametric benchmarks can underestimate recent warming dynamics
(Table~\ref{tab:sim_bias}). The quadratic model spreads curvature over the
post-1970 estimation window, while the changepoint model represents the
post-break period by a single slope. When the warming rate continues to rise
within these windows, the fitted summaries can fall below the endpoint rate.
All four estimators underestimate the endpoint warming rate and both acceleration estimands on average in these two designs. Each flexible estimator has less downward bias than either parametric benchmark. Indeed, under the NASA-smoothed design,
endpoint warming-rate biases are \(-0.211\) and \(-0.104\,^{\circ}\mathrm{C}\,\mathrm{decade}^{-1}\)
for the quadratic and changepoint estimators, respectively, compared with \(-0.063\) and
\(-0.032\,^{\circ}\mathrm{C}\,\mathrm{decade}^{-1}\) for the state-space and nonparametric estimators, respectively. Under the increasing-acceleration design, the changepoint estimator has the largest downward warming-rate bias, at \(-0.278\,^{\circ}\mathrm{C}\,\mathrm{decade}^{-1}\), compared with \(-0.049\) for the quadratic estimator and \(-0.024\) and \(-0.007\,^{\circ}\mathrm{C}\,\mathrm{decade}^{-1}\) for the state-space and nonparametric estimators, respectively. The state-space estimator also underestimates the endpoint warming rate in the changepoint design and has negligible rate bias in the two remaining designs. The downward bias in three designs, including the NASA-smoothed design, suggests that our state-space estimate of the mid-2026 warming rate could be conservative.

\begin{table}[p]
\centering
\caption{\textbf{Mean estimation errors (biases) across DGPs and estimators.} Results use 1,000 replications per DGP, with Monte Carlo standard errors in parentheses. Units, estimation windows, and kernel specification are as in Table~\ref{tab:simstudy}. Values are rounded to three decimal places.}
\label{tab:sim_bias}
\begingroup\scriptsize
\setlength{\tabcolsep}{4pt}
\begin{adjustbox}{max width=\linewidth}
\begin{tabular}{lcccc}\toprule
DGP & State-space (SS) & Kernel & Quadratic & Changepoint \\\midrule
\multicolumn{5}{l}{\textbf{Panel A: Estimand $x(T)$}}\\
\addlinespace[2pt]
Zero acceleration & 0.000 (0.001) & -0.001 (0.002) & -0.002 (0.001) & -0.002 (0.002) \\
Increasing acceleration & -0.011 (0.001) & -0.002 (0.002) & -0.022 (0.001) & -0.121 (0.001) \\
NASA-smoothed acceleration & -0.019 (0.001) & -0.007 (0.002) & -0.079 (0.001) & -0.020 (0.002) \\
Constant positive acceleration & 0.000 (0.001) & -0.001 (0.002) & -0.002 (0.001) & -0.010 (0.001) \\
Changepoint & -0.040 (0.001) & 0.015 (0.002) & -0.059 (0.001) & 0.007 (0.002) \\
\addlinespace[4pt]
\multicolumn{5}{l}{\textbf{Panel B: Estimand $v(T)$}}\\
\addlinespace[2pt]
Zero acceleration & 0.000 (0.000) & -0.001 (0.004) & -0.001 (0.001) & -0.001 (0.003) \\
Increasing acceleration & -0.024 (0.001) & -0.007 (0.004) & -0.049 (0.001) & -0.278 (0.001) \\
NASA-smoothed acceleration & -0.063 (0.002) & -0.032 (0.004) & -0.211 (0.001) & -0.104 (0.004) \\
Constant positive acceleration & 0.000 (0.000) & -0.001 (0.004) & -0.001 (0.001) & -0.024 (0.002) \\
Changepoint & -0.031 (0.002) & 0.086 (0.004) & -0.081 (0.001) & 0.027 (0.003) \\
\addlinespace[4pt]
\multicolumn{5}{l}{\textbf{Panel C: Estimand $a(T)$}}\\
\addlinespace[2pt]
Zero acceleration & 0.000 (0.000) & -0.001 (0.003) & 0.000 (0.000) & 0.000 (0.000) \\
Increasing acceleration & -0.027 (0.001) & -0.017 (0.003) & -0.044 (0.000) & -0.229 (0.000) \\
NASA-smoothed acceleration & -0.073 (0.002) & -0.053 (0.003) & -0.192 (0.000) & -0.229 (0.000) \\
Constant positive acceleration & 0.000 (0.000) & -0.001 (0.003) & 0.000 (0.000) & -0.035 (0.000) \\
Changepoint & 0.070 (0.001) & 0.146 (0.003) & 0.032 (0.000) & 0.000 (0.000) \\
\addlinespace[4pt]
\multicolumn{5}{l}{\textbf{Panel D: Estimand $\overline{a_{T_0,T}}$}}\\
\addlinespace[2pt]
Zero acceleration & 0.000 (0.000) & -0.001 (0.003) & 0.000 (0.000) & -0.002 (0.003) \\
Increasing acceleration & -0.019 (0.001) & -0.009 (0.003) & -0.036 (0.000) & -0.222 (0.000) \\
NASA-smoothed acceleration & -0.067 (0.002) & -0.046 (0.003) & -0.183 (0.000) & -0.114 (0.006) \\
Constant positive acceleration & 0.000 (0.000) & -0.001 (0.003) & 0.000 (0.000) & -0.021 (0.002) \\
Changepoint & 0.070 (0.001) & 0.150 (0.003) & 0.032 (0.000) & 0.073 (0.005) \\
\bottomrule\end{tabular}
\end{adjustbox}
\endgroup
\end{table}

\FloatBarrier

\subsection{Interval coverage and inference}\label{app:coverage}
We assess coverage for all four estimators under the NASA-smoothed warming path and then for each parametric benchmark under its own fitted warming path. The interval procedures are described in Appendices~\ref{app:ssbootstrap}, \ref{app:kernel_bootstrap}, \ref{sec:regression}, and~\ref{app:changepoint_bootstrap}.

\subsubsection{Local NASA-calibrated coverage assessment}\label{app:ssboot_mc}
We assess the finite-sample coverage of the four estimators' intervals in a
Monte Carlo experiment calibrated to the NASA series. The results use
\(R=400\) independently seeded outer replications under DGP~3 from
Appendix~\ref{app:sim_designs}. The NASA-smoothed acceleration path is held
fixed and integrated to obtain the warming rate and level. Each synthetic
record adds the fitted NASA covariate contribution and independently simulated
ARMA(1,1) irregular variation with monthly innovation variances and a
6,000-month burn-in. As in the accuracy experiment, the synthetic records
are not individually rebased. All four estimators are applied to the same
synthetic record, and the fixed true warming quantities are used to evaluate
coverage. The sample spans January 1881--August
2026, with past-decade average acceleration measured from August 2016.
The flexible estimators use the full sample, while the parametric benchmarks
use observations from 1970 onwards, as in the empirical analysis.

The state-space model is refitted, including its smoothing and nuisance
parameters, and nominal \(90\%\) intervals are constructed using \(B=399\)
inner bootstrap replications. Nonparametric intervals also use 399 inner
replications, compared with the 1,999 used for the empirical bootstrap intervals. Quadratic intervals use the empirical HAC procedure with 24
monthly lags. Changepoint intervals use 399 parametric bootstrap replications,
re-estimating the trend and covariate coefficients and repeating the complete
breakpoint search with at least 60 observations on either side of each
candidate break. As in the empirical procedure, the changepoint inner
bootstrap uses stationary ARMA(1,1) errors with constant innovation variance
fitted to the regression residuals.

Table~\ref{tab:ss_bootstrap_coverage} reports the results. The state-space intervals
have less than their nominal two-sided \(90\%\) coverage for the level,
rate, and past-decade average acceleration, while endpoint-acceleration
coverage is \(95.75\%\). Where there is undercoverage, it is strongly
asymmetric: estimated lower-tail noncoverage is below the nominal
\(5\%\), whereas estimated upper-tail noncoverage is materially above
\(5\%\). For average acceleration over the past decade, two-sided coverage
is \(81.5\%\), with lower- and upper-tail noncoverage of \(2.25\%\) and
\(16.25\%\), respectively.

For the nonparametric estimator, coverage of the nominal
\(90\%\) intervals is \(85\%\) for \(x(T)\), \(86.5\%\) for \(v(T)\),
\(82.75\%\) for \(a(T)\), and \(84\%\) for past-decade average acceleration
(Table~\ref{tab:ss_bootstrap_coverage}).
The acceleration undercoverage is entirely due to upper-tail misses.
The true endpoint acceleration falls below the lower bound in none of the
replications and above the upper bound in \(17.25\%\); the corresponding
frequencies for average acceleration are \(0\%\) and \(16\%\), compared
with nominal frequencies of \(5\%\) in each tail.

The parametric benchmarks have substantially poorer coverage for the warming
rate and acceleration. Quadratic intervals lie below the true endpoint rate,
endpoint acceleration, and past-decade average acceleration in all 400
replications, giving zero coverage for each. The corresponding
changepoint coverage is \(49.5\%\) for the rate and \(55.25\%\) for average
acceleration; no endpoint-acceleration interval is reported because the
changepoint model imposes zero acceleration away from the break.
This poor coverage is consistent with the substantial negative biases under
DGP~3 documented in Table~\ref{tab:sim_bias}. Neither benchmark accommodates
the evolution of the warming rate and acceleration in this design, and their
intervals tend to lie below the true current rate and acceleration. We next
examine whether the benchmarks' poor coverage persists under warming paths
satisfying their trend restrictions. 

\clearpage

\begin{table}[!htbp]
\centering
\caption{\textbf{Monte Carlo coverage under the common NASA-calibrated design.}
Nominal \(90\%\) confidence intervals for all four estimators are evaluated
on the same 400 outer samples under the fixed NASA-smoothed warming path
of DGP~3 in Section~\ref{app:sim_designs}. Bootstrap intervals use 399 inner replications;
quadratic intervals use HAC standard errors. Average acceleration spans
August 2016--August 2026. ``Below'' is the estimated probability that the
known truth lies below the lower interval endpoint, and ``Above'' is the
probability that it lies above the upper endpoint; their nominal values are
both \(5\%\). Monte Carlo standard errors are shown for two-sided coverage.
A dash denotes an interval not reported by the procedure.}
\label{tab:ss_bootstrap_coverage}
\small
\begin{tabular}{lrrrr}\toprule
Estimand & Coverage & MC s.e. & Below & Above \\\midrule
\multicolumn{5}{l}{\textbf{Panel A: State-space (semiparametric)}}\\
\addlinespace[2pt]
\(x(T)\) & 0.850 & 0.018 & 0.035 & 0.115 \\
\(v(T)\) & 0.800 & 0.020 & 0.030 & 0.170 \\
\(a(T)\) & 0.958 & 0.010 & 0.010 & 0.033 \\
\(\overline{a}_{T-10,T}\) & 0.815 & 0.019 & 0.023 & 0.163 \\
\addlinespace[4pt]
\multicolumn{5}{l}{\textbf{Panel B: Nonparametric}}\\
\addlinespace[2pt]
\(x(T)\) & 0.850 & 0.018 & 0.033 & 0.118 \\
\(v(T)\) & 0.865 & 0.017 & 0.000 & 0.135 \\
\(a(T)\) & 0.828 & 0.019 & 0.000 & 0.173 \\
\(\overline{a}_{T-10,T}\) & 0.840 & 0.018 & 0.000 & 0.160 \\
\addlinespace[4pt]
\multicolumn{5}{l}{\textbf{Panel C: Quadratic}}\\
\addlinespace[2pt]
\(x(T)\) & 0.193 & 0.020 & 0.000 & 0.808 \\
\(v(T)\) & 0.000 & 0.000 & 0.000 & 1.000 \\
\(a(T)\) & 0.000 & 0.000 & 0.000 & 1.000 \\
\(\overline{a}_{T-10,T}\) & 0.000 & 0.000 & 0.000 & 1.000 \\
\addlinespace[4pt]
\multicolumn{5}{l}{\textbf{Panel D: Changepoint}}\\
\addlinespace[2pt]
\(x(T)\) & 0.750 & 0.022 & 0.045 & 0.205 \\
\(v(T)\) & 0.495 & 0.025 & 0.020 & 0.485 \\
\(a(T)\) & -- & -- & -- & -- \\
\(\overline{a}_{T-10,T}\) & 0.553 & 0.025 & 0.053 & 0.395 \\
\bottomrule\end{tabular}
\end{table}

\FloatBarrier

\newpage

\subsubsection{Coverage under parametric warming paths}\label{app:parametric_reference_coverage}
We repeat the coverage assessment for the parametric estimators under fixed
warming paths satisfying their trend restrictions. The quadratic estimator
is applied to the NASA-fitted quadratic path, and the changepoint estimator
to the NASA-fitted broken-linear path with a February 2014 break. These are
DGPs~4 and~5 in Appendix~\ref{app:sim_designs}.
Each design uses its corresponding fitted covariate contribution and the
NASA state-space ARMA(1,1) noise calibration with monthly innovation variances
and a 6,000-month burn-in. The sample spans January
1881--August 2026, with past-decade average acceleration measured from
August 2016. Each estimator is evaluated on 400 outer replications of its
corresponding design.

Both estimators are fitted from 1970 onwards, re-estimating all trend and
covariate coefficients. The interval procedures are those used in
Appendix~\ref{app:ssboot_mc}: quadratic intervals use HAC standard errors with
24 monthly lags, and changepoint intervals use 399 bootstrap replications
with the complete breakpoint search repeated.  

Table~\ref{tab:parametric_smooth_coverage} reports the results. Endpoint
warming-rate coverage is \(79\%\) for the quadratic estimator and
\(84.75\%\) for the changepoint estimator, compared with \(0\%\) and
\(49.5\%\), respectively, under DGP~3.
Under warming paths satisfying their trend restrictions, the benchmarks
therefore achieve substantially higher rate coverage, although both remain
below the nominal \(90\%\). Unlike under DGP~3, their rate intervals miss
the truth on both sides with similar frequencies.  

\clearpage

\begin{table}[!htbp]
\centering
\caption{\textbf{Monte Carlo coverage under parametric warming paths.}
Each benchmark is evaluated under its corresponding NASA-fitted trend and
covariate contribution, with NASA state-space ARMA(1,1) noise and monthly
innovation variances. Results use 400 outer replications per design.
Nominal \(90\%\) quadratic intervals use HAC standard errors; changepoint
intervals use 399 inner bootstrap replications with breakpoint selection
repeated. Average acceleration spans August 2016--August 2026. Coverage,
its Monte Carlo standard error, and the ``Below'' and ``Above'' probabilities
are defined as in Table~\ref{tab:ss_bootstrap_coverage}. A dash denotes an
interval not reported by the procedure.}
\label{tab:parametric_smooth_coverage}
\small
\begin{tabular}{lrrrr}\toprule
Estimand & Coverage & MC s.e. & Below & Above \\
\midrule
\multicolumn{5}{l}{\textbf{Panel A: Quadratic under quadratic truth}}\\
\addlinespace[2pt]
\(x(T)\) & 0.763 & 0.021 & 0.100 & 0.138 \\
\(v(T)\) & 0.790 & 0.020 & 0.098 & 0.113 \\
\(a(T)\) & 0.790 & 0.020 & 0.088 & 0.123 \\
\(\overline{a}_{T-10,T}\) & 0.790 & 0.020 & 0.088 & 0.123 \\
\addlinespace[4pt]
\multicolumn{5}{l}{\textbf{Panel B: Changepoint under changepoint truth}}\\
\addlinespace[2pt]
\(x(T)\) & 0.843 & 0.018 & 0.088 & 0.070 \\
\(v(T)\) & 0.848 & 0.018 & 0.073 & 0.080 \\
\(a(T)\) & -- & -- & -- & -- \\
\(\overline{a}_{T-10,T}\) & 0.985 & 0.006 & 0.015 & 0.000 \\
\bottomrule\end{tabular}
\end{table}

\clearpage

\section{Sensitivity analyses}
\label{app:robust}

This appendix reports sensitivity analyses for the main endpoint estimates
reported in Section~\ref{sec:results}. The baseline analysis uses monthly data,
models stochastic variation in the state-space specification through the
warming acceleration, uses baseline smoothing choices for both the
state-space and local-polynomial estimators, and includes covariates. These choices are natural for the
main specification, but the endpoint quantities, especially the warming rate
\(v(T)\) and warming acceleration \(a(T)\), may be sensitive to temporal
aggregation, smoothing assumptions, and covariates. We therefore examine several sources of
sensitivity.

First, we repeat the analysis using calendar-year averages instead of monthly
observations (Section~\ref{app:robust_yearly}). This checks whether the main conclusions depend on the use of
monthly data and on the explicit modelling of monthly persistence and
time-varying variance. Second, for the parametric estimators, we vary the starting point $t_0$ (Section~\ref{app:parametric_t0}). Third, for the local-polynomial estimator, we vary the
bandwidth \(h\), which controls the degree of local smoothing (Section~\ref{app:kernel_bandwidth}). Fourth, for the
state-space estimator, we vary the smoothing parameter \(\sigma\) around its
dataset-specific baseline estimate (Section~\ref{app:ss_q}). Fifth, we estimate an alternative
state-space specification in which stochastic variation enters through the
warming rate \(v(t)\), rather than through the warming acceleration \(a(t)\) (Section~\ref{app:ss_velocity}). Sixth, we investigate the sensitivity of the average acceleration estimates $\overline{a_{T_0,T}}$  to varying the starting year \(T_0\) (Section~\ref{app:avgacc_sensitivity}). Lastly, we examine the sensitivity of the estimates coming from the preferred state-space model to removing the covariates (Section~\ref{app:no_covariates}).
Together, these analyses assess whether the main empirical conclusions are
specific to the baseline modelling choices or are stable across reasonable
alternative specifications.

\subsection{Using annual averages}
\label{app:robust_yearly}

The baseline analysis uses monthly observations in order to exploit the full
information in the temperature records and to model short-run persistence and
seasonal variation in the innovation variance directly. As a robustness check,
we repeat the analysis using calendar-year averages. This aggregation removes
much of the high-frequency variation in the monthly records and therefore
provides a useful check on whether the endpoint estimates are driven by the
monthly treatment of short-run variability. The annual analysis ends in 2025, whereas the baseline monthly analysis includes observations through July or August 2026; differences between them therefore reflect both temporal aggregation and the information available at the endpoint.

For the annual-data robustness exercise, we specify the irregular component
separately from the monthly baseline. The monthly baseline uses an
ARMA\((1,1)\) irregular component to capture short-run persistence in monthly
temperature anomalies. After aggregation to annual means, much of this
short-run persistence is removed, and an ARMA\((1,1)\) irregular component
can become overly flexible relative to the information content of the
annual series. We therefore use an AR\((1)\) irregular component for NASA,
NOAA, HadCRUT5, and Berkeley Earth, and an IID irregular component for ERA5. These choices were supported by the BIC.\footnote{Further, following common practice, e.g. \citet{BHK2024}, the annual ENSO
variable is constructed from the four-month-lagged monthly data; that is, it is
a September--August average.}

We hold each annual state-space diffusion parameter \(\sigma\) fixed at its
corresponding monthly estimate, while re-estimating the remaining parameters.
This holds trend flexibility constant across sampling frequencies and avoids
relying on weakly identified annual diffusion estimates, which can exhibit the
variance-estimator pile-up at zero documented in related time-varying-parameter
models by \citet{StockWatson1998}. The annual time step is retained in the state
innovation covariance matrix.

Table~\ref{tab:annual_robustness_endpoint} reports endpoint and average-acceleration
estimates from the annual-data analysis. The results are broadly consistent with the monthly
baseline. The estimated warming level \(x(T)\) remains close to
\(1.4\,^\circ\mathrm{C}\) across methods. The annual state-space mean rate is
\(0.416\,^\circ\mathrm{C}\,\mathrm{decade}^{-1}\), above the quadratic and
changepoint means of \(0.275\) and \(0.358\), respectively; all five state-space
rates exceed their changepoint counterparts. Annual state-space estimates of
\(v(T)\) and \(a(T)\) remain below their monthly counterparts. Mean endpoint
acceleration is \(0.147\,^\circ\mathrm{C}\,\mathrm{decade}^{-2}\), and mean
average acceleration over 2015--2025 is
\(0.143\,^\circ\mathrm{C}\,\mathrm{decade}^{-2}\). The acceleration estimates from the flexible
estimators remain positive on average and exceed the long-window quadratic
estimates. The \(90\%\) annual state-space endpoint acceleration intervals
include zero for four of the five records; only the ERA5 interval excludes zero.
Panel~D shows that average acceleration is positive for all five annual
state-space fits, with \(90\%\) intervals excluding zero for all five records. Annual kernel average-acceleration estimates are also
positive throughout, and all five nominal intervals exclude zero.

\subsection{Sensitivity of parametric benchmarks to the estimation window}
\label{app:parametric_t0}

The parametric benchmarks reported in Section~\ref{sec:results} are estimated over the
post-1970 sample, with baseline starting year \(t_0=1970\). This choice is
motivated by the common interpretation that the modern period of sustained
global warming begins around 1970. Since the parametric benchmarks impose
restrictive functional forms over the full estimation window, their estimates
may depend on the choice of \(t_0\). We therefore repeat the quadratic and
changepoint analyses for alternative starting years
\[
    t_0 \in \{1950,1960,1970,1980,1990\}.
\]

Table~\ref{tab:quadratic_startdate_robustness} reports the resulting endpoint
warming level, warming-rate, and acceleration estimates for the quadratic
model, while Table~\ref{tab:changepoint_startdate_robustness} reports the
endpoint warming level, endpoint warming rate, slope change, and selected break
date for the changepoint model. The exercise is not intended to select an
optimal starting year. Rather, it illustrates the sensitivity of the
parametric benchmarks. The quadratic model estimates a single constant
acceleration over the full window from \(t_0\) to \(T\), while the changepoint
model estimates a single post-break slope. Changing \(t_0\) therefore changes
the long-window object being summarized. Separate average-acceleration panels
are unnecessary for these parametric specifications: under a quadratic trend,
past-decade average acceleration equals the constant acceleration already
reported in Panel~C of Table~\ref{tab:quadratic_startdate_robustness}. For the
changepoint model it is zero for every specification, conditional on the
selected breaks, all of which precede the past-decade averaging interval
(Table~\ref{tab:changepoint_startdate_robustness}).

For the quadratic model, the estimated endpoint warming level is stable across
starting years and datasets. The endpoint warming rate and acceleration are more
sensitive to \(t_0\), but the qualitative conclusion is unchanged: the estimated
quadratic acceleration is positive for all datasets and all starting years
considered. The cross-dataset mean acceleration is smallest for the baseline
choice \(t_0=1970\), at
\(0.030\,^\circ\mathrm{C}\,\mathrm{decade}^{-2}\), and increases for later
starting years, reaching
\(0.079\,^\circ\mathrm{C}\,\mathrm{decade}^{-2}\) when \(t_0=1990\). Thus, the
quadratic benchmark supports positive curvature, but the magnitude of that
curvature is window-dependent.

For the changepoint model, the sensitivity to \(t_0\) is more structural. When
the estimation window begins in 1950, the single-break model often selects a
break in the 1970s, thereby using its single changepoint to capture the
transition from the mid-century plateau to the modern warming period. The
resulting endpoint warming rates are consequently much lower. For starting
years \(t_0\geq 1970\), the selected break dates are instead concentrated
around 2013--2014 for all datasets, and the post-break warming rates are stable
at approximately
\(0.35\text{--}0.36\,^\circ\mathrm{C}\,\mathrm{decade}^{-1}\) on average. This pattern
supports the interpretation that the recent changepoint result is robust within
the post-1970 warming era, but also shows that a one-break model estimates
different features of the temperature record when applied over longer windows.

Overall, the results confirm that the parametric benchmarks are useful but
window-dependent summaries. They provide informative reference models, and in
particular they support the qualitative conclusion of positive recent
acceleration or elevated recent warming rates. Their endpoint estimates are conditional on the imposed path restrictions.
Flexible estimators provide an alternative that allows these restrictions to
be relaxed, with additional sensitivity to the degree of smoothing.

\subsection{Alternative local-polynomial bandwidths}
\label{app:kernel_bandwidth}

The local-polynomial estimator depends on the bandwidth \(h\). Smaller values
of \(h\) allow the fitted path to respond more strongly to recent fluctuations,
whereas larger values impose greater smoothness. Since endpoint derivative
estimates are particularly sensitive to the degree of smoothing, we examine a
range of bandwidths around the baseline value \(h=25\) years.

Table~\ref{tab:kernel_bandwidth_robustness} reports endpoint and average-acceleration estimates for
bandwidths \(h=15,20,\ldots,35\) years. The estimated warming level \(x(T)\) is
relatively stable across bandwidths. The warming-rate estimates \(v(T)\) are
more sensitive, but remain above the corresponding parametric long-window estimates except for ERA5 at $h=15$ years, where the rate is slightly below its changepoint counterpart. The acceleration estimates \(a(T)\) are the
most bandwidth-sensitive, as expected for endpoint second derivatives. Average acceleration is positive for all five records at bandwidths of 15--35 years. Endpoint acceleration is also positive throughout, except for ERA5 at $h=15$ years, where the estimate is $-0.024\,^{\circ}\mathrm{C}\,\mathrm{decade}^{-2}$.  The cross-dataset mean past-decade average acceleration ranges from \(0.155\) to \(0.205\,^{\circ}\mathrm{C}\,\mathrm{decade}^{-2}\). These panels report point estimates; positivity alone does not establish significance at each bandwidth.

\subsection{State-space smoothing parameter}
\label{app:ss_q}

In the baseline kinematic state-space model, stochastic variation enters
through the acceleration equation
\[
    \mathrm{d}a(t)=\sigma\,\mathrm{d}W(t).
\]
The implied smoothness of the latent state path is governed by the scale of the
state innovation variance. In the baseline analysis, the smoothing parameter
\(\sigma\) is estimated from the data by maximum likelihood, jointly with the
covariate coefficients and irregular-component parameters. To assess sensitivity
around the dataset-specific maximum-likelihood estimate \(\hat{\sigma}\), we
fix \(\sigma\) at multiples of this estimate and re-estimate the remaining
parameters. Specifically, we consider
\[
    \sigma \in
    \{0.5\hat{\sigma},0.75\hat{\sigma},\hat{\sigma},
      1.5\hat{\sigma},2\hat{\sigma}\}.
\]
Smaller values of \(\sigma\) impose smoother latent paths, while larger values allow
more rapid local variation in the warming rate and acceleration.

Table~\ref{tab:ss_q_robustness} reports the resulting endpoint and average-acceleration estimates. The
warming level \(x(T)\) is stable across a wide range of \(\sigma\)-values. The
warming rate \(v(T)\) increases as \(\sigma\) increases, reflecting the greater
ability of the model to adapt to recent changes. The acceleration estimates
\(a(T)\) are more sensitive, but remain positive for all datasets and all
values of \(\sigma\) considered. This supports the qualitative conclusion that the
recent endpoint acceleration is positive, while confirming that its precise
magnitude depends on the amount of smoothing imposed. Panel~D gives the same
qualitative result for past-decade average acceleration: estimates are positive
for every dataset and increase with \(\sigma\). The cross-dataset mean rises
from \(0.112\) at \(0.5\hat\sigma\) to \(0.222\) at \(2\hat\sigma\),
compared with \(0.166\,^\circ\mathrm{C}\,\mathrm{decade}^{-2}\) at the
maximum-likelihood estimate. These are point-estimate comparisons; confidence intervals
are not reported for the alternative smoothing scales.

\subsection{State-space model specified through the warming rate}
\label{app:ss_velocity}

The baseline state-space model specifies stochastic variation in the
acceleration process,
\[
    \mathrm{d}a(t)=\sigma\,\mathrm{d}W(t).
\]
As discussed in Section~\ref{sec:ssmodel}, this can be interpreted as a
higher-order stochastic-smoothing specification: the warming level \(x(t)\),
warming rate \(v(t)\), and warming acceleration \(a(t)\) are estimated jointly,
with \(a(t)\) treated as a continuous latent state. This is useful because
\(a(T)\) is one of the main endpoint estimands.

As an alternative, we estimate a lower-order state-space model in which
stochastic variation enters through the warming rate,
\[
    \mathrm{d}v(t)=\sigma\,\mathrm{d}W(t).
\]
In the spline-smoothing interpretation, this is the analogue of the classical
cubic-spline specification, whereas the baseline acceleration-random-walk model
corresponds to a higher-order, quintic-spline-type specification. The
velocity-random-walk model estimates the latent warming level and warming rate
without requiring an explicit stochastic model for the acceleration process. It
therefore provides a useful robustness check for the estimates of \(x(T)\) and
\(v(T)\), but it does not provide a directly comparable endpoint estimate of
\(a(T)\).

Table~\ref{tab:ss_velocity_robustness} compares endpoint estimates from the two
state-space specifications. Warming-level estimates are similar, with
cross-dataset means of \(1.413\,^{\circ}\mathrm{C}\) for the alternative model
and \(1.420\,^{\circ}\mathrm{C}\) for the baseline model. The alternative model
produces lower warming-rate estimates for every record, with a cross-dataset
mean of \(0.389\,^{\circ}\mathrm{C}\,\mathrm{decade}^{-1}\), compared with
\(0.446\,^{\circ}\mathrm{C}\,\mathrm{decade}^{-1}\) under the baseline specification.
Nevertheless, its warming-rate estimates exceed both parametric benchmarks
for all five records. Thus, the higher current rates relative to the benchmarks
persist under this lower-order specification, although their magnitudes are
sensitive to the order of smoothing. The lower endpoint rates are consistent with this difference in the treatment of warming dynamics: the baseline specification represents persistent acceleration explicitly, which can help capture a continuing increase in the warming rate near the sample endpoint.

\subsection{Sensitivity to the starting year for average acceleration}
\label{app:avgacc_sensitivity}
Table~\ref{tab:rate_change_alt_t0} varies the starting year over $T_0\in\{2010,2012,2014,2015,2016,2018,2020\}$. State-space average acceleration estimates are positive and all nominal 90\% intervals exclude zero. Nonparametric point estimates are also positive throughout, with cross-dataset means ranging from \(0.165\) to \(0.220\,^{\circ}\mathrm{C}\,\mathrm{decade}^{-2}\). Its nominal intervals do not uniformly exclude zero.

\subsection{Sensitivity of state-space estimates to removing climate covariates}\label{app:no_covariates}
Table~\ref{tab:no_covariates} reports state-space estimates without climate covariates and nominal 90\% bootstrap-error intervals from 1,999 replications per record. These fits use the same temperature samples and re-estimate the parameters as in the adjusted analysis. Endpoint and past-decade average acceleration estimates remain positive across all five records and the arithmetic-mean series, but the cross-dataset mean past-decade average acceleration falls from $0.166$ to $0.107\,^{\circ}$C decade$^{-2}$.

The past-decade average-acceleration intervals exclude zero for four of the five records and the arithmetic-mean series; the HadCRUT5 interval includes zero. For the arithmetic-mean series, average acceleration is $0.110$ with interval $[0.020, 0.291]\,^{\circ}$C decade$^{-2}$. Endpoint acceleration is less consistently resolved: its interval excludes zero only for NASA, Berkeley Earth and ERA5, with the lower bounds for NASA and Berkeley Earth close to zero. Thus, evidence of positive recent average acceleration persists without covariate adjustment, although its estimated magnitude and statistical significance depend on the specification.

The ERA5 fit without covariates has a state-innovation variance scale close to zero ($\hat\sigma^2\simeq4.4\times10^{-13}$), producing an almost constant acceleration path. Its comparatively narrow intervals should be interpreted alongside this fitted specification.

 \clearpage

 \newpage

\begin{table}[tbp]
\centering\scriptsize
\setlength{\tabcolsep}{3pt}
\caption{\textbf{Annual-data endpoint and average-acceleration robustness.} Calendar-year means through 2025; Panel D spans 2015--2025. All records use the same baseline as the monthly analysis, with a 1991--2020 mean of $0.88\,^{\circ}$C. State-space diffusion parameters are fixed at the corresponding monthly estimates and the remaining parameters are re-estimated. State-space intervals use 1,999 bootstrap-error draws, conditional on these diffusion parameters. Nonparametric intervals use 1,999 moving-block draws with two-year blocks. Quadratic intervals use HAC errors; changepoint intervals use 1,999 AR(1) full-search bootstrap draws. The annual search requires ten observations on each side, so every selected break is at or before 2015 and every 2015--2025 rate contrast is zero, using right-hand slopes. Units are $^{\circ}$C for level, $^{\circ}$C decade$^{-1}$ for rate, and $^{\circ}$C decade$^{-2}$ for both acceleration estimands. $\dagger$ denotes a zero under the original fitted changepoint path. Endpoint acceleration is identically zero across bootstrap fits and shown without an interval; average-acceleration intervals incorporate breakpoint selection. Cross-dataset means average the five separately fitted estimates and have no pooled interval.}\label{tab:annual_robustness_endpoint}
\begin{adjustbox}{max width=\linewidth}
\begin{tabular}{lcccccccc}
\toprule
 & \multicolumn{2}{c}{Semiparametric} & \multicolumn{2}{c}{Nonparametric} & \multicolumn{2}{c}{Quadratic} & \multicolumn{2}{c}{Changepoint} \\
Dataset & Est. & 90\% CI & Est. & 90\% CI & Est. & 90\% CI & Est. & 90\% CI \\\midrule
\multicolumn{9}{l}{\textbf{Panel A: Endpoint warming level $x(T)$}}\\
NASA & 1.417 & [1.338, 1.516] & 1.436 & [1.357, 1.495] & 1.356 & [1.306, 1.407] & 1.404 & [1.358, 1.449] \\
NOAA & 1.396 & [1.330, 1.480] & 1.412 & [1.347, 1.472] & 1.345 & [1.301, 1.389] & 1.397 & [1.357, 1.437] \\
HadCRUT5 & 1.366 & [1.307, 1.453] & 1.393 & [1.319, 1.451] & 1.323 & [1.278, 1.367] & 1.368 & [1.324, 1.407] \\
Berkeley Earth & 1.379 & [1.317, 1.470] & 1.404 & [1.330, 1.467] & 1.328 & [1.281, 1.376] & 1.382 & [1.339, 1.419] \\
ERA5 & 1.430 & [1.386, 1.499] & 1.467 & [1.386, 1.534] & 1.384 & [1.334, 1.435] & 1.446 & [1.376, 1.511] \\
\addlinespace[2pt]Cross-dataset mean & 1.398 &  & 1.422 &  & 1.347 &  & 1.399 &  \\\addlinespace[5pt]
\multicolumn{9}{l}{\textbf{Panel B: Endpoint warming rate $v(T)$}}\\
NASA & 0.453 & [0.312, 0.674] & 0.539 & [0.314, 0.629] & 0.283 & [0.249, 0.317] & 0.356 & [0.292, 0.425] \\
NOAA & 0.419 & [0.282, 0.602] & 0.471 & [0.307, 0.590] & 0.279 & [0.251, 0.308] & 0.359 & [0.305, 0.422] \\
HadCRUT5 & 0.386 & [0.286, 0.569] & 0.481 & [0.275, 0.579] & 0.256 & [0.226, 0.285] & 0.330 & [0.260, 0.384] \\
Berkeley Earth & 0.411 & [0.300, 0.612] & 0.495 & [0.294, 0.610] & 0.258 & [0.226, 0.289] & 0.343 & [0.278, 0.395] \\
ERA5 & 0.414 & [0.354, 0.553] & 0.531 & [0.310, 0.645] & 0.298 & [0.266, 0.330] & 0.402 & [0.291, 0.484] \\
\addlinespace[2pt]Cross-dataset mean & 0.416 &  & 0.503 &  & 0.275 &  & 0.358 &  \\\addlinespace[5pt]
\multicolumn{9}{l}{\textbf{Panel C: Endpoint acceleration $a(T)$}}\\
NASA & 0.180 & [-0.035, 0.492] & 0.256 & [0.031, 0.338] & 0.032 & [0.022, 0.043] & $0^{\dagger}$ & -- \\
NOAA & 0.143 & [-0.081, 0.404] & 0.180 & [0.027, 0.293] & 0.034 & [0.025, 0.043] & $0^{\dagger}$ & -- \\
HadCRUT5 & 0.134 & [-0.032, 0.369] & 0.216 & [0.014, 0.305] & 0.022 & [0.013, 0.032] & $0^{\dagger}$ & -- \\
Berkeley Earth & 0.151 & [-0.035, 0.416] & 0.219 & [0.022, 0.330] & 0.022 & [0.012, 0.032] & $0^{\dagger}$ & -- \\
ERA5 & 0.124 & [0.034, 0.285] & 0.227 & [0.008, 0.333] & 0.034 & [0.023, 0.044] & $0^{\dagger}$ & -- \\
\addlinespace[2pt]Cross-dataset mean & 0.147 &  & 0.220 &  & 0.029 &  & 0.000 &  \\\addlinespace[5pt]
\multicolumn{9}{l}{\textbf{Panel D: Past-decade average acceleration $\overline{a}_{T-10,T}$}}\\
NASA & 0.173 & [0.040, 0.387] & 0.251 & [0.038, 0.325] & 0.032 & [0.022, 0.043] & $0^{\dagger}$ & [0.000, 0.000] \\
NOAA & 0.141 & [0.001, 0.318] & 0.183 & [0.033, 0.283] & 0.034 & [0.025, 0.043] & $0^{\dagger}$ & [0.000, 0.000] \\
HadCRUT5 & 0.130 & [0.029, 0.300] & 0.213 & [0.021, 0.295] & 0.022 & [0.013, 0.032] & $0^{\dagger}$ & [0.000, 0.000] \\
Berkeley Earth & 0.147 & [0.036, 0.337] & 0.217 & [0.028, 0.316] & 0.022 & [0.012, 0.032] & $0^{\dagger}$ & [0.000, 0.000] \\
ERA5 & 0.122 & [0.067, 0.245] & 0.225 & [0.014, 0.319] & 0.034 & [0.023, 0.044] & $0^{\dagger}$ & [0.000, 0.000] \\
\addlinespace[2pt]Cross-dataset mean & 0.143 &  & 0.218 &  & 0.029 &  & 0.000 &  \\\addlinespace[5pt]
\bottomrule
\end{tabular}
\end{adjustbox}
\end{table}

\begin{table}[tbp]
\centering\scriptsize
\setlength{\tabcolsep}{3pt}
\caption{\textbf{Quadratic estimation-window sensitivity.} Point estimates under alternative sample start years. Level and rate have units of degrees and degrees per decade; acceleration is in degrees per decade squared. Quadratic endpoint and past-decade average acceleration coincide. Changepoint average acceleration is zero for every specification because every fitted break precedes the past decade.}\label{tab:quadratic_startdate_robustness}
\begin{adjustbox}{max width=\linewidth}
\begin{tabular}{lrrrrr}
\toprule
Dataset & 1950 & 1960 & 1970 & 1980 & 1990 \\\midrule
\multicolumn{6}{l}{\textbf{Panel A: Endpoint warming level $x(T)$}}\\
NASA & 1.392 & 1.393 & 1.392 & 1.411 & 1.415 \\
NOAA & 1.374 & 1.373 & 1.376 & 1.400 & 1.402 \\
HadCRUT5 & 1.370 & 1.360 & 1.347 & 1.363 & 1.362 \\
Berkeley Earth & 1.379 & 1.368 & 1.360 & 1.379 & 1.386 \\
ERA5 & 1.425 & 1.432 & 1.414 & 1.444 & 1.449 \\
Cross-dataset mean & 1.388 & 1.385 & 1.378 & 1.399 & 1.403 \\
\addlinespace[5pt]
\multicolumn{6}{l}{\textbf{Panel B: Endpoint warming rate $v(T)$}}\\
NASA & 0.297 & 0.296 & 0.292 & 0.330 & 0.365 \\
NOAA & 0.288 & 0.285 & 0.286 & 0.331 & 0.362 \\
HadCRUT5 & 0.295 & 0.278 & 0.259 & 0.292 & 0.315 \\
Berkeley Earth & 0.296 & 0.278 & 0.265 & 0.303 & 0.345 \\
ERA5 & 0.318 & 0.325 & 0.302 & 0.365 & 0.407 \\
Cross-dataset mean & 0.299 & 0.293 & 0.281 & 0.324 & 0.359 \\
\addlinespace[5pt]
\multicolumn{6}{l}{\textbf{Panel C: Endpoint acceleration $a(T)$}}\\
NASA & 0.037 & 0.036 & 0.035 & 0.055 & 0.080 \\
NOAA & 0.037 & 0.035 & 0.035 & 0.059 & 0.082 \\
HadCRUT5 & 0.038 & 0.031 & 0.023 & 0.041 & 0.059 \\
Berkeley Earth & 0.036 & 0.029 & 0.024 & 0.044 & 0.074 \\
ERA5 & 0.041 & 0.044 & 0.035 & 0.069 & 0.099 \\
Cross-dataset mean & 0.038 & 0.035 & 0.030 & 0.054 & 0.079 \\
\addlinespace[5pt]
\bottomrule
\end{tabular}
\end{adjustbox}
\end{table}

\begin{table}[tbp]
\centering\scriptsize
\setlength{\tabcolsep}{3pt}
\caption{\textbf{Changepoint estimation-window sensitivity.} Point estimates under alternative sample start years. Warming level is measured in degrees Celsius; warming rate and the slope change at the fitted break are measured in degrees Celsius per decade. Selected break dates are reported as year--month. Changepoint average acceleration is zero for every specification because every fitted break precedes the past decade.}\label{tab:changepoint_startdate_robustness}
\begin{adjustbox}{max width=\linewidth}
\begin{tabular}{lrrrrr}
\toprule
Dataset & 1950 & 1960 & 1970 & 1980 & 1990 \\\midrule
\multicolumn{6}{l}{\textbf{Panel A: Endpoint warming level $x(T)$}}\\
NASA & 1.310 & 1.445 & 1.443 & 1.441 & 1.431 \\
NOAA & 1.289 & 1.427 & 1.424 & 1.423 & 1.413 \\
HadCRUT5 & 1.277 & 1.282 & 1.390 & 1.388 & 1.378 \\
Berkeley Earth & 1.288 & 1.415 & 1.412 & 1.409 & 1.403 \\
ERA5 & 1.330 & 1.336 & 1.473 & 1.476 & 1.466 \\
Cross-dataset mean & 1.299 & 1.381 & 1.428 & 1.427 & 1.418 \\
\addlinespace[5pt]
\multicolumn{6}{l}{\textbf{Panel B: Endpoint warming rate $v(T)$}}\\
NASA & 0.202 & 0.378 & 0.366 & 0.366 & 0.360 \\
NOAA & 0.192 & 0.364 & 0.349 & 0.351 & 0.343 \\
HadCRUT5 & 0.191 & 0.192 & 0.322 & 0.323 & 0.318 \\
Berkeley Earth & 0.195 & 0.356 & 0.340 & 0.339 & 0.344 \\
ERA5 & 0.210 & 0.213 & 0.387 & 0.401 & 0.393 \\
Cross-dataset mean & 0.198 & 0.301 & 0.353 & 0.356 & 0.352 \\
\addlinespace[5pt]
\multicolumn{6}{l}{\textbf{Panel C: Slope change (degrees per decade)}}\\
NASA & 0.149 & 0.230 & 0.200 & 0.202 & 0.188 \\
NOAA & 0.147 & 0.223 & 0.192 & 0.200 & 0.182 \\
HadCRUT5 & 0.203 & 0.248 & 0.148 & 0.153 & 0.147 \\
Berkeley Earth & 0.202 & 0.196 & 0.163 & 0.169 & 0.180 \\
ERA5 & 0.177 & 0.215 & 0.211 & 0.241 & 0.223 \\
Cross-dataset mean & 0.175 & 0.222 & 0.183 & 0.193 & 0.184 \\
\addlinespace[5pt]
\multicolumn{6}{l}{\textbf{Panel D: Selected break date}}\\
NASA & 1978-04 & 2013-05 & 2014-02 & 2014-02 & 2014-07 \\
NOAA & 1977-12 & 2013-01 & 2013-05 & 2013-04 & 2013-08 \\
HadCRUT5 & 1971-12 & 1971-03 & 2014-07 & 2014-05 & 2014-07 \\
Berkeley Earth & 1971-02 & 2013-12 & 2014-07 & 2014-02 & 2014-06 \\
ERA5 & 1976-11 & 1976-09 & 2014-04 & 2014-02 & 2014-06 \\
\addlinespace[5pt]
\bottomrule
\end{tabular}
\end{adjustbox}
\end{table}

 \begin{table}[tbp]
\centering\scriptsize
\setlength{\tabcolsep}{3pt}
\caption{\textbf{Nonparametric bandwidth sensitivity.} Point estimates under bandwidths $h=15,20,\ldots,35$ years. Panel D uses $T_0=T-10$. Units are $^{\circ}$C for level, $^{\circ}$C decade$^{-1}$ for rate, and $^{\circ}$C decade$^{-2}$ for both acceleration estimands. }\label{tab:kernel_bandwidth_robustness}
\begin{adjustbox}{max width=\linewidth}
\begin{tabular}{lrrrrr}
\toprule
Dataset & 15 & 20 & 25 & 30 & 35 \\\midrule
\multicolumn{6}{l}{\textbf{Panel A: Endpoint warming level $x(T)$}}\\
NASA & 1.498 & 1.484 & 1.482 & 1.470 & 1.454 \\
NOAA & 1.426 & 1.431 & 1.441 & 1.441 & 1.434 \\
HadCRUT5 & 1.418 & 1.420 & 1.420 & 1.412 & 1.399 \\
Berkeley Earth & 1.444 & 1.440 & 1.442 & 1.434 & 1.422 \\
ERA5 & 1.470 & 1.483 & 1.502 & 1.498 & 1.485 \\
Cross-dataset mean & 1.451 & 1.451 & 1.457 & 1.451 & 1.439 \\
\addlinespace[5pt]
\multicolumn{6}{l}{\textbf{Panel B: Endpoint warming rate $v(T)$}}\\
NASA & 0.631 & 0.561 & 0.556 & 0.514 & 0.462 \\
NOAA & 0.368 & 0.399 & 0.453 & 0.455 & 0.433 \\
HadCRUT5 & 0.449 & 0.466 & 0.476 & 0.443 & 0.404 \\
Berkeley Earth & 0.498 & 0.485 & 0.502 & 0.477 & 0.439 \\
ERA5 & 0.373 & 0.430 & 0.513 & 0.496 & 0.455 \\
Cross-dataset mean & 0.464 & 0.468 & 0.500 & 0.477 & 0.438 \\
\addlinespace[5pt]
\multicolumn{6}{l}{\textbf{Panel C: Endpoint acceleration $a(T)$}}\\
NASA & 0.391 & 0.259 & 0.255 & 0.206 & 0.155 \\
NOAA & 0.016 & 0.070 & 0.152 & 0.156 & 0.135 \\
HadCRUT5 & 0.149 & 0.183 & 0.202 & 0.165 & 0.127 \\
Berkeley Earth & 0.208 & 0.183 & 0.212 & 0.183 & 0.146 \\
ERA5 & -0.024 & 0.069 & 0.193 & 0.177 & 0.136 \\
Cross-dataset mean & 0.148 & 0.153 & 0.203 & 0.177 & 0.140 \\
\addlinespace[5pt]
\multicolumn{6}{l}{\textbf{Panel D: Past-decade average acceleration $\overline{a}_{T-10,T}$}}\\
NASA & 0.328 & 0.254 & 0.254 & 0.219 & 0.171 \\
NOAA & 0.065 & 0.099 & 0.157 & 0.165 & 0.148 \\
HadCRUT5 & 0.173 & 0.188 & 0.202 & 0.176 & 0.141 \\
Berkeley Earth & 0.205 & 0.192 & 0.213 & 0.194 & 0.163 \\
ERA5 & 0.053 & 0.111 & 0.198 & 0.187 & 0.149 \\
Cross-dataset mean & 0.165 & 0.169 & 0.205 & 0.188 & 0.155 \\
\addlinespace[5pt]
\bottomrule
\end{tabular}
\end{adjustbox}
\end{table}

\begin{table}[tbp]
\centering\scriptsize
\setlength{\tabcolsep}{3pt}
\caption{\textbf{State-space smoothing sensitivity.} Diffusion parameters are fixed at the indicated multiples of the baseline estimate; remaining parameters are refitted. Point estimates only. Panel D uses $T_0=T-10$. Units are $^{\circ}$C for level, $^{\circ}$C decade$^{-1}$ for rate, and $^{\circ}$C decade$^{-2}$ for both acceleration estimands. }\label{tab:ss_q_robustness}
\begin{adjustbox}{max width=\linewidth}
\begin{tabular}{lrrrrr}
\toprule
Dataset & $0.5\hat\sigma$ & $0.75\hat\sigma$ & $1\hat\sigma$ & $1.5\hat\sigma$ & $2\hat\sigma$ \\\midrule
\multicolumn{6}{l}{\textbf{Panel A: Endpoint warming level $x(T)$}}\\
NASA & 1.414 & 1.428 & 1.437 & 1.449 & 1.458 \\
NOAA & 1.396 & 1.406 & 1.410 & 1.415 & 1.417 \\
HadCRUT5 & 1.361 & 1.373 & 1.381 & 1.390 & 1.396 \\
Berkeley Earth & 1.388 & 1.405 & 1.414 & 1.426 & 1.433 \\
ERA5 & 1.435 & 1.448 & 1.457 & 1.470 & 1.478 \\
Cross-dataset mean & 1.399 & 1.412 & 1.420 & 1.430 & 1.436 \\
\addlinespace[5pt]
\multicolumn{6}{l}{\textbf{Panel B: Endpoint warming rate $v(T)$}}\\
NASA & 0.432 & 0.474 & 0.504 & 0.550 & 0.582 \\
NOAA & 0.397 & 0.420 & 0.433 & 0.446 & 0.450 \\
HadCRUT5 & 0.338 & 0.373 & 0.394 & 0.423 & 0.445 \\
Berkeley Earth & 0.398 & 0.444 & 0.473 & 0.509 & 0.533 \\
ERA5 & 0.368 & 0.400 & 0.427 & 0.464 & 0.486 \\
Cross-dataset mean & 0.387 & 0.422 & 0.446 & 0.479 & 0.499 \\
\addlinespace[5pt]
\multicolumn{6}{l}{\textbf{Panel C: Endpoint acceleration $a(T)$}}\\
NASA & 0.152 & 0.195 & 0.229 & 0.285 & 0.324 \\
NOAA & 0.122 & 0.141 & 0.151 & 0.159 & 0.154 \\
HadCRUT5 & 0.086 & 0.116 & 0.137 & 0.168 & 0.190 \\
Berkeley Earth & 0.129 & 0.170 & 0.197 & 0.236 & 0.261 \\
ERA5 & 0.083 & 0.109 & 0.132 & 0.164 & 0.184 \\
Cross-dataset mean & 0.114 & 0.146 & 0.169 & 0.202 & 0.223 \\
\addlinespace[5pt]
\multicolumn{6}{l}{\textbf{Panel D: Past-decade average acceleration $\overline{a}_{T-10,T}$}}\\
NASA & 0.148 & 0.189 & 0.221 & 0.273 & 0.312 \\
NOAA & 0.121 & 0.140 & 0.152 & 0.166 & 0.170 \\
HadCRUT5 & 0.084 & 0.114 & 0.135 & 0.166 & 0.190 \\
Berkeley Earth & 0.126 & 0.166 & 0.193 & 0.230 & 0.257 \\
ERA5 & 0.082 & 0.108 & 0.130 & 0.162 & 0.182 \\
Cross-dataset mean & 0.112 & 0.143 & 0.166 & 0.199 & 0.222 \\
\addlinespace[5pt]
\bottomrule
\end{tabular}
\end{adjustbox}
\end{table}

\begin{table}[tbp]
\centering\small
\setlength{\tabcolsep}{6pt}
\caption{\textbf{Sensitivity to the order of the state-space trend specification.}
Endpoint point estimates for the baseline model, $\mathrm{d}a(t)=\sigma\,\mathrm{d}W(t)$,
and the alternative model, $\mathrm{d}v(t)=\sigma\,\mathrm{d}W(t)$.
Both models use the same monthly samples and climate covariates, with an
ARMA(1,1) irregular component and calendar-month-specific innovation variances.
The diffusion parameter and remaining parameters are estimated by maximum
likelihood separately for each model and record. Endpoints are August 2026,
except July 2026 for HadCRUT5. Level is in $^{\circ}$C and rate in
$^{\circ}$C decade$^{-1}$. Only point estimates are reported; bootstrap intervals
were not computed for the alternative model. The cross-dataset mean averages
the five individual estimates.}\label{tab:ss_velocity_robustness}
\begin{adjustbox}{max width=\linewidth}
\begin{tabular}{lrrrr}
\toprule
 & \multicolumn{2}{c}{Warming level $x(T)$} & \multicolumn{2}{c}{Warming rate $v(T)$}\\
Dataset & Baseline & Alternative & Baseline & Alternative\\\midrule
NASA & 1.437 & 1.442 & 0.504 & 0.465 \\
NOAA & 1.410 & 1.401 & 0.433 & 0.364 \\
HadCRUT5 & 1.381 & 1.369 & 0.394 & 0.337 \\
Berkeley Earth & 1.414 & 1.398 & 0.473 & 0.389 \\
ERA5 & 1.457 & 1.455 & 0.427 & 0.390 \\
\addlinespace[2pt]Cross-dataset mean & 1.420 & 1.413 & 0.446 & 0.389 \\
\bottomrule
\end{tabular}
\end{adjustbox}
\end{table}

 \FloatBarrier
\begin{landscape}
\begingroup\scriptsize\setlength{\tabcolsep}{3pt}
\begin{longtable}{lcccccccc}
\caption{\textbf{Alternative averaging intervals.} Average acceleration from January of the indicated year to the monthly endpoint, in $^{\circ}$C decade$^{-2}$. Both flexible estimators use 1,999 paired bootstrap draws. Quadratic intervals use HAC errors; changepoint intervals use 1,999 ARMA(1,1) draws with breakpoint selection repeated. Rates at a break use the right-hand slope, so a break at the starting date is excluded from the contrast.}\label{tab:rate_change_alt_t0}\\
\toprule Dataset & \multicolumn{2}{c}{Semiparametric} & \multicolumn{2}{c}{Nonparametric} & \multicolumn{2}{c}{Quadratic} & \multicolumn{2}{c}{Changepoint}\\
 & Est. & 90\% CI & Est. & 90\% CI & Est. & 90\% CI & Est. & 90\% CI\\\midrule
\endfirsthead
\toprule Dataset & \multicolumn{2}{c}{Semiparametric} & \multicolumn{2}{c}{Nonparametric} & \multicolumn{2}{c}{Quadratic} & \multicolumn{2}{c}{Changepoint}\\
 & Est. & 90\% CI & Est. & 90\% CI & Est. & 90\% CI & Est. & 90\% CI\\\midrule
\endhead
\multicolumn{9}{l}{\textbf{Starting January 2010}}\\*
NASA & 0.186 & [0.127, 0.330] & 0.198 & [0.054, 0.264] & 0.035 & [0.024, 0.045] & 0.120 & [0.000, 0.191]\\*
NOAA & 0.140 & [0.066, 0.248] & 0.137 & [0.037, 0.227] & 0.035 & [0.026, 0.044] & 0.116 & [0.000, 0.170]\\*
HadCRUT5 & 0.119 & [0.052, 0.240] & 0.158 & [0.022, 0.236] & 0.023 & [0.014, 0.033] & 0.090 & [0.000, 0.182]\\*
Berkeley Earth & 0.170 & [0.106, 0.303] & 0.172 & [0.033, 0.260] & 0.024 & [0.014, 0.033] & 0.098 & [0.000, 0.189]\\*
ERA5 & 0.121 & [0.076, 0.250] & 0.160 & [0.029, 0.247] & 0.035 & [0.024, 0.045] & 0.127 & [0.000, 0.217]\\*
Cross-dataset mean & 0.147 &  & 0.165 &  & 0.030 &  & 0.110 & \\\addlinespace[5pt]
\multicolumn{9}{l}{\textbf{Starting January 2012}}\\*
NASA & 0.199 & [0.136, 0.366] & 0.213 & [0.051, 0.284] & 0.035 & [0.024, 0.045] & 0.137 & [0.000, 0.217]\\*
NOAA & 0.146 & [0.066, 0.269] & 0.144 & [0.031, 0.244] & 0.035 & [0.026, 0.044] & 0.132 & [0.000, 0.194]\\*
HadCRUT5 & 0.125 & [0.051, 0.260] & 0.170 & [0.017, 0.256] & 0.023 & [0.014, 0.033] & 0.102 & [0.000, 0.207]\\*
Berkeley Earth & 0.179 & [0.110, 0.327] & 0.184 & [0.029, 0.281] & 0.024 & [0.014, 0.033] & 0.112 & [0.000, 0.215]\\*
ERA5 & 0.125 & [0.076, 0.264] & 0.170 & [0.022, 0.266] & 0.035 & [0.024, 0.045] & 0.145 & [0.000, 0.247]\\*
Cross-dataset mean & 0.155 &  & 0.176 &  & 0.030 &  & 0.125 & \\\addlinespace[5pt]
\multicolumn{9}{l}{\textbf{Starting January 2014}}\\*
NASA & 0.210 & [0.137, 0.397] & 0.230 & [0.046, 0.306] & 0.035 & [0.024, 0.045] & 0.159 & [0.000, 0.252]\\*
NOAA & 0.150 & [0.059, 0.287] & 0.151 & [0.024, 0.262] & 0.035 & [0.026, 0.044] & 0.000 & [0.000, 0.225]\\*
HadCRUT5 & 0.130 & [0.047, 0.280] & 0.184 & [0.011, 0.280] & 0.023 & [0.014, 0.033] & 0.118 & [0.000, 0.240]\\*
Berkeley Earth & 0.186 & [0.109, 0.350] & 0.197 & [0.021, 0.306] & 0.024 & [0.014, 0.033] & 0.129 & [0.000, 0.249]\\*
ERA5 & 0.128 & [0.070, 0.277] & 0.182 & [0.013, 0.288] & 0.035 & [0.024, 0.045] & 0.168 & [0.000, 0.287]\\*
Cross-dataset mean & 0.161 &  & 0.189 &  & 0.030 &  & 0.115 & \\\addlinespace[5pt]
\multicolumn{9}{l}{\textbf{Starting January 2015}}\\*
NASA & 0.215 & [0.137, 0.413] & 0.240 & [0.044, 0.321] & 0.035 & [0.024, 0.045] & 0.000 & [0.000, 0.273]\\*
NOAA & 0.151 & [0.050, 0.296] & 0.154 & [0.017, 0.274] & 0.035 & [0.026, 0.044] & 0.000 & [0.000, 0.243]\\*
HadCRUT5 & 0.132 & [0.045, 0.290] & 0.191 & [0.006, 0.293] & 0.023 & [0.014, 0.033] & 0.000 & [0.000, 0.261]\\*
Berkeley Earth & 0.189 & [0.107, 0.361] & 0.203 & [0.018, 0.320] & 0.024 & [0.014, 0.033] & 0.000 & [0.000, 0.271]\\*
ERA5 & 0.129 & [0.068, 0.282] & 0.188 & [0.008, 0.302] & 0.035 & [0.024, 0.045] & 0.000 & [0.000, 0.311]\\*
Cross-dataset mean & 0.163 &  & 0.195 &  & 0.030 &  & 0.000 & \\\addlinespace[5pt]
\multicolumn{9}{l}{\textbf{Starting January 2016}}\\*
NASA & 0.219 & [0.135, 0.429] & 0.249 & [0.040, 0.336] & 0.035 & [0.024, 0.045] & 0.000 & [0.000, 0.298]\\*
NOAA & 0.152 & [0.045, 0.307] & 0.156 & [0.013, 0.286] & 0.035 & [0.026, 0.044] & 0.000 & [0.000, 0.256]\\*
HadCRUT5 & 0.134 & [0.041, 0.298] & 0.198 & [0.0004, 0.308] & 0.023 & [0.014, 0.033] & 0.000 & [0.000, 0.285]\\*
Berkeley Earth & 0.192 & [0.105, 0.373] & 0.210 & [0.013, 0.335] & 0.024 & [0.014, 0.033] & 0.000 & [0.000, 0.295]\\*
ERA5 & 0.130 & [0.064, 0.289] & 0.194 & [0.001, 0.314] & 0.035 & [0.024, 0.045] & 0.000 & [0.000, 0.340]\\*
Cross-dataset mean & 0.165 &  & 0.201 &  & 0.030 &  & 0.000 & \\\addlinespace[5pt]
\multicolumn{9}{l}{\textbf{Starting January 2018}}\\*
NASA & 0.225 & [0.125, 0.459] & 0.267 & [0.030, 0.370] & 0.035 & [0.024, 0.045] & 0.000 & [0.000, 0.000]\\*
NOAA & 0.153 & [0.027, 0.326] & 0.157 & [0.001, 0.311] & 0.035 & [0.026, 0.044] & 0.000 & [0.000, 0.000]\\*
HadCRUT5 & 0.136 & [0.033, 0.313] & 0.211 & [-0.011, 0.338] & 0.023 & [0.014, 0.033] & 0.000 & [0.000, 0.345]\\*
Berkeley Earth & 0.195 & [0.092, 0.394] & 0.221 & [-0.001, 0.365] & 0.024 & [0.014, 0.033] & 0.000 & [0.000, 0.354]\\*
ERA5 & 0.131 & [0.062, 0.299] & 0.205 & [-0.017, 0.344] & 0.035 & [0.024, 0.045] & 0.000 & [0.000, 0.377]\\*
Cross-dataset mean & 0.168 &  & 0.212 &  & 0.030 &  & 0.000 & \\\addlinespace[5pt]
\multicolumn{9}{l}{\textbf{Starting January 2020}}\\*
NASA & 0.228 & [0.105, 0.480] & 0.284 & [0.015, 0.411] & 0.035 & [0.024, 0.045] & 0.000 & [0.000, 0.000]\\*
NOAA & 0.153 & [0.003, 0.342] & 0.154 & [-0.016, 0.346] & 0.035 & [0.026, 0.044] & 0.000 & [0.000, 0.000]\\*
HadCRUT5 & 0.137 & [0.016, 0.328] & 0.223 & [-0.028, 0.380] & 0.023 & [0.014, 0.033] & 0.000 & [0.000, 0.000]\\*
Berkeley Earth & 0.197 & [0.077, 0.410] & 0.230 & [-0.022, 0.406] & 0.024 & [0.014, 0.033] & 0.000 & [0.000, 0.000]\\*
ERA5 & 0.131 & [0.058, 0.309] & 0.211 & [-0.035, 0.377] & 0.035 & [0.024, 0.045] & 0.000 & [0.000, 0.000]\\*
Cross-dataset mean & 0.169 &  & 0.220 &  & 0.030 &  & 0.000 & \\\addlinespace[5pt]
\bottomrule\end{longtable}\endgroup
\end{landscape}\FloatBarrier

\begin{table}[htbp]
\centering\small
\setlength{\tabcolsep}{5pt}
\caption{Semiparametric estimates without climate covariates, with nominal 90\% bootstrap-error intervals from 1,999 replications for each record. Level is in $^{\circ}$C, rate in $^{\circ}$C decade$^{-1}$, and both acceleration estimands in $^{\circ}$C decade$^{-2}$. The arithmetic-mean row fits the mean temperature series; it is not an average of the individual estimates.}\label{tab:no_covariates}
\begin{adjustbox}{max width=\linewidth}
\begin{tabular}{lrrrr}\toprule
 & \multicolumn{2}{c}{Warming level} & \multicolumn{2}{c}{Warming rate}\\
Dataset & Estimate & 90\% CI & Estimate & 90\% CI\\\midrule
NASA & 1.477 & [1.403, 1.588] & 0.461 & [0.347, 0.699]\\
NOAA & 1.452 & [1.380, 1.553] & 0.410 & [0.304, 0.610]\\
HadCRUT5 & 1.434 & [1.349, 1.545] & 0.362 & [0.236, 0.554]\\
Berkeley Earth & 1.466 & [1.390, 1.583] & 0.442 & [0.339, 0.684]\\
ERA5 & 1.487 & [1.411, 1.570] & 0.348 & [0.283, 0.415]\\
Arithmetic mean & 1.457 & [1.382, 1.568] & 0.408 & [0.301, 0.612]\\
\addlinespace[6pt]
 & \multicolumn{2}{c}{Endpoint acceleration} & \multicolumn{2}{c}{Past-decade average}\\
Dataset & Estimate & 90\% CI & Estimate & 90\% CI\\\midrule
NASA & 0.158 & [0.003, 0.450] & 0.152 & [0.058, 0.376]\\
NOAA & 0.116 & [-0.023, 0.343] & 0.114 & [0.025, 0.291]\\
HadCRUT5 & 0.084 & [-0.074, 0.312] & 0.082 & [-0.020, 0.253]\\
Berkeley Earth & 0.143 & [0.001, 0.419] & 0.138 & [0.053, 0.355]\\
ERA5 & 0.048 & [0.015, 0.082] & 0.048 & [0.018, 0.078]\\
Arithmetic mean & 0.112 & [-0.035, 0.348] & 0.110 & [0.020, 0.291]\\
\addlinespace[6pt]
\bottomrule\end{tabular}
\end{adjustbox}\end{table}

\clearpage

\section{Conditional extrapolations to higher warming levels}
\label{app:extrapolation}

To illustrate the implications of differences in the estimated current warming rate and acceleration, we calculate when simple continuations of the fitted paths would reach \(1.5\,^{\circ}\mathrm{C}\) and \(2\,^{\circ}\mathrm{C}\) above the adopted pre-industrial reference. We use the separately fitted arithmetic-mean series, ending in July 2026, and the endpoint estimates reported in Table~\ref{tab:main}. The calculation concerns the smooth warming component \(x(t)\), without adding future covariate effects or irregular variation.

Let \(h\geq0\) denote time after the sample endpoint, measured in decades, and write \(\hat x_T\), \(\hat v_T\) and \(\hat a_T\) for the estimated endpoint level, rate and acceleration, with derivatives expressed in degrees Celsius per decade and per decade squared. A continuation holding endpoint acceleration constant is
\begin{equation}
 \tilde x(T+10h)=\hat x_T+\hat v_T h+\tfrac12\hat a_T h^2.
 \label{eq:extrapolation_path}
\end{equation}
For any target warming level \(L>\hat x_T\), the crossing time for the estimates considered here is $\tilde T_L=T+10h_L$ with $h_L$ given by
\begin{equation}
 h_L=\begin{cases}
 (L-\hat x_T)/\hat v_T, & \hat a_T=0,\\[6pt]
 \displaystyle\frac{2(L-\hat x_T)}{\hat v_T+\sqrt{\hat v_T^2+2\hat a_T(L-\hat x_T)}}, & \hat a_T>0.
 \end{cases}
 \label{eq:extrapolation_crossing}
\end{equation}
We evaluate Equation~\eqref{eq:extrapolation_crossing} at \(L=1.5\,^{\circ}\mathrm{C}\) and \(L=2\,^{\circ}\mathrm{C}\), using the same endpoint estimates for both thresholds. The fitted quadratic continues with its constant acceleration, whereas the changepoint model continues at its post-break rate without another break. For the semiparametric and nonparametric estimates, we consider both a constant-rate continuation and a continuation maintaining endpoint acceleration. We use endpoint acceleration, not the past-decade average, in Equation~\eqref{eq:extrapolation_path}. Figure~\ref{fig:continuations_constant_rate} shows the conditional paths under both continuation rules.

Table~\ref{tab:15c_continuations} reports the conditional crossings of \(1.5\,^{\circ}\mathrm{C}\). The semiparametric continuations reach this level around mid-2028, compared with late 2028 under the changepoint continuation and late 2030 under the quadratic continuation. The nonparametric continuations reach it around mid-2027. Maintaining semiparametric endpoint acceleration advances the crossing by less than one month relative to holding the current rate constant. At this short horizon, the estimated warming level and rate therefore dominate the calculation, with little time for acceleration to accumulate.

Table~\ref{tab:2c_continuations} reports the conditional crossings of \(2\,^{\circ}\mathrm{C}\). Holding the semiparametric endpoint rate constant gives a crossing around 2040, approximately three years earlier than the changepoint continuation. Maintaining the semiparametric endpoint acceleration advances the crossing by more than two years, to around mid-2037. Thus, much of the difference from the changepoint continuation arises from the higher current rate, without requiring acceleration to persist. The nonparametric continuation maintaining endpoint acceleration reaches the level around 2036, while the quadratic continuation reaches it around 2047.

The primary calculation starts each continuation at its own estimated warming level. To isolate differences in the derivatives in the \(2\,^{\circ}\mathrm{C}\) comparison, we also repeat it from the common semiparametric endpoint level, \(\hat x_T=1.414593\,^{\circ}\mathrm{C}\), retaining each method's rate and acceleration. The quadratic, changepoint and nonparametric constant-acceleration crossing dates then become approximately 2046, 2043 and 2036, respectively; the two semiparametric dates are unchanged. The ordering and broad differences therefore do not depend on using different initial warming levels.

These calculations are conditional point illustrations, not probabilistic forecasts or emissions scenarios. They hold the estimated rate or acceleration fixed and do not propagate estimation uncertainty or uncertainty about future changes in the dynamics. They also refer to the smooth observational warming component, rather than an attributed response or the 20-year-average warming level used in assessments of threshold crossings \citep{KirchengastPichler2025}. Their purpose is to quantify the implications of the estimated current dynamics, not to establish that those dynamics will persist.

\clearpage

\begin{table}[htbp]
\centering
\small
\caption{\textbf{Conditional continuations to \(1.5\,^{\circ}\mathrm{C}\) for the arithmetic-mean series.} Initial warming levels are in degrees Celsius above the adopted pre-industrial reference; rates and accelerations are in degrees Celsius per decade and per decade squared. All continuations begin in July 2026. The constant-rate calculations set future acceleration to zero while retaining each estimator's fitted endpoint level and rate.}
\label{tab:15c_continuations}
\begin{adjustbox}{max width=\linewidth}
\begin{tabular}{p{6.5cm}rrrr}
\toprule
Continuation & Level & Rate & Acceleration & Crossing year \\
\midrule
Quadratic continuation & 1.374 & 0.280 & 0.030 & 2030.9 \\
Changepoint continuation & 1.424 & 0.351 & 0.000 & 2028.7 \\
Semiparametric, constant rate & 1.415 & 0.444 & 0.000 & 2028.5 \\
Semiparametric, constant acceleration & 1.415 & 0.444 & 0.166 & 2028.4 \\
Nonparametric, constant rate & 1.453 & 0.500 & 0.000 & 2027.5 \\
Nonparametric, constant acceleration & 1.453 & 0.500 & 0.204 & 2027.5 \\
\bottomrule
\end{tabular}
\end{adjustbox}
\end{table}

\begin{table}[htbp]
\centering
\small
\caption{\textbf{Conditional continuations to \(2\,^{\circ}\mathrm{C}\) for the arithmetic-mean series.} Initial warming levels are in degrees Celsius above the adopted pre-industrial reference; rates and accelerations are in degrees Celsius per decade and per decade squared. All continuations begin in July 2026. The constant-rate calculations set future acceleration to zero while retaining each estimator's fitted endpoint level and rate.}
\label{tab:2c_continuations}
\begin{adjustbox}{max width=\linewidth}
\begin{tabular}{p{6.5cm}rrrr}
\toprule
Continuation & Level & Rate & Acceleration & Crossing year \\
\midrule
Quadratic continuation & 1.374 & 0.280 & 0.030 & 2046.7 \\
Changepoint continuation & 1.424 & 0.351 & 0.000 & 2042.9 \\
Semiparametric, constant rate & 1.415 & 0.444 & 0.000 & 2039.7 \\
Semiparametric, constant acceleration & 1.415 & 0.444 & 0.166 & 2037.5 \\
Nonparametric, constant rate & 1.453 & 0.500 & 0.000 & 2037.5 \\
Nonparametric, constant acceleration & 1.453 & 0.500 & 0.204 & 2035.7 \\
\bottomrule
\end{tabular}
\end{adjustbox}
\end{table}

\end{document}